\pdfoutput=1
\documentclass[aps,prd,superscriptaddress,amsmath,amssymb,twocolumn,fleqn,nofootinbib]{revtex4-1} 
\usepackage{xcolor,colortbl}
\usepackage{amsmath,graphicx}

\usepackage{xcolor}
\usepackage{graphicx}
\usepackage{colordvi}
\usepackage{mathtools}
\usepackage{empheq}
\usepackage{bm}%

\usepackage{floatrow}
\usepackage{braket}
\usepackage{soul}

\usepackage{amssymb}
\usepackage{amsmath}

\usepackage{hyperref}

\def\CL{{\cal L}}

\def\high{\vphantom{\Biggl(}\displaystyle}

\def\mpl{M_{\rm Pl}}
\def\half{\frac{1}{2}}

\definecolor{Gray}{gray}{0.85}
\definecolor{LightCyan}{rgb}{0.88,1,1}

\newcolumntype{a}{>{\columncolor{Gray}}c}
\newcolumntype{b}{>{\columncolor{white}}c}

\begin{document}
\newcommand{\be}{\begin{equation}}
\newcommand{\ee}{\end{equation}}
\newcommand{\bea}{\begin{eqnarray}}
\newcommand{\eea}{\end{eqnarray}}
\newcommand{\barr}{\begin{array}}
\newcommand{\earr}{\end{array}}
\def\bal#1\eal{\begin{align}#1\end{align}}

\newcommand{\calR}{{\cal{R}}}
\newcommand{\calP}{{\cal{P}}}
\newcommand{\calH}{{\cal{H}}}
\newcommand{\calI}{{\cal{I}}}




\title{Primordial Standard Clock Models and CMB Residual Anomalies}

\author{Matteo~Braglia}\email{matteo.braglia@csic.es}
\affiliation{Instituto de Fisica Teorica, Universidad Autonoma de Madrid, Madrid, 28049, Spain}

\affiliation{INAF/OAS Bologna, via Gobetti 101, I-40129 Bologna, Italy}

\author{Xingang~Chen}\email{xingang.chen@cfa.harvard.edu}
\affiliation{Institute for Theory and Computation, Harvard-Smithsonian Center for Astrophysics, 60
Garden Street, Cambridge, MA 02138, USA}

\author{Dhiraj~Kumar~Hazra}\email{dhiraj@imsc.res.in}
\affiliation{The  Institute  of  Mathematical  Sciences,  HBNI,  CIT  Campus, Chennai  600113,  India}
\affiliation{INAF/OAS Bologna, via Gobetti 101, I-40129 Bologna, Italy}




\date{\today}
\begin{abstract}
The residuals of the power spectra of WMAP and Planck's cosmic microwave background (CMB) anisotropies data are known to exhibit a few interesting anomalies at different scales with marginal statistical significance.
Combining bottom-up and top-down model-building approaches and using a pipeline that efficiently compares model predictions with data, we construct a model of primordial standard clock that is able to link and address the anomalies at both the large and small scales. This model, and its variant, provide some of the best fits to the feature anomalies in CMB. According to Bayes evidences, these models are currently statistically indistinguishable from the Standard Model. We show that the difference between them will soon become statistically significant with various higher quality data on the CMB polarization.
We demonstrate that such a model-building and data-analyses process may be used to uncover a portion of detailed evolutionary history of our universe during its primordial epoch.
\end{abstract}

\pacs{Valid PACS appear here}
\maketitle

\section{Introduction}
\label{Sec:Introduction}
\setcounter{equation}{0}

Current observations of cosmic microwave background (CMB) and large-scale structure (LSS) show that the structure in our Universe is seeded by primordial density perturbations that are approximately scale-invariant, Gaussian and adiabatic.
This has become part of the experimental foundation of the Standard Model of Cosmology and has provided valuable information about the primordial Universe that started the Big Bang.
Future observations of CMB polarization and galaxy surveys are expected to significantly improve the precisions of the measurements of the density perturbations, from which we hope to see signs of beyond-Standard-Model physics that help us understand the nature of the primordial Universe.
One of the most important such possibilities is signals of primordial features.

Primordial feature signals are strongly-scale-dependent deviations from the overall approximately scale-invariant spectra. They may carry rich information about the primordial Universe, see \cite{Chen:2010xka,Chluba:2015bqa,Slosar:2019gvt} for reviews.
Although currently there is no evidence for any primordial feature signals in CMB anisotropies, the Planck data exhibits several feature-like anomalies at both large and small scales that have marginal statistical significance \cite{Akrami:2018odb}. This has generated great interest in the literature because they may be candidates of primordial features.

There are roughly two types of anomalies. One appears at large scales around the multipole $\ell \sim 25$ in the TT spectrum. This has been noticed since the WMAP results \cite{Peiris:2003ff}. Another appears at small scales, most significantly around $\ell \sim 750$ in the TT and TE spectra, observed by the Planck satellite \cite{Akrami:2018odb,Hazra:2014jwa}. Many models have been proposed as possible explanations to these anomalies. So far, the proposed models are only able to address either one of these anomalies. See \cite{Braglia:2021sun} for a brief summary. On the other hand, it has been pointed out in \cite{Chen:2014joa,Chen:2014cwa} that these two anomalies may be connected by the classical primordial standard clock (CPSC) effect \cite{Chen:2011zf,Chen:2011tu,Chen:2012ja}.

CPSC models are a class of models in which the inflaton\footnote{CPSC models can exist in both inflation and alternative-to-inflation scenarios. In this paper, we only study the inflationary models.} gets temporarily displaced from the attractor solution by some sharp feature in its trajectory. Under certain conditions, the displacement can trigger oscillation of a massive field that has a mass much larger than the Hubble parameter. In the primordial spectrum of the density perturbations, the sharp feature generates a signal located at relatively larger scales with sinusoidal scale-dependence, and the oscillation of the massive field generates a resonant-running clock-signal at smaller scales.

Despite the proposal, such a possibility has not been explicitly realized in a full model and the CPSC models constructed previously \cite{Chen:2014joa,Chen:2014cwa} only attempted to address the small scale anomaly. Recently, in \cite{Braglia:2021sun} we have constructed such a model and demonstrated how it provides interesting fits to both feature anomalies in the Planck data across the wide scales. Not only does such a model contain the clock signal that directly measures the evolutionary history of the background scale factor $a(t)$, but also it shows a clip of very detailed evolutionary history of the inflationary dynamics. In this companion paper, we will provide more extensive and detailed analyses of the model, including motivations for the model-building, analytical and numerical predictions from the model, comparison with different Planck likelihoods, and comparison with other models. We will demonstrate that such a model, and its variant, provide some of the best fits to the feature anomalies observed in the Planck data, although statistically they are as significant as the Standard Model due to the introduction of many extra parameters. We will also demonstrate the future prospects of such model-building efforts by showing how several future observations in CMB polarization will be able to constrain the main characteristic properties of such models.
 
This paper is organized as follows.  
In Sec.~\ref{Sec:Models}, we review general properties of the CPSC models, the model-building requirements, motivations for the ingredients of the model that we will present in the paper. We then present a full model and a special limit of this model (i.e.~the restricted model), with full numerical predictions. In Sec.~\ref{Sec:data_analysis_method} we describe our method of data analysis. In Sec.~\ref{sec:results} and \ref{sec:results_restricted} we present the results of data analyses for both the full and restricted model. 
In Sec.~\ref{sec:forecasts}, we forecast how the constraints will be improved with future observations of CMB polarization. We conclude in Sec.~\ref{Sec:Conclusion}. The Appendices contain several additional details of the comments mentioned in the main text.

\section{The models}
\label{Sec:Models}

\subsection{Model-building requirements and CPSC signals}
\label{Sec:CPSC models}

In this subsection, we briefly summarize the basic model-building requirements for CPSC models, different degrees of model-dependency in various properties of the full CPSC signal, and their usages in probing the physics of the primordial Universe. More details can be found in \cite{Chen:2011zf,Chen:2011tu,Chen:2014cwa}.

CPSC models have two simple model-building requirements. First, there are two observable inflationary\footnote{Or whichever mechanism used as an alternative to inflation.} stages joined by a sharp feature. Second, a massive field is excited classically by the sharp feature and oscillating. Note that, during the whole process, the background is always dominantly inflation; the sharp feature and oscillating mass field are both small perturbations.

The full CPSC signal has two components. The sharp feature signal generated by the sharp feature is located at larger scales, while the clock signal generated by the oscillating massive field is located at shorter scales. The smooth connection between the two components can usually only be described by numerical results.

The properties of the clock signal, including both its oscillatory resonant running and the envelop of the oscillation as functions of the wavenumber $k$, is quite model-independent once the scenario of the primordial Universe is given. With the assumptions of approximately constant mass and coupling, these properties are determined by the background scale factor $a(t)$ only. The oscillatory running of the sharp feature signal, being an overall sinusoidal function of $k$, is also robust and even universal for different scenarios.

On the other hand, the envelop and phase of the sinusoidal running in the sharp feature signal are highly model-dependent. They depend on detailed properties of the sharp feature, including its type, size and sharpness. Also, the relative size between the amplitudes of the sharp feature and clock signal is highly model-dependent. It depends on model configurations at the sharp feature and in the trajectory where the massive field is oscillating.

The different levels of model-dependencies of these components lead to their usages in probing different properties of the primordial universe. The model-independent clock signal can be used to measure $a(t)$ directly and distinguish the inflation and alternative scenarios in a model-independent fashion. The high model-sensitivity of the sharp feature signal can be used to extract a clip of rather detailed evolutionary history of the primordial Universe.
This will be an emphasis of this paper.

\subsection{Sensitivity on background evolution}
\label{Sec:Sensitivity}

As we have just summarized, the full predictions of primordial feature models sensitively depend on the details of the background evolution. To build a model starting from the Lagrangian and place various model ingredients into place, it often helps to combine both the bottom-up approach, in which we examine the properties of feature-signal candidates in data \cite{Peiris:2003ff,Akrami:2018odb}, and the top-down approach, in which we classify the properties of various feature models \cite{Chen:2010xka}.

Before we present our model, we would like to review and discuss the past efforts related to the model-building of the CPSC models and their relevance in explaining the anomalies in the CMB spectra. This also helps to illustrate the points we mentioned above.

As discussed, with the two characteristically different signals, CPSC models offer an interesting model-building possibility to address both the large and small-scale anomalies in the CMB power spectra. To build such a model, we need to construct certain sharp features that can make a massive field oscillate classically. We also need to compute the imprints of these features in the density perturbations to test whether they belong to the type that is more likely to explain the anomalies appearing in the data.

One can imagine a variety of ways the inflaton can get temporarily disturbed from its eventual attractor evolution, and subsequently excite the oscillation of a massive field generating a clock signal. Several such possibilities have been explored in the literature. 

One possibility is to have two straight potential valleys connect to each other at an angle. If this junction is sharp enough and the massive field is sufficiently heavy, the inflaton would climb onto a side of cliff near the junction point and excite the oscillation of the massive field. In such a model, Ref.~\cite{Gao:2013ota,Noumi:2013cfa} have shown that, due to the lack of a direct coupling between the inflaton and massive field, the amplitude of the clock signal in this model is too small and, phenomenologically in the power spectrum, we end up with just a pure sharp feature model. In addition, such a sharp feature produces a prominent bump instead of dip in the power spectrum. The lessons to be learned from these results are that, for the sharp feature, a smoother turning of the trajectory than sharp bend is needed to avoid the bump feature in power spectrum; for the clock signal, a direct coupling, such as a curved path (or derivative couplings) \cite{Chen:2009we,Chen:2009zp,Saito:2012pd,Saito:2013aqa}, needs to be introduced.

Regarding the dip feature in the power spectrum that is present near $\ell\sim 25$ in CMB, the simplest way to generate that is to have a step in the potential \cite{Peiris:2003ff,Adams:2001vc, Bean:2008na, Mortonson:2009qv,Hazra:2010ve, Hazra:2014goa, Miranda:2014fwa}. Although all sharp features (e.g.~\cite{Starobinsky:1992ts,Adams:2001vc,Chen:2006xjb,Bean:2008na,Hazra:2010ve,Achucarro:2010da,Adshead:2011jq,Gao:2012uq,Miranda:2012rm, Miranda:2014fwa, Bartolo:2013exa, Hazra:2014goa, Fumagalli:2020nvq,Braglia:2020taf,Hazra:2021eqk}) generate a characteristic sinusoidal-running signal which extends to smaller scales, what is phenomenologically important here is the phase and envelop at the largest scales. The brief acceleration of the inflaton induced by the step in the potential generates a dip in the power spectrum, with its depth determined by model details. In contrast, a bump or kink in the potential would generate different types of large-scale profiles, which may be also compatible with the current data but does not explain the dip feature at the large scale.

Based on these results, we naturally would like to construct a model in which the inflaton first falls down a step in the potential as (part of) the sharp feature, and then enters a curved path through a smooth junction.

The kind of models that combine the above ingredients have indeed been tested in Ref.~\cite{Chen:2014joa,Chen:2014cwa}. However, it turns out that the placement of the step in the whole model configuration is phenomenologically also very important. 
In \cite{Chen:2014joa,Chen:2014cwa}, this step potential is placed at the side of a curved potential valley. Although the clock signal indeed becomes more prominent (and so the model is useful as a candidate for addressing the small-scale anomaly), the dip feature that the step generates is too shallow to provide any meaningful fit to the large-scale anomaly.

Therefore, none of the explicit CPSC models constructed previously has succeeded in addressing the CMB anomalies in both scales.  In this work, and the companion paper \cite{Braglia:2021sun}, we make some further adjustments to this type of models. We rearrange the relative position of the step potential and make it directly face the entrance of the curved path. We also leave a short distance between the end of the step and the entrance of the curved path, because even such a subtle detail can generate a difference that improves the fit to the current data and may be detectable in future observations.

Equally importantly, an efficient data-analysis pipeline \cite{Braglia:2021ckn} plays an important role in making possible the progress presented in this work.
Since it is usually very difficult to capture the results of complicated multi-field feature models in terms of analytical templates, in previous efforts only a very small subset of predictions of such models could be compared with data. 
The model selection process was inefficient given the high sensitivity of the feature model predictions to the background model-building details, as illustrated above.
The methodological pipeline tested in \cite{Braglia:2021ckn} assembles several methods that have been developed in literature on numerical \cite{Lalak:2007vi,Braglia:2020fms} and data analysis methods \cite{Handley:2015fda,Handley:2015vkr,BOBYQA}, and adapts them for the case of oscillatory features. 
The pipeline also introduces the usage of the effective parameters as a bridge between a model and data.
This pipeline enables us to directly compare the full numerical predictions of complicated feature models with data very efficiently.

\subsection{A model-building guideline}
\label{Sec:guideline}

Because feature models usually come with many adjustable details, there is a model-building question on which of these details should be described by free parameters and which can be built in without introducing any parameters. We find it reasonable to follow the following guideline. 

If a property of a beyond-Standard-Model signal (or candidate signal) in data needs to be addressed by a parameter taking a finite value, such a parameter should be made freely variable. If such a property can be explained as long as this parameter satisfies an upper or a lower bound, we can take a special limit or fix the value of this parameter, and do not treat it as a free parameter.

This guideline is also supported by the way the Bayesian evidence of a particular model is determined. When varying a parameter within its prior, if there are some values that reduce or increase the likelihood of the model, there will be a penalty or award, respectively, in the Bayesian evidence for introducing this extra parameter. On the other hand, if this variation has no effect on the likelihood of the model, the introduction of this extra parameter does not change the Bayesian evidence of the model.

This guideline will also be illustrated in our model-building.

\subsection{A full model}

\begin{figure*}
\includegraphics[width=.9\columnwidth]{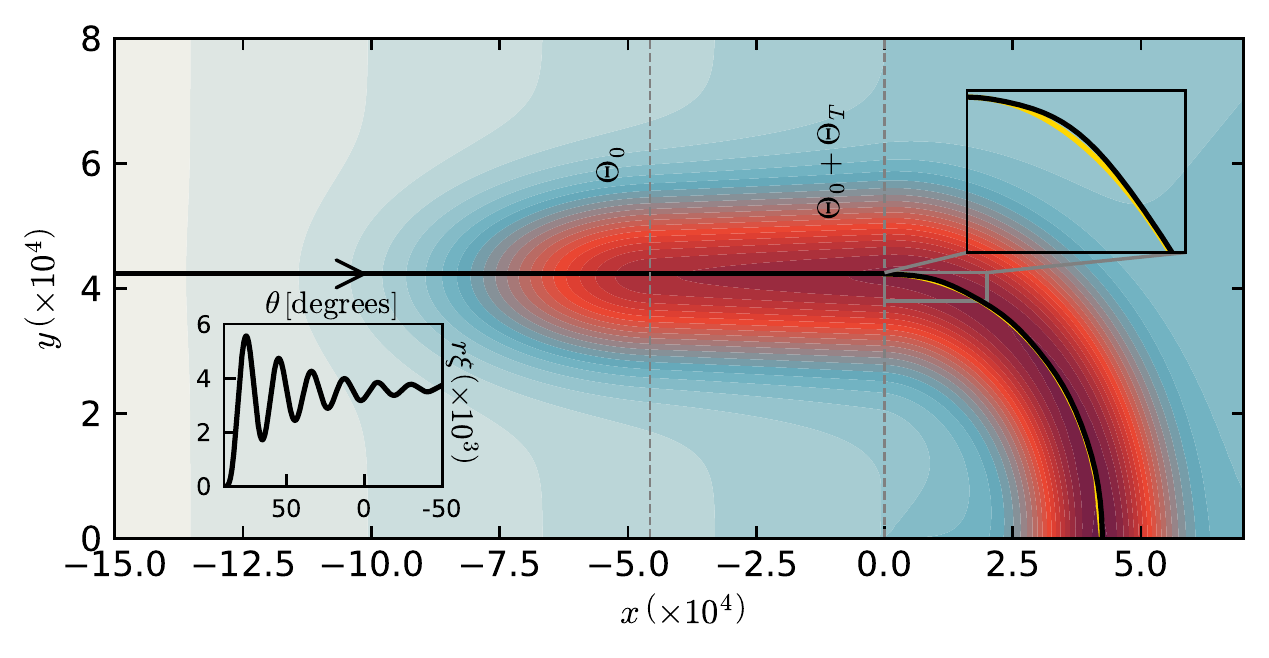}
\caption{\label{fig:trajectory}
A bird's eye view of the model configuration and the background trajectory, plotted over equipotential surfaces (the redder color corresponds to lower potential). In terms of the Cartesian coordinates $x$ and $y$ shown here, for $x<0$, the $\Theta$ and $\sigma$ are Cartesian with $x=\Theta-(\Theta_T+\Theta_0)$ and $y=\sigma+\xi^{-1}$, so the straight solid line at the top of the plateau represents $\sigma=0$. For $x>0$, the $\Theta$ and $\sigma$ become radial coordinates with $r=\sigma+\xi^{-1}$ and $\theta=\pi/2-\xi(\Theta-\Theta_0-\Theta_T)$. Together with the non-canonical coupling \eqref{eq:noncanonical_coupling}, they describe a curved path at the bottom of the potential valley, as in \cite{Chen:2009we,Chen:2009zp}. The insets show how the inflaton overshoots the bottom of the valley (yellow line) and climbs onto a side of the cliff [top-right], and the clock oscillation [bottom-left].
}
\end{figure*}

Our full model is described by the following Lagrangian
\bal
\CL =& -\half \left(1+ \Xi(\Theta) \sigma\right)^2 (\partial\Theta)^2
-\half (\partial \sigma)^2
- V(\Theta, \sigma) ~,
\eal
where the potential $V(\Theta,\sigma)$ takes the following form,
\begin{widetext}
\bal
V (\Theta, \sigma) = V_{\rm inf} 
\begin{cases}	\label{eq:potdip}
\high{ 
	1- \half C_\Theta \Theta^2 
	+ C_{\sigma} \left[ 1-\exp \left( -  \frac{(\Theta-\Theta_0)^2}{\Theta_f^2}  -\frac{\sigma^2}{\sigma_f^2} \right) \right]
}
& \text{for} ~~ \Theta \le \Theta_0 ~, \\
\high{ 1 - \half C_\Theta \Theta^2 
	+ C_{\sigma} \left[ 1- \exp \left( -\sigma^2/\sigma_f^2 \right) \right]}
& \text{for} ~~ \Theta>\Theta_0 ~,
\end{cases}
\eal
\end{widetext}
and the non-canonical coupling depends on the field $\Theta$,
\bal
\Xi (\Theta)=
\begin{cases}
	0 
	& \text{for} ~~ \Theta \le \Theta_0 + \Theta_T ~, \\
	\high{
		\xi
	}
	& \text{for} ~~ \Theta > \Theta_0 + \Theta_T ~.
\end{cases}
\label{eq:noncanonical_coupling}
\eal

These potentials and couplings are constructed to model the simple geometric picture of a potential landscape illustrated in Fig.~\ref{fig:trajectory}. The choices of the analytical functions are not unique as long as they capture the same picture to a good approximation. With the construction, we have incorporated the lessons learned from the previous attempts reviewed in Sec.~\ref{Sec:Sensitivity}.

This potential landscape consists of two parts. The first part is a potential step at $\Theta=\Theta_0$. The inflaton is initially rolling at the top of the step along the line $\sigma=0$, as long as we start with the initial condition $\sigma_i=0$ with zero velocity. One can also stabilize the value $\sigma_i=0$ by building a potential around it in the $\sigma$-direction. The parameters of such a potential does not enter the observables. The inflaton then falls into a lower valley over the cliff of the step.

The second part of the landscape consists of a curved path at the bottom of the valley, starting at $\Theta=\Theta_0+\Theta_T$. The inflaton enters the path tangentially with an initial velocity. It overshoots the bottom of the valley, climbs onto a side of the cliff and excites the oscillation of the massive $\sigma$-field. 
Such a curved path has two benefits.
It is a much milder way of exciting an oscillating massive field than the sharp bending trajectory mentioned earlier. Once the massive field starts oscillating, the curved path also provides a direct coupling between the massive field and the inflaton field, which enhances the amplitude of the clock signal compared to a straight path.

It is very natural to connect the above two parts of the landscape into a full model. Nonetheless, the model predictions are still very sensitive to the details on how this connection is done. As mentioned, having the inflaton fall down a step at a side of the curved path does not produce a dip that is deep enough to explain the large-scale dip. We have to make the step directly face the entrance of the curved path. Also, the locations of the step and the curved-path entrance may coincide, in which case we have one sharp feature at $\Theta=\Theta_0$ with the parameter $\Theta_T=0$. The two parts can also be separated and connected by a straight path, in which case we have two sharp features or an extended sharp feature, from $\Theta=\Theta_0$ to $\Theta=\Theta_0+\Theta_T$. We will proceed with the more general case by leaving $\Theta_T$ as a free parameter, and we shall see that the data marginally prefer a non-zero $\Theta_T$. 

Besides the full model, in Sec.~\ref{sec:restricted model} we will also build a simpler model that only makes use of the second part of the potential landscape described above. We use this simpler model to address only the features present in the small scales of CMB ($\ell\gtrsim 500)$.

Perhaps the most important message we can learn from the above discussion and this model-building exercise is how many details of the inflationary history we can potentially learn from the primordial feature signals.

\begin{figure}
\includegraphics[width=\columnwidth]{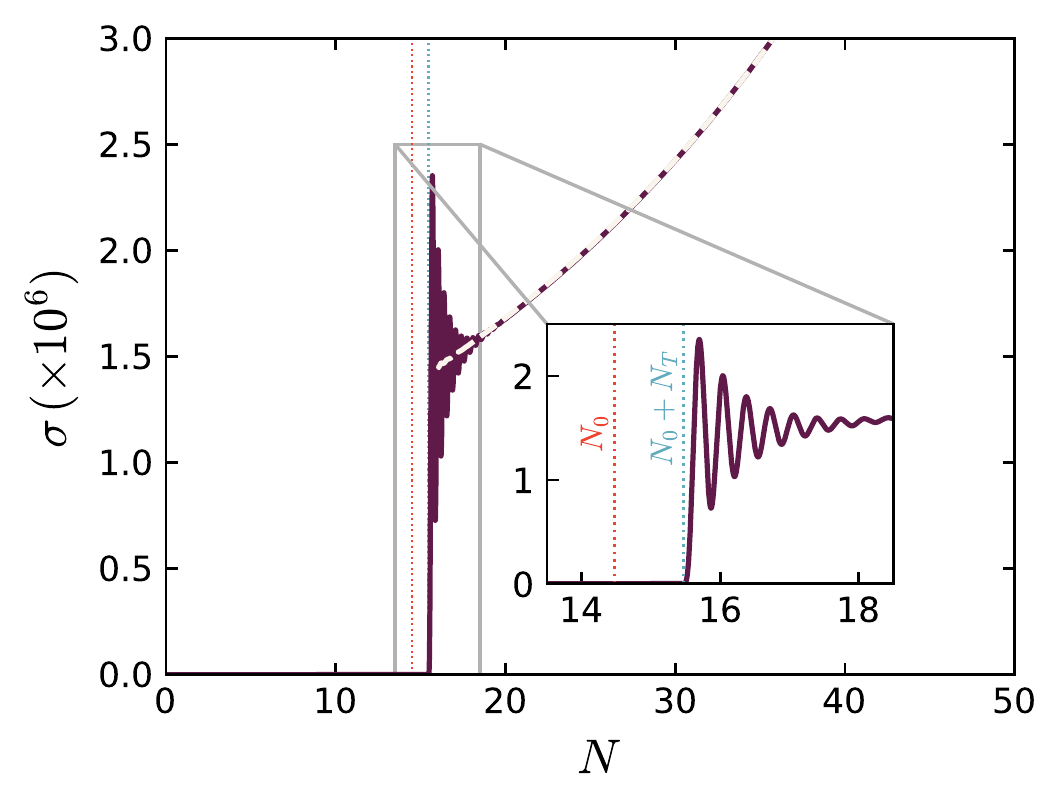}
\includegraphics[width=\columnwidth]{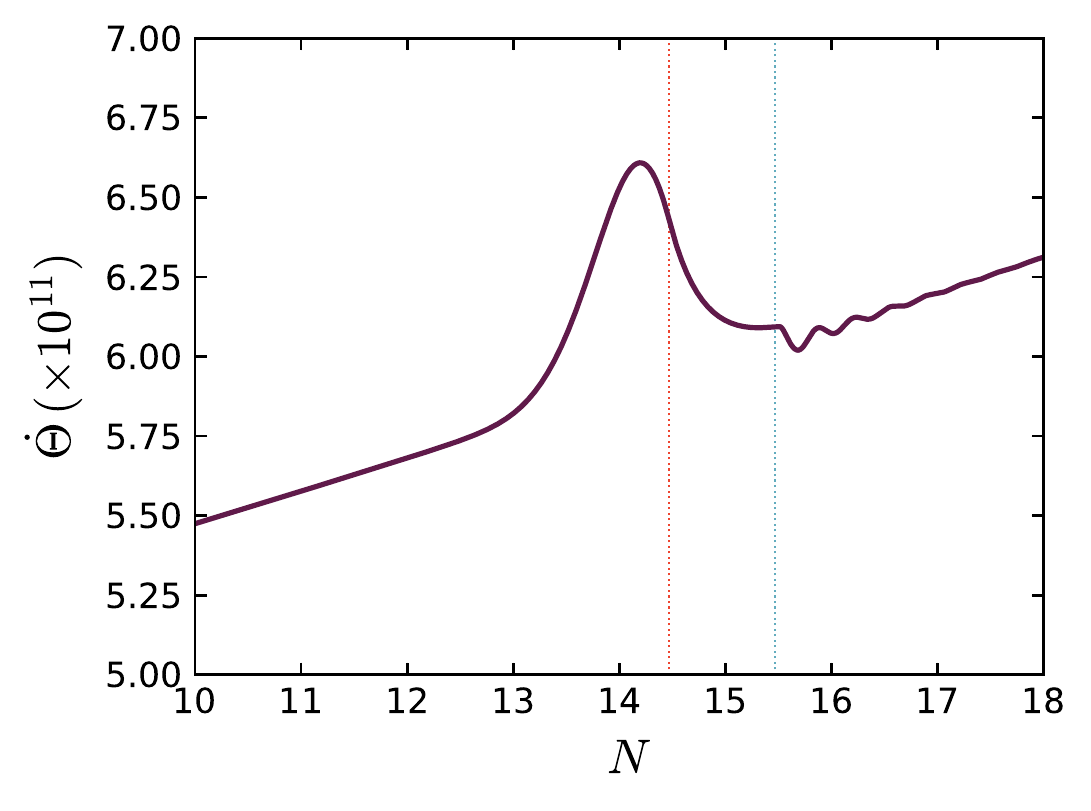}
\caption{\label{fig:background_intro} Example of the evolution of the clock field  [top] and the time derivative of the inflaton $\dot{\Theta}$ [bottom]. In the top panel, we also plot the effective minimum of $\sigma$ given by Eq.~\eqref{eq:effmin} in white dashed line. }
\end{figure}

We give an example of the evolution of the massive field $\sigma$ and of the velocity of the inflaton $\Theta$ in  Fig.~\ref{fig:background_intro}.
The effective minimum of the $\sigma$-potential after the turn is found by solving the following equation
\begin{equation}
\partial V/\partial \sigma =\xi(1+\xi\sigma)\dot{\Theta}^2,
\end{equation}
where the $\xi\sigma$ term, being very small, can be safely disregarded, and $\partial V/\partial \sigma \approx 2 V_{\rm inf} C_\sigma \sigma/\sigma_f^2$ near the bottom of the potential valley.
This leads to
\begin{equation}
\label{eq:effmin}
\sigma_{\rm min} 
\approx \frac{\xi \dot\Theta^2 \sigma_f^2}{2 V_{\rm inf} C_\sigma} ~.
\end{equation}
As can be seen, after the oscillatory transient, $\sigma$ tracks the effective minimum described by Eq.~\eqref{eq:effmin}.

\subsection{Effective parameters}
\label{sec:effectiveparameters}

From model parameters we construct six effective parameters, each of which directly characterizes a property of the full signal appearing in the power spectrum. These effective parameters serve as the connection between the model and data analyses \cite{Braglia:2021ckn}. The derivation below will also help us better understand the phenomenological properties of the model.

The sharp feature signal of this model is dominated by the step potential. We first characterize this part of the signal.

$\bullet$ {\em Depth of dip: $\Delta P_{\rm dip}/P$.}
The velocity of the inflaton increases as it rolls down a step. Since the amplitude of the power spectrum is related to the inflaton velocity through $P_\zeta=H^2/(4\pi^2 \dot\Theta^2)$, this generates a dip in the power spectrum and the depth of the dip can be estimated using the maximum velocity. For a narrow step which the inflaton rolls across in much less than an $e$-fold, this velocity can be estimated as $\dot\Theta^2 \approx (2C_\sigma+2\epsilon/3)V$, where the first term comes from the height of the step and the second is the attractor velocity. (For a wider step, the maximum velocity reduces due to the Hubble friction.) We can estimate the depth of the dip as
\bea
\frac{\Delta P_{\rm dip}}{P} \approx
1- \left( 1+\frac{3C_\sigma}{\epsilon} \right)^{-1/2} ~,
\label{DP_dip}
\eea
where $P$ is the spectrum value without the dip and $\Delta P_{\rm dip}$ is the change caused by the step, both evaluated at the scale of horizon-crossing at the time of the step.
As will be explained at the end of this subsection, we fix $\epsilon=10^{-7}$. So, the parameter $C_\sigma$ characterizes the depth of the dip. Or, more straightforwardly as will be done in the main text, we choose $\Delta P_{\rm dip}/{P}$ defined in \eqref{DP_dip} as one of the effective parameters.

For the candidate signal near $\ell\sim 25$, a best-fit large field model suggests that $C_\sigma/\epsilon \sim 0.4$ \cite{Peiris:2003ff}. We note two interesting facts associated with this value \cite{Bean:2008na}. First, from the above analyses we can see that this estimate should apply to both small and large field inflation models, and is independent of the value of $\epsilon$. Second, for a narrow step, this value suggests that the initial attractor velocity and the height of the potential happen to contribute similarly to the maximum velocity in terms of order of magnitude.

$\bullet$ {\em Extension of sinusoidal running: $C_\sigma/\Theta_f^2$.}
The extensiveness of the sinusoidal running of the sharp feature signal in the $k$-space following the dip is determined by the sharpness of the step, and can be estimated as \cite{Bean:2008na}
\bea
\Delta k \sim k_0\sqrt{\frac{C_\sigma}{\Theta_f^2}} ~,
\label{Dk_dip}
\eea
where $k_0$ is the starting $k$-location of the sharp feature signal, and roughly it is around the location of the dip.
Therefore, we use the effective parameter $C_\sigma/\Theta_f^2$ to characterize the extension of the sinusoidal running.

$\bullet$ {\em Transition to curved path: $N_T$.}
In this model, there is a transition period after the inflaton falls down the potential till it encounters a curved path. We can also regard this as an approximation of a gradual transition from a straight trajectory to a full curved path.
The velocity of $\Theta$ after falling down the step is dominated either by the attractor velocity or by the contribution from the step potential. As mentioned above, for the type of feature interesting to this work, the two happens to be of the same order of magnitude. So let us use the one from the step and estimate $\dot\Theta \approx \sqrt{2 C_\sigma V_{\rm inf}}$. The number of e-folds the $\Theta$ field travels, after falling down the step and before entering to the curved path, is $N_T=H\Theta_T/\dot\Theta$, which is
    \bea
    N_T = \frac{1}{\sqrt{6}} \frac{\Theta_T}{C_\sigma^{1/2} \mpl} 
    =\frac{\Theta_T}{\sigma_f} \frac{H}{m_\sigma} ~.
    \label{Eq:NT}
    \eea
$N_T$ is much more model-independent than the field extension $\Theta_T$, so we use this definition as an effective parameter to characterize the length of the transition period.

Next, let us characterize the clock signal part, that is induced by the oscillating $\sigma$ field excited as the inflaton enters the curved path at $(\Theta, \sigma)=(\Theta_0+\Theta_T,0)$.

$\bullet$ {\em Frequency of clock signal: $m_\sigma/H$.}
The curved path excites the oscillation of the massive field $\sigma$ that lies perpendicularly to the path, and at the same time introduces a coupling between the massive field and the inflaton field.
The $\sigma$ field oscillates classically with an amplitude decaying as $a^{-3/2}$. Through the coupling between $\sigma$ and $\Theta$, this induces a clock signal that runs as a function of $k$ as follows \cite{Chen:2014joa,Chen:2014cwa}:
\bea
\frac{\Delta P_{\rm clock}}{P_{\zeta0}} \propto
\left( \frac{2k}{k_r} \right)^{-3/2}
\sin\left( \frac{m_\sigma}{H} \ln \frac{2k}{k_r} + {\rm phase} \right)
~. \nonumber \\
\eea
So the effective parameter
\bea
\frac{m_\sigma}{H} = \frac{V_{\sigma\sigma}(\sigma=0)}{H}
\approx \frac{\sqrt{6 C_\sigma}}{\sigma_f}
\label{Eq:m/H}
\eea
characterizes the frequency of the clock signal.

$\bullet$ {\em Amplitude of clock signal: $\Delta P_{\rm clock}/{P}$.}
We first estimate the initial oscillation amplitude of the $\sigma$ field and then use it to compute the amplitude of the clock signal.
To estimate the initial amplitude of $\sigma$ excited by the curved path, it is sufficient to ignore the background metric because the excitation happens within a time period much less than an $e$-fold. The equation of motion for $\sigma$ is
\bea
\ddot\sigma + m_\sigma^2 \sigma - (\xi + \xi^2 \sigma) \dot\Theta^2 =0 ~.
\eea
We will work under the approximation that the oscillation amplitude is small comparing to the radius of the curved path, $\xi\sigma \ll 1$, as we will show self-consistently later.
Also, $\dot\Theta$ can be approximated by its attractor value $\dot\Theta_{\rm attr}^2 = 2H^2 \epsilon$ at the injection point $(\Theta, \sigma)=(\Theta_0,0)$, because the step potential contribution is either of the same order or gets dissipated during the transition.
The equation of motion then simplifies to
\bea
\ddot\sigma + m_\sigma^2 \sigma - \xi \dot\Theta_{\rm attr}^2 =0 ~.
\label{sigmaEOM_simple}
\eea
Since $\dot\sigma=0$ at the injection point, from \eqref{sigmaEOM_simple} we see that $\sigma$ oscillates around the new average $\langle \sigma \rangle = \xi\dot\Theta_{\rm attr}^2/m_\sigma^2$ with an amplitude 
\bea
\sigma_A=\xi\dot\Theta_{\rm attr}^2/m_\sigma^2
~,
\eea
(which will then decay if we include the background metric). This initial amplitude can be used in the estimation \cite{Chen:2014cwa}
\bea
\frac{\Delta P_\zeta |_{\rm clock}}{P}\Big|_{\rm amp} \approx 
\sqrt{2\pi} \sigma_A \xi \left( \frac{m_\sigma}{H} \right)^{1/2}
\eea
to compute the amplitude of the clock signal generated through the resonance between the $\sigma$ oscillation and the inflaton field, and we get
\bea
\frac{\Delta P_\zeta |_{\rm clock}}{P} \Big|_{\rm amp} \approx
\frac{\sqrt{2\pi}}{3} \frac{\epsilon}{C_\sigma} (\sigma_f \xi)^2 \left( \frac{m_\sigma}{H} \right)^{1/2} ~.
\label{DP_amp}
\eea
For example, for the best fit model LFCII with $m_\sigma/H\sim 18$ (see Section~\ref{sec:BF_full}), we get $\Delta P_{\rm clock}/P_{\zeta 0} \approx 0.038$. Comparing to the numerical result $0.0255$, we over-estimate by a factor of $1.5$. At $m_\sigma/H\sim 50$, the over-estimation is a factor of $1.1$.  Also, $\xi\sigma\sim \xi\sigma_A \sim 0.003$ satisfies the aforementioned working assumption.

Therefore, we can introduce the parameter $\xi\sigma_f$, or more straightforwardly as will be done in the main text, the parameter $\Delta P_{\rm clock}/{P}$ defined in \eqref{DP_amp}, as one of the effective parameters.

$\bullet$ {\em Scale location of full signal: $N_0$.}
Finally, the effective parameter $N_0$ characterizes the overall scale-location of the full standard clock signal, defined as the $e$-fold from the beginning of inflation till the inflaton crosses the location of the step $\Theta(N_0)\equiv\Theta_0$. 
(In the simulation, We fix a total of $68$ $e$-folds and we choose $k=0.05\,{\rm Mpc}^{-1}$  for the mode that exits the horizon $50$ $e$-folds towards the end of the inflation. Both these overall $e$-folds may be shifted by shifting the cutoff point of the ending of the inflation without any consequence to the results.)

\begin{figure*}
\includegraphics[width=\columnwidth]{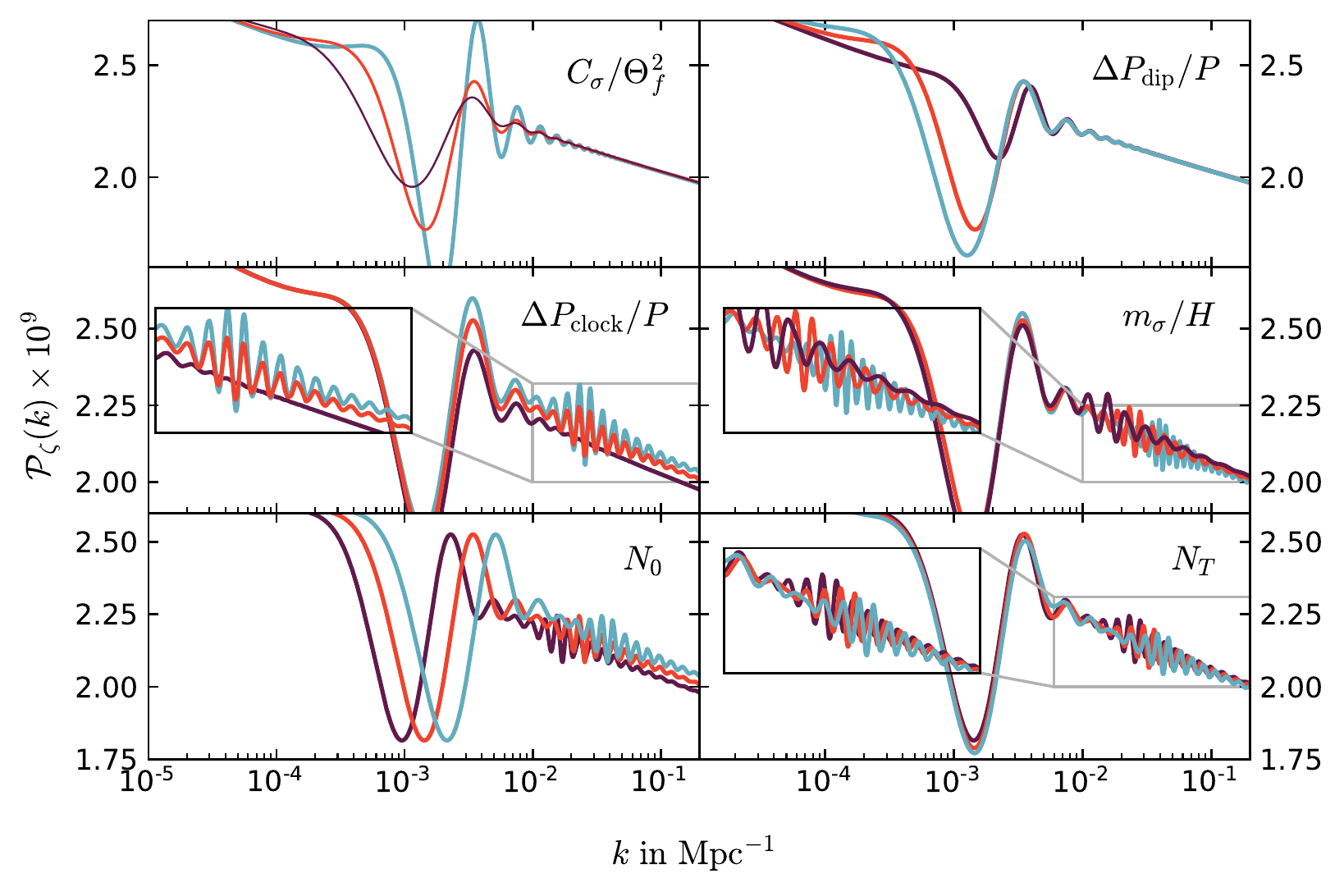}
\caption{\label{fig:PK_intro} Effects of varying the effective parameters of the model on the primordial scalar power spectrum. We fix the parameters of the baseline model to $C_\Theta\times 10^2=1.85,\,\log_{10} P_{\zeta *}=-8.64,\, C_\sigma/\Theta_f^2=0.15,\,\Delta P_{\rm dip}/P=0.24,\,\Delta P_{\rm clock}/P=0,\,m_\sigma/H=30,\,N_0=14.3,\,N_T=0$ and vary one parameter at a time. Specifically, we vary $C_\sigma/\Theta_f^2\in\{0.07,\,0.15,\,0.45\}$ in the top-left panel,  $\Delta P_{\rm dip}/P\in\{0.07,\,0.24,\,0.33\}$ in the top-right panel, $\Delta P_{\rm clock}/P\in\{0,\,0.046,\,0.077\}$ in the center-left panel, $m_\sigma/H\in\{18,\,30,\,50\}$ in the center-right panel, $N_0\in\{13.9,\,14.3,\,14.7\}$ in the bottom-left panel and $N_T\in\{0,\,0.3,\,0.6\}$ in the bottom-right panel. In the last three panel, we fix $\Delta P_{\rm clock}/P=0.046$. In each plot, the convention is that purple, dark orange and light blue lines represent the smallest, the central and the largest value of the parameter which is being varied respectively.}
\end{figure*}

The effect of each of these parameters are illustrated in Fig.~\ref{fig:PK_intro}.

We end this subsection by a discussion of several properties of the model that are not described by free parameters. The overall guideline on which properties do not have to be described by free parameters is stated in Sec.~\ref{Sec:guideline}. Here are some detailed illustrations in the context of this model. 

The first one is the overall energy scale of the model, or equivalently the value of the slow-roll parameter $\epsilon$. As long as $\epsilon$ is less than $10^{-3}$, the phenomenology of the model on feature predictions is kept the same if we scale $\epsilon \to C^2 \epsilon$ together with $V_{\rm inf} \to C^2 V_{\rm inf}$, $\sigma_f \to C \sigma_f$, $\Theta_{0} \to C \Theta_{0}$, $\Theta_{T} \to C \Theta_{T}$, $\xi \to C^{-1} \xi$, and $\Theta_f \to C \Theta_f$. The slope parameter, $C_\Theta$, of the slow-roll potential is kept the same to maintain the same spectrum index. This property implies that we can only determine the mass of the massive field $\sigma$ in unit of $H$ unless the tensor mode is determined independently. For larger $\epsilon$ values, the parameter changes are more complicated.
These properties are illustrated in more details in Appendix~\ref{Sec:several details}.
In this work, we set $\epsilon=10^{-7}$.

The second one is the straight-to-curved-path transition at $\Theta=\Theta_0+\Theta_T$. Because the values of the potential and its derivatives are continuous at this transition, this sharp feature is rather mild and the sharp feature signal is dominated by the step potential. 
Nonetheless, the sharpness of this transition may affect the amplitude of the $\sigma$-field oscillation. E.g., if the transition time scale is greater than $1/m_\sigma$, the amplitude will be reduced. On the other hand, as long as the time scale is less than $1/m_\sigma$, the sharpness of this transition has no effect on the observables and this is the parameter space we use in this work.

Lastly, we place the clock field initially at rest at $\sigma_i=0$. As long as this parameter stays within $\sigma_i\lesssim0.7\sigma_f$, the initial value of the clock field has no effect on the observables, see Appendix~\ref{Sec:several details}.  A simple way to stabilize the value of $\sigma_i$ is to extend the potential well of $\sigma$-field to the plateau and make the path of the inflaton field at the top of the plateau also a valley. The details of such a potential well has no effect on the observables.

\subsection{A restricted model: $\Theta_f \to \infty$ (i.e.~no dip)}
\label{sec:restricted model}

In this subsection, we consider a special limit of the model (and will be called the restricted model) in which only the small-scale features are significant. We remove the step potential by taking the limit $\Theta_f \to \infty$, so that the large-scale dip is no longer present; and consequently the remaining sharp feature signal becomes a sinusoidal running signal with a much smaller amplitude, as the result of the sharp transition at $\Theta=\Theta_0+\Theta_T$. This is an interesting limit because the model now has five inflationary parameters (i.e.~three more than the Standard Model). This is one parameter fewer than a previous CPSC model introduced in \cite{Chen:2014cwa}, and has the same number of parameters as the simplest example of the resonant model \cite{Chen:2008wn}; but nonetheless, as we will show, this restricted model provides slightly better fits than either model.

In this limit, the height of the potential cliff for the $\sigma$ field is irrelevant, because the inflaton is always rolling in the valley. So the parameterization of the $\sigma$ potential can be simplified as
\bal
V (\Theta, \sigma) = V_{\rm inf} 
\left(	1- \half C_\Theta \Theta^2  \right)
	+ \half M_\sigma^2 \sigma^2 ~,
\label{eq:pot_simplelimit}
\eal
with the same non-canonical coupling,
\bal
\Xi (\Theta)=
\begin{cases}
	0 
	& \text{for} ~~ \Theta \le \Theta_{0T} ~, \\
	\xi
	& \text{for} ~~ \Theta > \Theta_{0T} ~.
\end{cases}
\label{eq:noncanonical_coupling_simple}
\eal

The first stage of inflation is slow-roll, until it enters the same curved path as in the full model.
The inflaton overshoots the bottom of the valley and excites a classical oscillation of a massive field. Same as explained previously, 
the full signal of this restricted model is also a combination of a sharp feature signal and a clock signal. The difference from the previous full model is that, because
the junction between the straight trajectory and the curved path is a mild sharp feature, the amplitude of the induced sharp feature signal is relatively small and, in particular, there is no large dip. On the other hand, the clock signal remains as prominent as in the full model.

We can see that, besides the standard slow-roll inflationary parameters $V_{\rm inf}$ and $C_\Theta$, there are three extra model parameters, $m_\sigma$, $\xi$ and $\Theta_{0T}$. The three effective parameters can be chosen as follows.

The frequency of the clock signal is characterized by
$m_\sigma/H = \sqrt{3} m_\sigma/\sqrt{V_{\rm inf}}$, which can be taken directly as an effective parameter. The amplitude of the clock signal is now expressed as
\bea
\frac{\Delta P_{\rm clock}}{P}\Big|_{\rm amp} \approx 
2\sqrt{2\pi}~ \epsilon \xi^2 \left( \frac{m_\sigma}{H} \right)^{-3/2} ~,
\eea
which can be used as the second effective parameter. The last effective parameter is still the overall scale of the feature, parameterized by $N_0$ and chosen as the $e$-fold from the beginning of inflation till the inflaton crosses any reference point in the model.

In Sec.~\ref{sec:results} \& \ref{sec:results_restricted}, we will be interested in comparing the constraints on this restricted model with the ones on the full model. To this concern, it is useful to see Eqs.~\eqref{eq:pot_simplelimit} and \eqref{eq:noncanonical_coupling_simple} as a nested limit of the full model. Only in this way, can we adopt the same priors on the parameters shared by the two models and perform a fair Bayesian comparison. This can be simply achieved by sending $\Theta_f\to \infty$ and setting $\Theta_0+\Theta_T\equiv \Theta_{0T}$ and $2 V_{\rm inf} C_\sigma/\sigma_f^2\equiv m_\sigma^2$, where parameters on the left (right) are full (restricted) model parameters. In terms of the full model effective parameters in Section~\ref{sec:effectiveparameters}, this corresponds to fixing $N_0$ or $N_T$ and $C_\sigma$ to an arbitrary value within the prior volume of the full model. With this convention, the free extra parameters describing the frequency and amplitude of the clock signal in the restricted model are $\log_{10} m_\sigma/H$ and $\frac{\Delta P_{\rm clock}}{P}$, as in the full model.  Note that, for a fixed value of $m_\sigma/H$, $\sigma_f$ is simply computed from Eq.~\eqref{Eq:m/H}.  

Regarding the location of the feature, it is governed by either $N_0$ or $N_T$. Since $N_0$ fixes $\Theta_0$ through a shooting algorithm and in the restricted model we get rid of the step in the potential by sending $\Theta_f\to\infty$, $N_0$ effectively sets the pivot scale and it is thus expected to be degenerate with both the amplitude and the tilt of the power spectrum. Therefore, in order not to introduce such a degeneracy, we choose to fix $N_0$ and vary $N_T$ instead, using a prior that allows the clock signal to span the same range of scales as in the full model.

\section{Method of data analysis}
\label{Sec:data_analysis_method}

We now discuss the methodology of our data analysis and describe in details the datasets that we analyze as well as the sampling techniques and the priors adopted. The reader only interested in the results may skip this Section and go directly to Sec.~\ref{sec:results}.

$\bullet$ {\em Data.} We use publicly available CMB temperature and E-mode polarization anisotropies data from the latest release of the Planck mission \cite{Planck:2018nkj}. The likelihoods delivered by Planck separately address large and small angular scales offering the opportunity to constrain the dip and clock feature in our model respectively. Specifically, the large scale likelihoods operating in the multipole range $\ell=2-29$ are the $\tt{commander\_dx12\_v3\_2\_29}$ likelihood (lowT) for temperature anisotropies and the $\tt{simall\_100x143\_offlike5\_EE\_Aplanck\_B}$ (lowE) for E-mode ones. For smaller angular scales, crucial to constrain the clock signal, we consider two different likelihoods. The first one is the official Planck likelihood, also used in the feature analysis of the Planck Inflation paper~\cite{Akrami:2018odb}, and is called {\tt Plik bin1} (hereafter P18). This likelihood spans the range $\ell=30-2508$ in TT and $\ell=30-1996$ in TE and EE. In addition to P18, we test the robustness of our results against a second likelihood, qualitatively different from P18, i.e.~{\tt 2020 CamSpec release v12.5}\footnote{\href{https://people.ast.cam.ac.uk/~stg20/camspec/index.html}{https://people.ast.cam.ac.uk/~stg20/camspec}} (hereafter EG20, from the name of its authors).
EG20 spans a slightly different multipole range than P18, corresponding to $\ell=30-2500$ in TT and $\ell=30-2000$ in TE and EE. We stress that both likelihoods are completely unbinned, as required to compare oscillatory features with data. The results presented in the next Sections are obtained by combining both low-$\ell$ likelihoods with either P18 or EG20 high-$\ell$ ones. The feature analysis for each separate likelihood is presented in Ref.~\cite{Braglia:2021ckn} for a different CPSC model.

Since our prior allows the feature signal to be outside the range probed by Planck, we could also employ in our analysis recent small scale measurements of the temperature and polarization power spectra from the Atacama
Cosmology Telescope (ACT)~\cite{Thornton:2016wjq} and the South Pole Telescope~\cite{SPT-3G:2021eoc}. As a representative example we tested the ACT data. However, our findings suggest that adding such data, at the present stage, does not significantly change our conclusions.  
There are two main reasons. Firstly, the feature signals that provide a better fit to Planck data significantly deviate from scale invariance at $\ell\lesssim1500$ (see next Sections). Although from $900\lesssim\ell\lesssim1500$ the error bars on the EE power spectra are smaller than Planck ones, the ones on TT (and thus TE) are worse. It turns out that Planck's TT and TE currently have the most constraining power on our model. Secondly, ACT data are binned and not efficient in constraining oscillatory feature signals (see above) with high frequencies. Nevertheless, for completeness, we present our analysis including ACT data in Appendix~\ref{Appendix:ACT}.

$\bullet$ {\em Sampling technique.} Given the plethora of primordial power spectra that can be generated within our model, the posterior distribution is expected to be highly multi-modal. Thus, the optimal method to sample the vast CPSC parameter space is nested sampling and we resort to {\tt PolyChord}\footnote{\href{https://github.com/PolyChord/PolyChordLite}{https://github.com/PolyChord}}~\cite{Handley:2015fda,Handley:2015vkr} through its implementation in {\tt CosmoChord} ({\tt PolyChord} add-on of {\tt CosmoMC}\footnote{\href{https://cosmologist.info/cosmomc/}{https://cosmologist.info/cosmomc/}} \cite{Lewis:2002ah}). As customary, the convergence criterion adopted by the nested sampling procedure is to stop the likelihood exploration when the variation of the evidence is much smaller than the error on it. In our case, this corresponds to variations of the evidence of  $\Delta\ln Z=0.05$. From the samples produced by the data analysis, we can plot the posterior distributions for the parameters and read off their constraints. Furthermore, we can compute the Bayesian evidence for the model and compare it to the one for the baseline featureless one. As a result, we get the Bayes factor defined as $B\equiv Z_1/Z_2$, where $Z_1$ ($Z_2$) is the evidence for the CPSC (featureless) model. The natural logarithm of the Bayes factor can be compared to the so-called Jeffreys' scale~\cite{Kass:1995loi} to assess which model is preferred by data. In our conventions a positive $\ln B$ corresponds to a preference of the CPSC model over the baseline.

 \begin{table}
 	\centering
 	\begin{tabular}{|l|l|}
 		\hline
 		Parameters               & Priors      \\ \hline
 		$\Omega_\mathrm{b}h^2$   & $0.02,\, 0.0265$ \\ \hline
 		$\Omega_\mathrm{CDM}h^2$ & $0.1,\, 0.135$   \\ \hline
 		$100*\theta_s$              & $1.03,\, 1.05$   \\ \hline
 		$\tau$                    & $0.03,\, 0.08$      \\ \hline
 		$C_\Theta$               & $0.002,\, 0.04$  \\ \hline
 		$\log_{10} P_{\zeta *}$              & $-8.8,\, -8.1$   \\ \hline
 		$C_{\sigma}/\Theta_f^2$             & $0.0,\, 2.5$   \\ \hline
 		$\Delta P_{\rm dip}/P_0$             & $0.0,\, 0.5$   \\ \hline
 		$N_0$             & $13,\, 15.5$   \\ \hline
 		$\log_{10} m_\sigma/H$             & $0,\, \log_{10} 75$        \\ \hline
 		$\Delta P_{\rm clock}/P_0$            & $0,\, 0.35$       \\ \hline
 		$N_T$                    & $0,\, 1.2$     \\ \hline
 	\end{tabular}
 	\caption{\label{tab:priors} Priors on the background and effective parameters used in our analysis. The priors on foreground and calibration parameters are kept unchanged from their default values in the Polychord analysis.}
 \end{table}
 
$\bullet$ {\em Priors on the cosmological parameters.} The choice of the prior volume is crucial to correctly compare our model to data. The guideline we follow to arrive at the priors summarized in Table~\ref{tab:priors} is that they should be large enough for both the dip and the clock features to have amplitudes that range from zero to a value well larger than Planck residuals and for their location to span the entire range of scales probed by Planck. Using $n_s\approx1-2C_\Theta$ and $P_{\zeta *} \approx V_{\rm inf}/24\pi^2 \epsilon$, the priors on $P_{\zeta *}$ and $C_\Theta$ can be translated to ranges of variation of the amplitude and tilt of the primordial power spectrum (PPS). The latter are the only inflationary parameters that we vary in the analysis for the featureless model, which is obtained by fixing $\Delta P_{\rm clock}/P_0=\Delta P_{\rm dip}/P_0=C_\sigma/\Theta_f^2=0$ and $N_0,\,\log_{10} m_\sigma/H,\,N_T$ to arbitrary values within the prior range in Table~\ref{tab:priors}. 

We note that our results (parameter estimation and Bayesian evidence) are robust to the specific choice of the priors and parameters describing the feature properties, as explicitly shown in Appendix~\ref{sec:LinPar}. As a last point, we mention that a small difference between our procedure and those in \cite{Akrami:2018odb,Canas-Herrera:2020mme} is that we also vary the nuisance parameters along with the cosmological ones.

$\bullet$ {\em Best-fit candidates.} In addition to the statistics of our model, we are also interested in finding the set of parameters which best fit the observational data and comparing the resulting feature model to the featureless one. With such best-fit candidates in hand, we can plot the CMB residuals and inspect which of their features our model is addressing.
Since the number of cosmological and nuisance parameters involved in the analysis is large, our sampling procedure is not efficient in properly identifying best-fit candidates. Therefore, we identify the better likelihood regions in our samples and starting from these combinations of parameters we search for best-fit candidates using the likelihood maximizer BOBYQA~\cite{BOBYQA}. Then, in order to correctly understand the fit to each data set, we break down the total $\chi^2$ to data as follows:
\begin{equation}
\chi^2_\mathrm{TOT}=\chi^2_\mathrm{high-\ell}+\chi^2_\mathrm{low-T}+\chi^2_\mathrm{low-E}+\chi^2_\mathrm{prior} ~,
\end{equation}
and report the variation of each term in the sum with respect to the baseline model. Note that $\chi^2_\mathrm{prior}$ corresponds to the Gaussian priors on particular nuisance parameters, see the discussion in Ref.~\cite{Braglia:2021ckn}. In this paper, to obtain the best-fit we use the method-III outlined in~\cite{Braglia:2021ckn}.

\section{Results of the data analysis: full model}
\label{sec:results}

\begin{figure*}
	\includegraphics[width=\columnwidth]{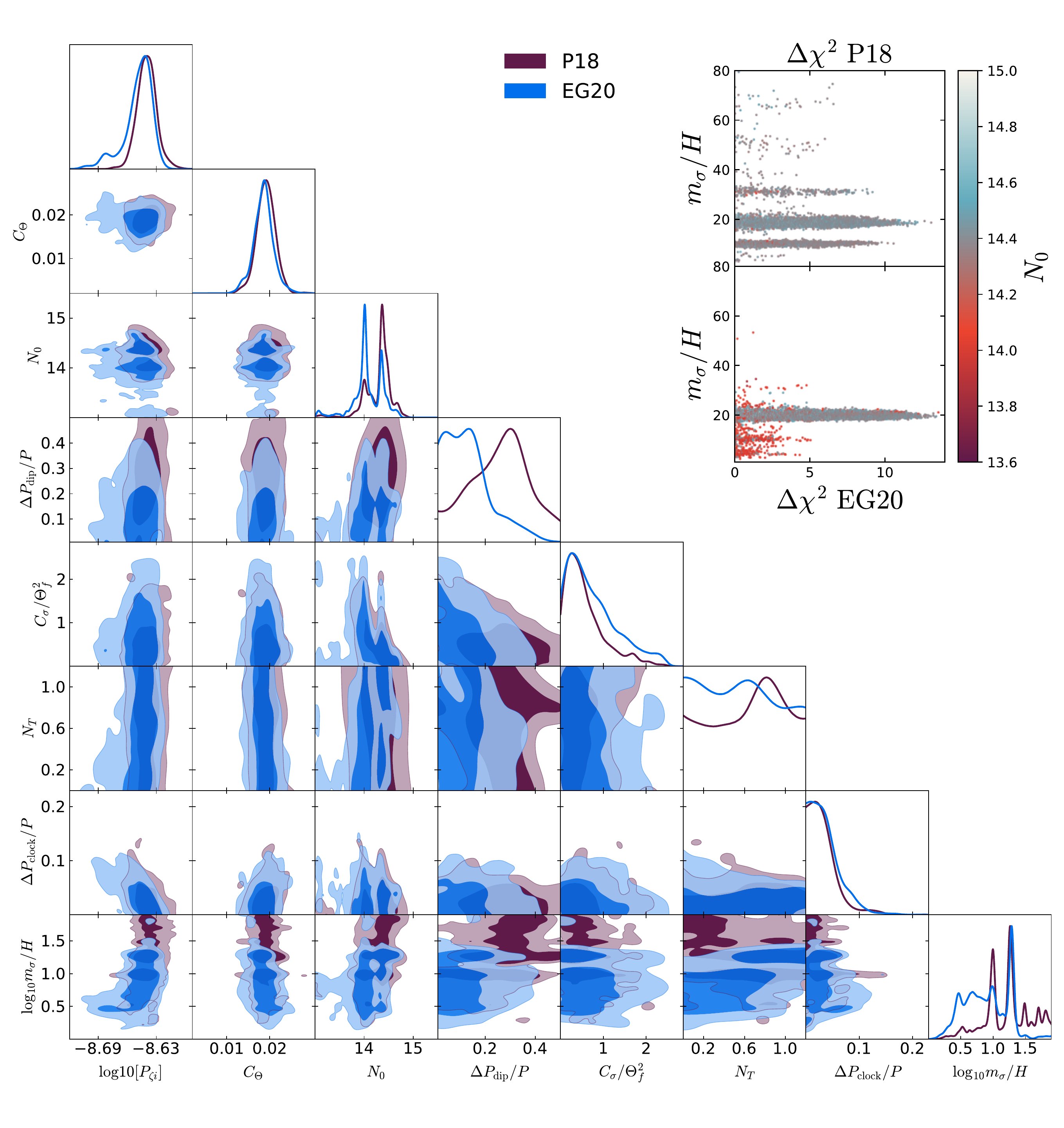}
	\caption{\footnotesize\label{fig:triangle_log} Triangle plot of the full model for PlikHM bin1 (P18, purple) and Clean CamSpec v12.5HMcln (EG20, light blue) datasets. We also plot better likelihood samples in the top-right corner of the plot ($\Delta\chi^2$ in the colorbar is defined as $\Delta\chi^2\equiv\Delta\chi^2_{\rm featureless}-\Delta\chi^2_{\rm CPSC}$).}
\end{figure*}

The results of our data analysis are shown in Fig.~\ref{fig:triangle_log}, where we plot the 1 and 2-dimensional posterior distributions for the parameters describing the CPSC model, as well as better likelihood samples in the top-right corner. We now comment on these results.

The figure shows that data can place tight constraints only on the amplitude and tilt of the PPS, as expected. In particular, the posterior distribution of the latter is slightly different for P18 and EG20. Using $n_s\sim1-2C_\Theta$ we get the constraints $n_s = 0.9617\pm 0.0044$ and $n_s = 0.9630\pm 0.0047$ at 68\% CL for P18 and EG20 respectively. The $\sim0.0013$ difference in the mean of values $n_s$ of EG20 and P18 is consistent with the findings of Ref.~\cite{Efstathiou:2019mdh}. We also notice a slight decrease in the mean value of $n_s$ compared to the featureless case, for which we get results in agreement with Refs.~\cite{Akrami:2018odb} and \cite{Efstathiou:2019mdh}. 
This is expected as constraints on $n_s$ for the baseline model are obtained by assuming constant tilt from the largest to the smallest scales while the full model allows a dip at the large scales. Since the primordial power spectrum has different tilts for the full model at large and small scales, the spectral tilt essentially refers to the average tilt of the spectrum at the intermediate and small scales. Therefore it is constrained essentially by high-$\ell$ data. Although, for brevity, we do not show this explicitly, we do not find any appreciable shift in the standard cosmological parameters $\omega_b,\, \omega_c,\, \tau_{\rm reio}$ and $100*\theta_s$ with respect to the baseline model.

Regarding the parameters describing the dip feature at large scales, we observe that the tail of the posterior distribution of $N_0$ extends toward the lower limit of the $N_0$ prior range, which corresponds to the dip feature being located at unobservably large scales, but it also shows an upper limit, consistently with the dip feature not being supported by data for $\ell\gtrsim30$. Also for $\Delta P_{\rm dip}/P_0$ and $C_\sigma/\Theta_f^2$, we only get upper bounds, since the dip feature and the subsequent oscillations cannot be too large.  All these arguments point to the dip feature not being detected. However, if we limit ourselves to the 1$\sigma$ level, i.e. 68\% CL, we do constrain all  of the three dip feature parameters as follows: $N_0 = 14.29^{+0.25}_{-0.33},\,\,$$\Delta P_{\rm dip}/P_0 = 0.26\pm 0.11,\,\,$$C_\sigma/\Theta_f^2 = 0.586^{+0.098}_{-0.57}$ for P18 and $N_0 = 14.08^{+0.37}_{-0.21},\,\,$$\Delta P_{\rm dip}/P_0 = 0.138^{+0.028}_{-0.13},\,\,$$C_\sigma/\Theta_f^2 = 0.66^{+0.14}_{-0.63}$ for EG20.

It is also interesting to note a weak degeneracy between $C_\sigma/\Theta_f^2$ and $\Delta P_{\rm dip}/P_0$ since not only the latter, but also the former slightly affects the depth of the dip, as can be appreciated from Fig.~\ref{fig:PK_intro}. However, the primary effect of $C_\sigma/\Theta_f^2$ is to change the width of the dip as well as the amplitude of subsequent peak and oscillations, which leads to another weak degeneracy between $C_\sigma/\Theta_f^2$ and $N_0$ seen in their 2-dimensional posterior where larger values of $N_0$ correspond to smaller values of $C_\sigma/\Theta_f^2$, so that the large sharp feature oscillations do not spoil the fit to high-$\ell$ data. As a last comment, we note that the bimodal posterior of $N_0$ has peaks around $N_0\sim14.00$ and $N_0\sim14.45$ corresponding to a sharp feature at $k_0\sim4.6\times10^{-4}\,{\rm Mpc}^{-1}$
and $k_0\sim7.4\times10^{-4}\,{\rm Mpc}^{-1}$ respectively, therefore providing a similar fit to the $\ell\sim20-30$ feature in TT data. The prominence of each peak changes depending on the high-$\ell$ data used, the one at $N_0\sim14.00$ being less significant for P18 and vice versa for EG20. This is also seen in top-right corner of Fig.~\ref{fig:triangle_log}, which depicts how the samples populate the $N_0$ prior range.

We now go on and discuss the parameters describing the location, amplitude and frequency of the clock signal. The parameter $N_T$, describing the location of the clock signal relative to the dip feature, is currently unconstrained. 
This is mostly due to the large uncertainties in data between the dip feature and small-scale oscillation.

The posterior distribution of the clock amplitude and frequency parameters are instead much more informative. On one hand, we get upper bounds on the amplitude of the clock signal at 95\% CL, i.e. $\Delta P_{\rm clock}/P < 0.0847$ for P18 and $\Delta P_{\rm clock}/P < 0.0915$ for EG20.\footnote{We remind the reader that the parameter $\Delta P_{\rm clock}/P$ is an estimate for the maximum amplitude of the clock signal which differs from the numerical result by a factor of $\mathcal{O}(1-1.7)$. See the discussion below \eqref{DP_amp}. } On the other hand, several frequencies of the clock signals are allowed by data. This multimodality comes because, given one of these frequencies, it is possible to tune the starting point and the maximum amplitude of the clock signal to fit some of the CMB residuals to some extent. This fact is also observed in the 2-dimensional contours in the plane spanned by $\Delta P_{\rm clock}/P$ and $\log_{10} m_\sigma/H$ where different frequencies can be seen to correspond to different amplitudes. 

However, a notable property of the distribution of $\log_{10} m_\sigma/H$ is that, despite its multimodality, it clearly shows two dominant local peaks around $m_\sigma/H\sim10$ and $18$. In particular, the latter is the most prominent for both datasets and it corresponds to the region where better-likelihood samples locally accumulate. It is interesting to note that such a peak corresponds to the best-fit value of the frequency found in other studies on resonant features \cite{Akrami:2018odb,Braglia:2021ckn}. We will come back to this point again in the next sections. 

Finally, let us  comment on the evidence of the model. We report the natural logarithm of the Bayes factor, computed using the featureless model as a reference, in Table~\ref{tab:evidence_log}. Because of the large number of extra parameters (6 in total), despite the better fit to data,  our model is not preferred over the baseline, though, according to the Jeffreys scale, the extent to which the baseline model is preferred is not worth more than a bare mention for both P18 an EG20~\cite{Kass:1995loi}. Higher quality CMB data are therefore needed to assess the statistical significance of our model. We address this point in Section~\ref{sec:forecasts}, performing a forecast for future CMB experiments. In order to be able to do so, we need to select a feature signal to be used as fiducial cosmology, which we pick up from the best-fit search of the next subsection.

\begin{table}
	\centering
	\begin{tabular}{|l|l|l|}
		\hline
		Dataset          & P18 (Plik bin1) & EG20 (CamSpec v12.5HMcln) \\ \hline
		$\ln B$     &      $-0.13\pm0.38$           & $-1.08    \pm0.36$        \\         \hline
	\end{tabular}
	\caption{Bayes factors ($\ln B$) obtained for the CPSC model {\it w.r.t.} the baseline model. The baseline model has the same prior volume as in the featureless part of the CPSC model. }
	\label{tab:evidence_log}
\end{table}

\subsection{Best-fit candidates}
\label{sec:BF_full}
\begin{table*}[t]
	\begin{center}
		\begin{tabular}{|c||c|a|c|c||c|c|a|c|}
			\hline
			& \multicolumn{4}{c||}{P18 } &\multicolumn{4}{c|}{EG20 }\\ \hline
			& LFC I & LFC IIa & LFC IIb & HFC& LFC I   & LFC IIa & LFC IIb & HFC   \\ \hline 	
			$\log_{10} P_{\zeta *}$  & -8.639 & -8.638 &-8.642&-8.638&-8.639&-8.639&  -8.643&-8.638\\				 \hline
			$C_\Theta$     &  0.0212  &0.0189  & 0.0194 &0.0177&0.0213 &0.0189&0.0186&0.0180\\ \hline  
			$N_{0}$ & 14.40 &14.38 & 14.47&14.00& 14.40&14.38& 14.47& 14.00\\ \hline
			$C_\sigma/\Theta_f^2$   & 0.359   & 0.495 &0.232& 0.882& 0.353&0.481& 0.208& 0.899\\ \hline
			$N_T$& 0.95 &1.17  &1.19& 0.39&0.95 &1.17& 1.16 &0.39\\ \hline   
			${m_\sigma}/{H}$&9.23 &18.48  &18.01&51.25 &9.21&18.47& 18.60& 51.26 \\ 
			\hline
			$\Delta P_{\rm dip}/P$  & 0.29   &0.28 &0.24&0.17 &0.30 &0.28 &0.25 &0.17\\    \hline
			$\Delta P_{\rm clock}/P$  &0.037    &0.038 &0.039&0.07 &0.032 &0.035 & 0.035&0.061\\    \hline
			\hline
			$C_\sigma\times10^8$   & 3.228  & 3.151 & 2.378 & 1.491 & 3.489 &3.177 &2.583  &1.544  \\ \hline
			$\xi\sigma_f$  & 0.069    &0.058  &0.051 & 0.041& 0.066 &0.056& 0.05&0.040\\ \hline 
			$k_0\,\times10^{3} {\rm Mpc}$        &    0.683  & 0.666 &0.735 &0.458&0.683&0.666&  0.735 &0.458\\ \hline

		\end{tabular}
		\begin{tabular}{|c||c|c|c|c|c|c||}
			\hline
			& $\Delta \chi^2_{\rm TOT}$   & $\Delta\chi^2_\mathrm{prior}$    & $\Delta\chi^2_\mathrm{high-\ell}$  & $\Delta\chi^2_\mathrm{low-T}$ & $\Delta\chi^2_\mathrm{low-E}$  \\ \hline
			& \multicolumn{5}{c|}{P18 (Plik bin1)} \\  \hline  
			LFC I       & 14.0& 0.05      & 7.30 & 5.14 & 1.48  \\\hline \rowcolor[gray]{0.8}				LFC IIa &   19.8    & 0.01    &13.31 & 5.34 & 1.11  \\ \hline
			LFC IIb &  18.7     & -0.02    &11.84 & 5.26& 1.61  \\ \hline				HFC &   15.5    & 0.07    &10.9 & 2.28 & 1.11 \\  \hline  
			\hline
			& \multicolumn{5}{c|}{EG20 (CamSpec v12.5HMcln)} \\ \hline
			LFC I       &6.5 & 0.22   & 1.63   & 4.49  & 0.15  \\\hline
			LFC IIa       &16.7& -0.05  &11.99   &4.93   & -0.22 \\\hline
			\rowcolor[gray]{0.8}	LFC IIb  & 17.7 & -0.09  &12.96   &4.65  & 0.21  \\
			\hline
			HFC      &14.5& 0.17  &11.28   &3.35   & -0.27\\\hline
		\end{tabular}
		\caption{[Top] Best-fit candidates found in P18 and EG20 likelihood using TTTEEE+lowT+lowE. $k_0$ is defined as $k_0\equiv a(N_0) H(N_0)/2$.
		In the first table, the three parameters below the double horizontal line are not independent of the first eight parameters.
			[Bottom] Individual $\chi^2$'s for each candidate.}
		\label{tab:candidates}
	\end{center}
\end{table*}

\begin{figure}
	\includegraphics[width=\columnwidth]{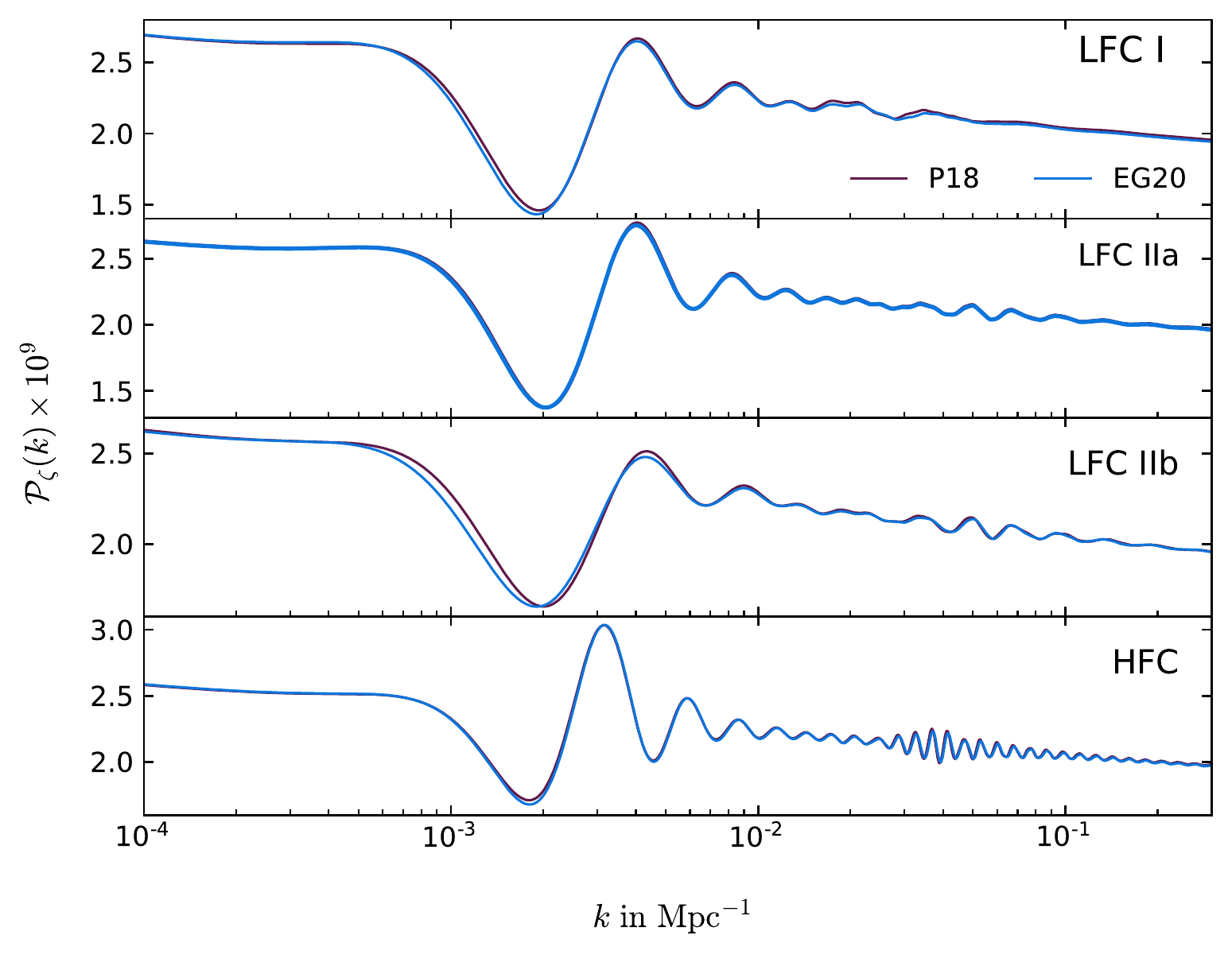}
	\caption{\label{fig:bestfit} PPS of the low-frequency best-fit candidates (LFCI, LFC IIa and LFC IIb), and high-frequency best-fit candidate (HFC). }
\end{figure}

\begin{figure*}
	\centering
	\includegraphics[width=\columnwidth]{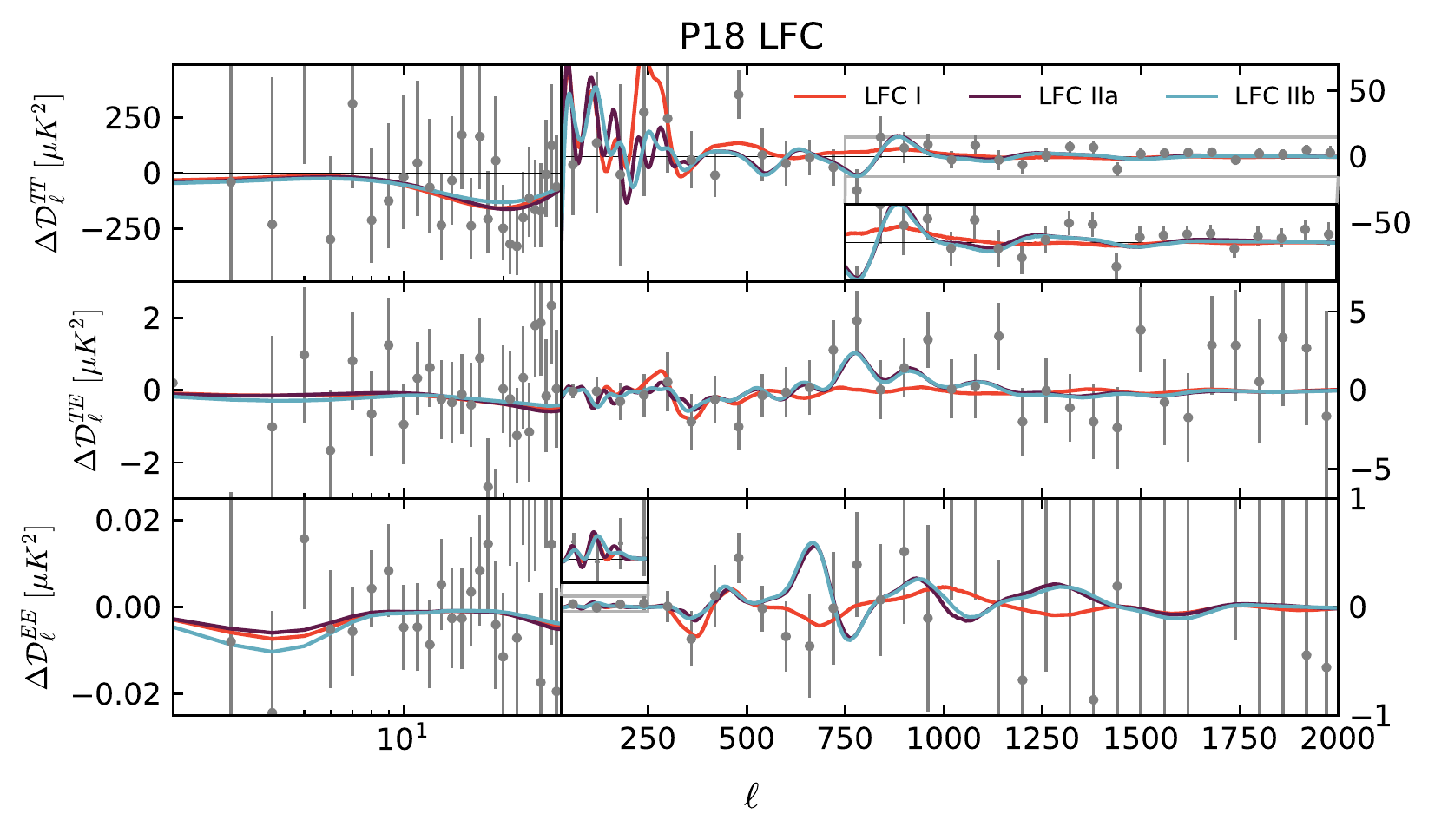}
	\includegraphics[width=\columnwidth]{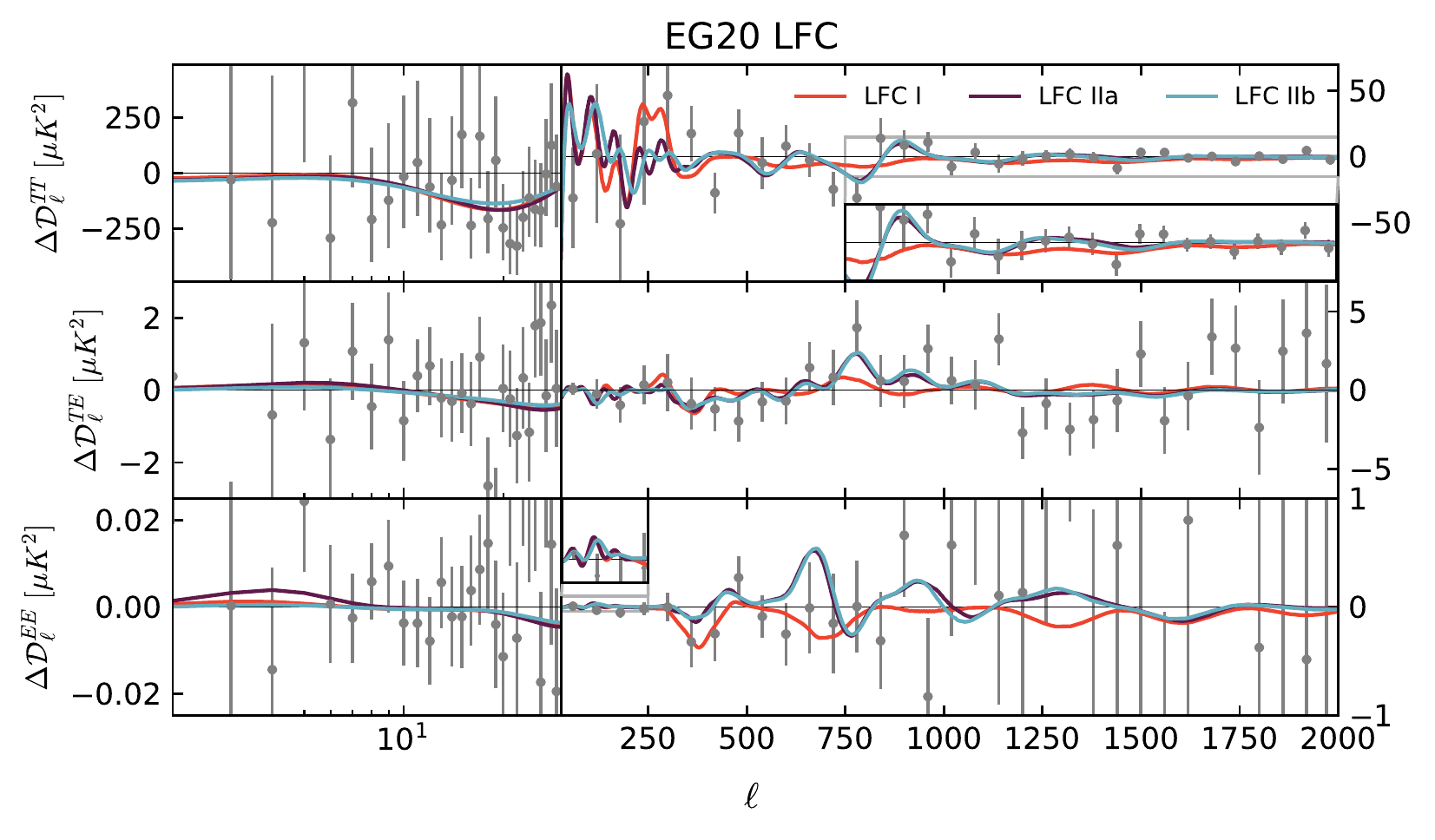}
	\caption{\footnotesize\label{fig:LFC_res} Residual plots of the full model for the Low-Frequency best-fit candidates to P18 [top] and EG20 [bottom] likelihood. Orange, purple and light blue lines correspond to LFC I, LFC IIa and LFC IIb respectively.}
\end{figure*}

\begin{figure*}
	\centering
	\includegraphics[width=\columnwidth]{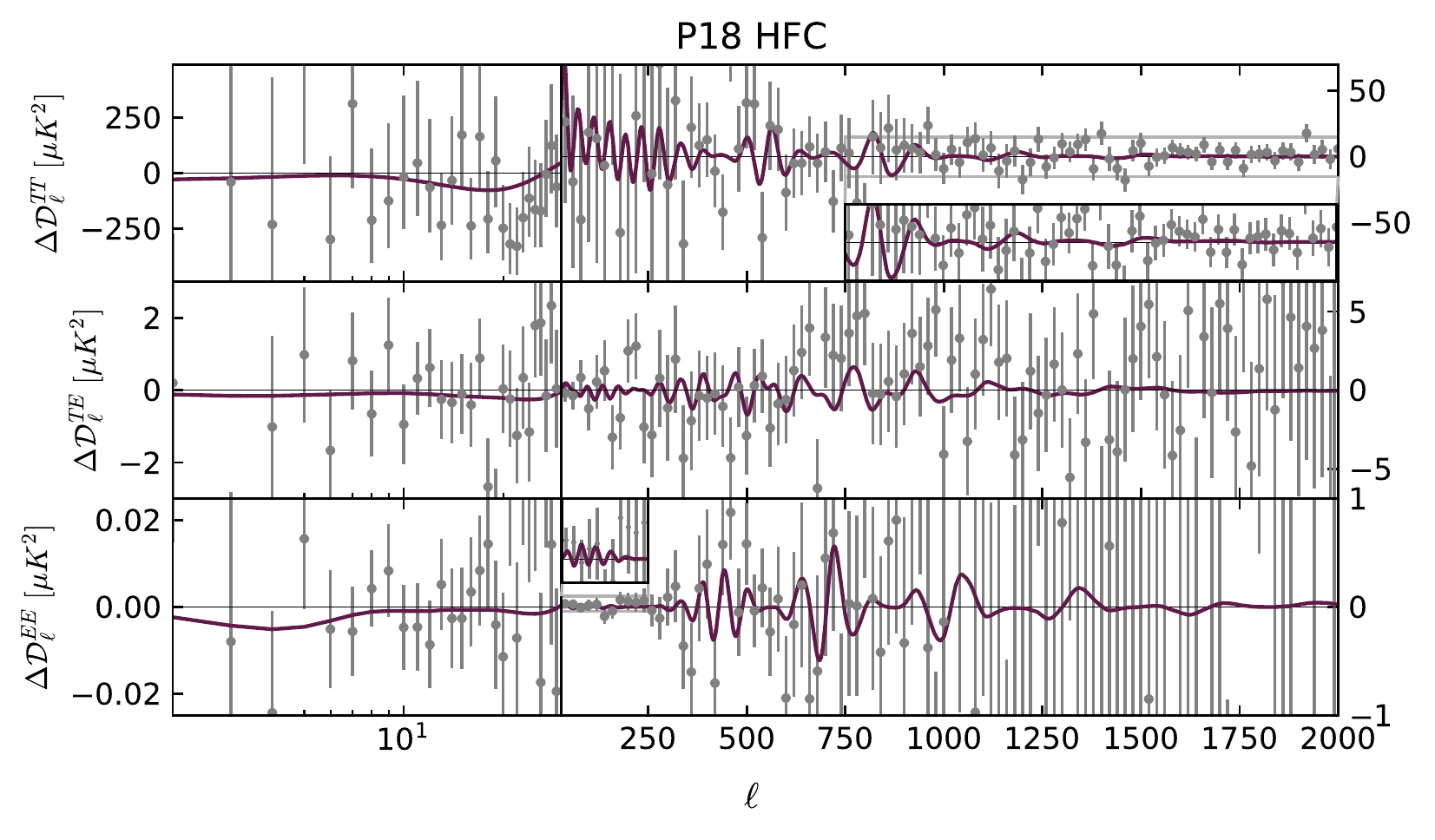}
	\includegraphics[width=\columnwidth]{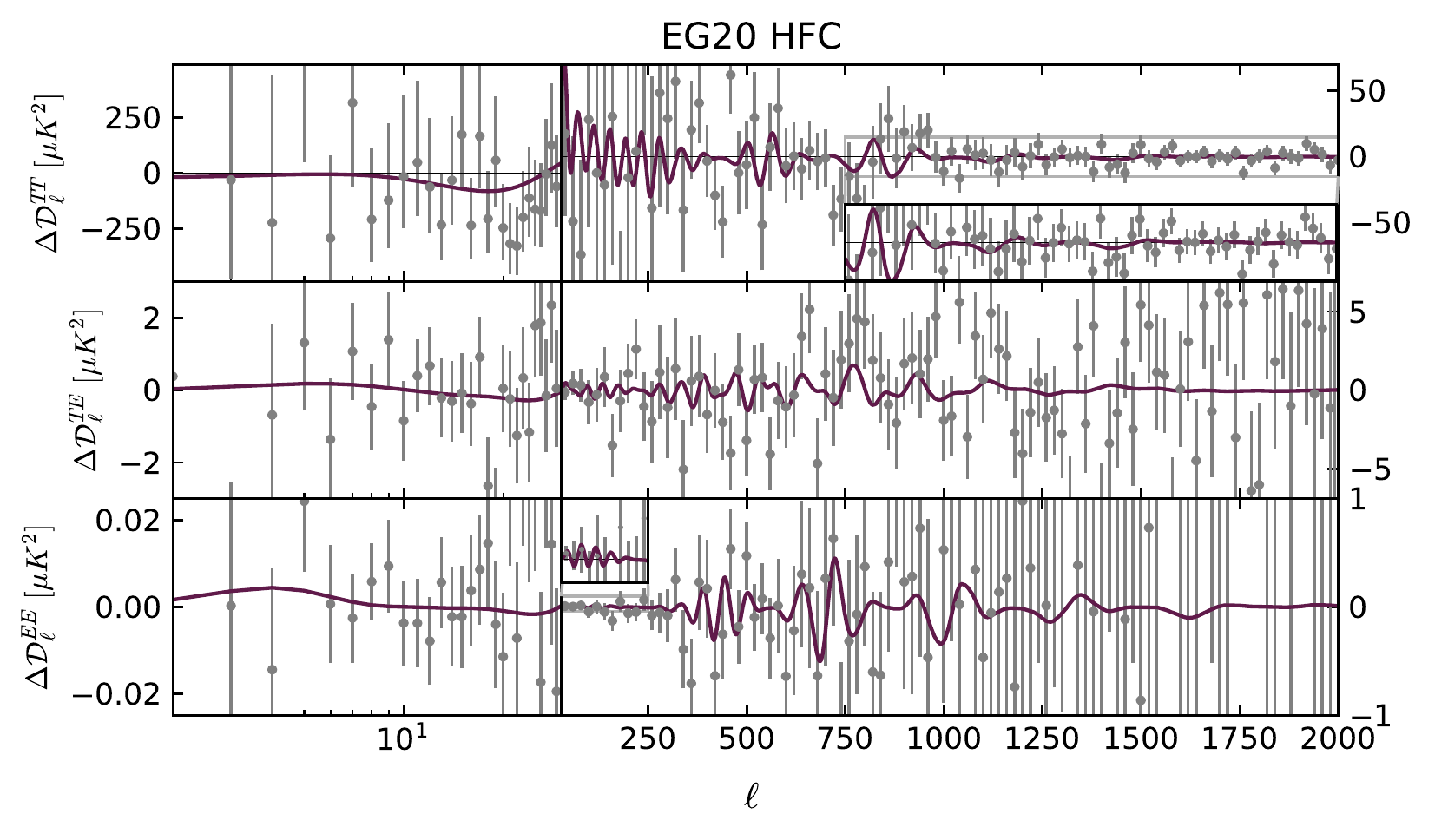}
	\caption{\footnotesize\label{fig:HFC_res} Residual plots of the full model for the High-Frequency best-fit candidate to P18 [top] and EG20 [bottom] likelihood.}
\end{figure*}

In order to better characterize our feature model we now look for combinations of parameters that provide the best fit to the CMB data we consider. Unlike the quantities shown in the previous section, that describe the statistical properties of the model and are used to perform model selection, such combinations are point estimators with zero measure in the large CPSC prior volume. Nevertheless, they are important for mainly two reasons. First, especially for models that are not very well constrained by data,  best-fit parameters do not necessarily lie on the peaks of the posterior distributions and therefore may not be straightforwardly inferred by simply inspecting the samples.\footnote{A recent and illustrative example is the Early Dark Energy model, where CMB and LSS data place a tight bound on the energy injected by scalar field into the cosmic fluid, but still large values of the injected energy provide the best-fit to these datasets \cite{Smith:2020rxx,Braglia:2020auw}. } This is particularly important in our model where we vary a large number of cosmological and nuisance parameters in the nested sampling and observe multimodal posterior distributions. 
Second, the best-fit model can be used to demonstrate how some of the CMB residual anomalies are addressed. Due to the low statistical significance of such anomalies, however, these are only best-fit {\em candidates} and only future experiments will tell us whether they really represent the true model of our Universe or not. 

As in Ref.~\cite{Braglia:2021ckn}, we classify best-fit candidates into low, medium and high frequency according to the value of the clock field effective mass as LFC: $m_\sigma/H\in[1,\,25]$, MFC: $m_\sigma/H\in[25,\,40]$ and HFC: $m_\sigma/H>40$. Since the clock frequency does not show a single peak posterior (see discussion above), a plethora of spectra with different values of $m_\sigma/H$ provide a better fit to data with respect to a featureless spectra. Presenting all of them would make the discussion cumbersome, so here we focus on best-fit candidates with $m_\sigma/H\sim10,\,18$ and $50$. This is motivated by the following reasons. The first two values, i.e. $m_\sigma/H\sim10$ and $18$ are the locations of the two most prominent modes in the posterior distribution of $m_\sigma/H$, as seen in Fig.~\ref{fig:triangle_log},\footnote{This also holds if $m_\sigma/H$ is used as free parameter in the nested sampling analysis, instead of $\log_{10} m_\sigma/H$, as shown by Fig.~\ref{fig:triangle} in Appendix~\ref{sec:LinPar}.} 
especially the value $m_\sigma/H\sim18$ that shows up more consistently in both P18 and EG20.
In addition, we also consider $m_\sigma/H\sim50$ as a representative value for high frequency spectra that improve the fit to data. Although many other frequencies can fit data better than the baseline (see for example samples around $m_\sigma/H\sim30$ in the right corner of Fig.~\ref{fig:triangle_log}), of all the medium and high frequency spectra, we find that the candidate with $m_\sigma/H\sim50$ provides the best fit to both P18 and EG20. In particular, we find that spectra with $m_\sigma/H\gtrsim52$ do not fit EG20 data as good as those with $m_\sigma/H\lesssim50$, as also proved in Ref.~\cite{Braglia:2021ckn}.
With this in mind, we now go on and comment on each candidate in turn. The parameters describing them as well as the $\Delta\chi^2$ improvement are given in Table~\ref{tab:candidates} and their spectra in Fig.~\ref{fig:bestfit}.

For residual comparison with P18, we make use of the unbinned data from the Planck Legacy Archive and use the inverse variance binning using only the diagonal components of the covariance matrix. Neglecting the off-diagonal terms will marginally underestimate the errors. However, we stress that since the residual plot is just for visualization purposes,  results of our analysis are not affected. For EG20, we extract the angular power spectrum data and the covariance matrices for different spectra obtained from auto and cross-correlations from different frequency channels. We subtract the baseline best-fit foreground angular power spectra from these spectra. Then the spectra are co-added following the procedure outlined in Section C.4 of~\cite{Planck:2015bpv}. TE and EE spectra in EG20 liklelihood are already used in the co-added form. The co-added spectra are then binned using inverse variance binning. We use different bin-widths for plotting purposes so that the fits to the outliers by the features of different frequencies can be noticed.

$\bullet$ {\em LFC I.}
The PPS for this candidate is shown in the upper panel of Fig.~\ref{fig:bestfit} and the corresponding residuals are shown in orange lines in Fig.~\ref{fig:LFC_res} for both P18 and EG20. Given the low frequency of the clock oscillations, we bin data in bins of width $\Delta\ell=60$ in order to better appreciate the residuals addressed by the candidate.

As for the other best-fit candidates, LFC I greatly improves the fit to lowT data, fitting the dip around $\ell\sim20-30$. However, the fit to high-$\ell$ data is significantly different depending on whether P18 or EG20 is considered. This can be qualitatively understood as follows. The spectrum is characterized by a clock signal with very broad oscillations that become almost negligible at scales $k\gtrsim0.04 {\rm Mpc}^{-1}$ due to the very low frequency $m_\sigma/H=9.23$. Most of the improvement to the fit to P18 thus comes from the fit to low-$\ell$ residuals. In fact, we see that the orange line in the P18 TT residuals of Fig.~\ref{fig:LFC_res} crosses many data points 
and also  addresses the binned data point around $\ell\sim60$ in EE which seems to be 2$\sigma$ away from zero (see the zoomed inset). On the other hand, in the case of EG20, these data points become more consistent with zero. In particular, we see that the EE data point around $\Delta\ell\sim60$ becomes consistent with zero at less than 2$\sigma$.

$\bullet$ {\em LFC II.} We now consider candidates with frequency corresponding to the main peak of the posterior distribution at $m_\sigma/H\sim18$. 
These candidates provide the global best-fit to data. We consider two different candidates which we label as IIa and IIb, that share the same clock signal, but a slightly different dip feature one, in order to examine in details what specific part of the signal best fits P18 and EG20.
Their PPS are shown in the middle two panels in Fig.~\ref{fig:bestfit} and the corresponding residuals are shown in purple and light blue lines in Fig.~\ref{fig:LFC_res}.
In particular, LFC IIa is characterized by a  smaller value of $N_0$ compared to LFC IIb, which corresponds to a dip feature located at slightly larger scales, and a larger value of $\Delta P_{\rm dip}/P$ and $C_\sigma/\Theta_f^2$ corresponding to a deeper and thinner dip followed by sharp feature oscillations with a larger amplitude.

Since the amplitude of the dip is larger, the fit to lowT is slightly better in the case of LFC IIa. The $\Delta\chi^2_\mathrm{high-\ell}$ of the two candidates differs mainly because of the different fit to $30\lesssim\ell\lesssim300$, as the residuals are essentially the same for larger multipoles. We see that in the case of P18 this leads to LFC IIa being the global best-fit whereas the global best-fit to EG20 is instead LFC IIb. Indeed for EG20, as discussed above, residuals at those scales are more consistent with 0, leading to a slightly better fit when the dip feature is not too sharp. Nevertheless, the differences between LFC IIa and LFC IIb and their $\Delta\chi^2$s are not very significant and, within the error bars, they can be safely considered as the same candidate and we refer collectively to them as LFC II in the following.

The most interesting feature in LFC II, however, is the clock signal at $\ell\gtrsim600$. As can be seen, the features address the dip around $\ell\sim750$ followed by a small bump in TT residuals and the corresponding bump in TE.
This improvement in the fit to high-$\ell$ data is robust to the change in the large-scale dip feature. In fact, as we will demonstrate in the following Section, it is almost unchanged when the amplitude of the dip feature is set to zero. 

Overall, our global best-fit candidates  provide an excellent fit to the CMB data, with an improvement of $\Delta \chi^2_{\rm TOT}=19.4$ and $17.7$  when considering P18 and EG20 respectively.

$\bullet$ {\em HFC.} Finally, we describe a candidate with a higher frequency in clock signal. The PPS for this candidate is shown in the lower panel in Fig.~\ref{fig:bestfit} and the corresponding residuals are shown in Fig.~\ref{fig:HFC_res}, where we now bin the data using a smaller bin width $\Delta\ell=20$.

In this case, the fit to lowT is slightly worse than the cases considered above because of a smaller amplitude of the dip, due to the following reason. With a larger $m_\sigma/H$, the separation between the scale of the sharp feature and that of the clock signal increases. This pushes the dip to the larger scale or the clock signal to smaller scale, away from the actual anomalies. 

HFC does not address the dip around $\ell\sim 750$. Its clock signal seems to provide some fit to the oscillations in the range $750\lesssim\ell\lesssim900$, as well as the bump in TE right after $\ell\sim750$.

As a final remark, we note that all the P18 candidates provide a fit to lowE data which is roughly $\Delta\chi_{\rm low-E}^2\sim1.5-2$ better than their EG20 counterparts. The reason is that the best-fit baselines for P18 and EG20, as already discussed, have different values for $A_s$ and $n_s$ (and consequently for the optical depth $\tau_{\rm reio}$), therefore providing a slightly different fit to lowE.

\begin{figure}
	\includegraphics[width=\columnwidth]{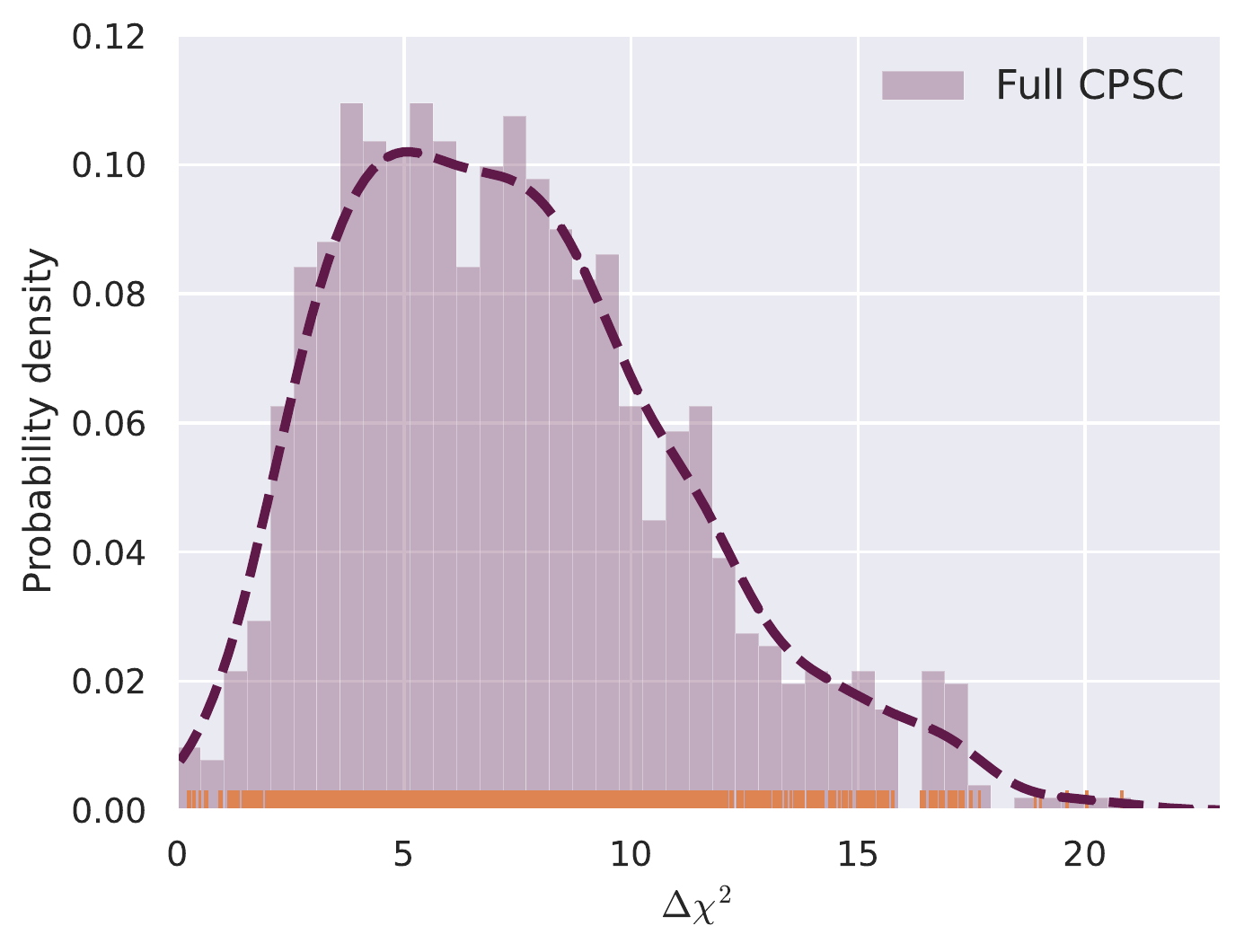}
	\caption{\footnotesize\label{fig:null_full_model} $\Delta\chi^2$ histogram for our model from fitting 1000 simulations of Planck noise. The dashed line is the probability distribution function estimated from the data points, that are shown with yellow ticks on the x-axis. Details about the simulation are provided in the main text. }
\end{figure}

	It was pointed out in earlier works \cite{Meerburg:2013dla,Peiris:2013opa} that spurious fit to noise typically leads to improvements of $\Delta\chi^2\sim10$ for some 3-parameter feature models. In order to test whether this is indeed the case for our model, we generate 1000 mock CMB angular power spectrum data using Planck noise. Our random realizations are drawn from the baseline best fit using a Gaussian multivariate distribution. In order to make the simulation as realistic as possible, we use the CamSpec covariance matrix to draw the mock realizations of the data. Note that here the correlations between multipoles are taken into account. We generate TT, TE and EE mock data for the multipole ranges probed by Planck. For each realization, we run an MCMC analysis for the baseline and for the CPSC model and extract the correspondent best fits to data.\footnote{Note that, while  MCMC is not able to capture the posterior distribution of feature parameters, here we are simply interested in finding the bestfit to the simulated data. We find that, in the absence of the foreground and calibration parameters, the best fit obtained with MCMC converges to the global best fit obtained with MCMC followed by BOBYQA minimization. } Due to the absence of any features in the mock signals, any positive $\Delta\chi^2$ will be entirely attributed to noise fitting.
	
	The histogram of $\Delta\chi^2$ is shown in Fig.~\ref{fig:null_full_model}, together with its estimated probability distribution function\footnote{The estimated probability distribution is computed with the Seaborn software \href{https://seaborn.pydata.org/}{https://seaborn.pydata.org/}}. Note that, as the fit to data is highly non-linear, the distribution is not exactly a $\chi^2$-distribution, as also noted in Ref.~\cite{Meerburg:2013dla}. Although improvements of $\Delta\chi^2\sim 20$, such as for LFCII, are possible, we find that the distribution has a broad peak around  $\Delta\chi^2\sim5$  which extends to  $\Delta\chi^2\sim5$ to $12$ and then falls off for larger values.  Compared to previous findings (see again Ref.~\cite{Meerburg:2013dla}) our results are more optimistic and show that noise fitting is expected to typically lead to $\Delta\chi^2\sim5-10$ for our 6-parameter model. Besides the different feature model considered, there are three main reasons for this to happen. First, we consider here the simultaneous fit to TT, TE and EE, which are simulated using the appropriate covariance matrices; while at the time when the work ~\cite{Meerburg:2013dla} was done, the uncertainties in the published polarization data were still too large to be relevant. While noise fitting to each dataset generically leads to a large improvement, when considered together it decreases significantly. Second, in our MCMC we also vary the background cosmological parameters for both the baseline and the CPSC model. Since each realization is characterized by  different bestfit values for $(\omega_{\rm b}h^2,\,\omega_{\rm CDM}h^2,\,100*\theta_s,\,\tau)$, keeping them fixed as in Ref.~\cite{Meerburg:2013dla} always leads to a larger $\Delta\chi^2$ as the feature fit compensates from the bad fit resulting from fixing wrong cosmological parameters. Third, the dip feature, which improves the fit to real Planck data to around $\Delta\chi^2\sim6-7$, is a very strong feature which is unlikely to fit the noise on top of a featureless signal. 
	
	We stress that this null test is only intended to understand what is the typical noise fitting expected in our model, but the correct way to assess  the evidence for our feature model is only through a Bayesian model selection, which we have presented in Sec.~\ref{sec:results}.

\section{Results of the data analysis: restricted model with $\Theta_f\to\infty$}
\label{sec:results_restricted}

\begin{figure*}[t]
	\includegraphics[width=\columnwidth]{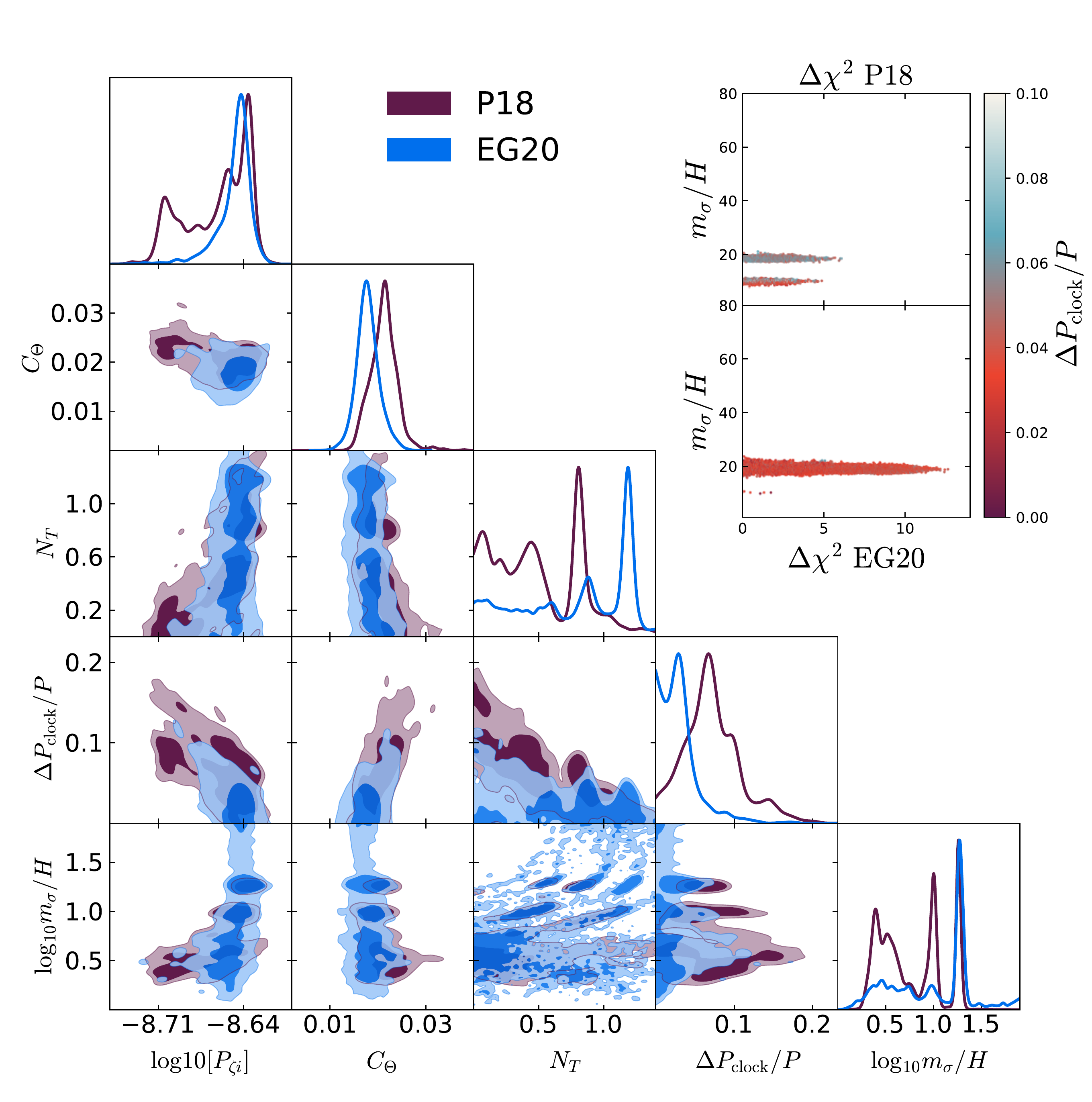}
	\caption{\footnotesize\label{fig:triangle_log_3params} Triangle plot for the restricted model for PlikHM bin1 (P18, purple) and Clean CamSpec v12.5HMcln (EG20, light blue) datasets.}
\end{figure*}

The best-fit analysis performed in the previous Section shows that the improvement in $\Delta\chi^2$ is mainly driven by the fit to high-$\ell$ data. Therefore, it is interesting to repeat the data analysis for the simpler model, introduced in Section~\ref{sec:restricted model}, in which the dip feature is removed by sending $\Theta_f\to\infty$. As mentioned above, a nice aspect of this restricted model is that the power spectrum is only described by  3 extra parameters, namely the frequency, maximum amplitude and location of the clock signal, described by $\log_{10}m_\sigma/H,\,\Delta P_{\rm clock}/P$ and $N_T$, respectively. 	
In this analysis, we keep using the priors provided in Table~\ref{tab:priors}, except for the location parameter for which we use a wider prior $N_T\in[0,\,1.4]$, in order for the clock signal to span the  whole range of scales, since we fix $N_0=14.3$ (see Section~\ref{sec:restricted model}).

The constraints on the model parameters are shown in Fig.~\ref{fig:triangle_log_3params}. Considering the baseline parameters first, we see that the posterior distributions of the amplitude $\log_{10} P_\zeta$ now show a tail at large negative values, which is correlated with small values of the frequency parameter $\log_{10}m_\sigma/H\lesssim0.6$. The reason is that for those frequencies $V(\sigma)$ is shallower making it easier for the centrifugal force to shift the inflaton in the $\sigma$-direction. This results in a larger correction to the kinetic term of $\Theta$ and, to compensate for this effect, the parameter $\log_{10} P_\zeta$, which we remind the reader is defined as $\log_{10} P_\zeta\equiv\log_{10} H^2/(4\pi^2\dot{\Theta}^2)$,	needs to acquire a smaller value. 
For such small values of $\log_{10}m_\sigma/H$, the power spectrum exhibits some suppression of power at large scales, followed by a small bump and a near scale invariant power spectrum with a spectral index that, as usual, depends on $C_\Theta$. (A detailed discussion of these points is presented in Appendix~\ref{Sec:several details}.)
Since the spectral index is not significantly affected by the specific value of $m_\sigma/H$, the constraints on $C_\Theta$ are similar to those in the full model scenario.

We now comment on the clock signal parameters. Starting form the location parameter $N_T$, we see that, although its posterior shows some peaks, it is not constrained, similarly to the previous Section. We again get only upper bounds on the amplitude, that is $\Delta P_{\rm clock}/P < 0.144$ and $\Delta P_{\rm clock}/P  < 0.0807$ at 95\% CL for P18 and EG20 respectively. As can be seen from Fig.~\ref{fig:triangle_log_3params}, amplitudes close to this upper bound are allowed only for models with small values of $m_\sigma/H$ which, as discussed before, only show a bump after the large scales suppression. 

Inspecting the 2-dimensional P18 contours in the amplitude-frequency plane, it is very interesting to notice an island around $m_\sigma/H\sim18$ which is more than 2$\sigma$ away from zero, explaining the peak in the posterior distribution of $\Delta P_{\rm clock}/P$. A similar feature is observed also in the EG20 contours, where, however, the island is only 1$\sigma$ away from zero. Regarding the frequency of the clock signal, the situation is very similar to the one presented in the previous Section and we see again a multimodal posterior which is however strongly peaked at the frequency of the best-fit candidate presented before, i.e. $m_\sigma/H\approx 18$.

Finally, we report the Bayes factor of the model with respect to the baseline in Table~\ref{tab:evidence_nodip_log}. Despite the reduced parameter space compared to the full model, the evidence of the model is still inconclusive and consistent, within the errors reported, with that of the full model, showing that the reduction in the number of parameters is compensated by the worsening of the fit to high-$\ell$ data.

\subsection{Comparison with simple resonant model}
Before we go on to inspect the best-fit candidates in the restricted model, it is instructive to compare our findings with a simple resonant (SiRe) model extensively studied in the literature, also described by three parameters. The power spectrum of the model is described by the following template \cite{Chen:2008wn,Akrami:2018odb}: 
\begin{equation}
\label{eq:simple_template}
P(k)=A_s \kappa^{n_s-1}\left[1+\frac{\Delta P}{P}\sin\left(\omega \ln 2 k+\varphi\right)\right] ~,
\end{equation}
where $\kappa\equiv k/k_*$ and $k_*=0.05\,{\rm Mpc}^{-1}$. This PPS can arise in the so-called resonant models where there are periodic ripples in the inflaton potential \cite{Chen:2008wn,Flauger:2009ab,Flauger:2010ja,Chen:2010bka}. Here we denote it as ``simple resonant model" to distinguish it from the resonant mechanism involved in the clock signal. Although this power spectrum also describes a resonant feature, it is significantly different from the clock signal considered above for two reasons. The first is that it does not contain any sharp feature signal and the second is that the resonant oscillations extend over the full range of scales probed by Planck, and their amplitude is constant over this range.
In order to compare this model to ours, we identify the parameters $\Delta P/P$ and $\log_{10}\omega$ with $\Delta P_{\rm clock}/P$ and $\log_{10}m_\sigma/H,$ respectively, and use the priors given in Table~\ref{tab:priors}. For the phase $\varphi$, we simply adopt the prior $\varphi\in[0,\,2\pi]$.

\begin{figure}
	\includegraphics[width=.9\columnwidth]{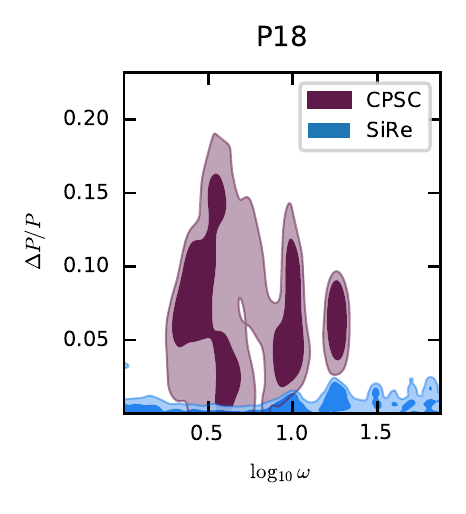}
	\includegraphics[width=.9\columnwidth]{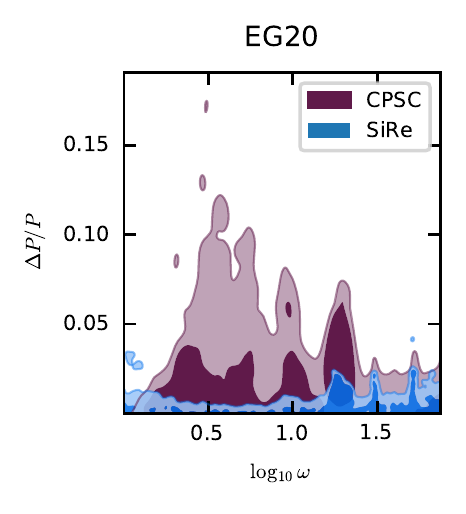}
	\caption{\label{fig:CPSC_vs_TEMPLATE_LOG} Comparison of the amplitude and frequency 2-dimensional posterior contours of the restricted CPSC (purple) and SiRe (blue) model. We show contours for P18 and EG20 in the upper and lower panels respectively.}
\end{figure}

We show the 2-dimensional posterior distribution of frequency and amplitude parameters in Fig.~\ref{fig:CPSC_vs_TEMPLATE_LOG} for both P18 and EG20 and we compare them to those obtained for our restricted CPSC model. Since the phase $\varphi$ cannot be compared to $N_T$  and is unconstrained by current data, we do not show its posterior.

First, we note that the contours for the SiRe model shown in the upper panel for P18 are consistent with the findings of Ref.~\cite{Akrami:2018odb}.\footnote{\label{ft:SiRe}
We note that, for the same SiRe model and using the same P18 likelihood, the value of our Bayes factor relative to the featureless model (presented in Table \ref{tab:evidence_nodip_log}) is less than that obtained in \cite{Akrami:2018odb}, although the difference ($\sim 1.4$) is not significant. This is due to our more conservative analysis (here we allow nuisance parameters to vary in contrast to Planck analysis) and the use of different sampler (we use Polychord compared to MultiNest used in Planck). This suggests that our method gives a more pessimistic Bayes factor than that by the Planck team. Since we use the same method throughout the paper, this difference does not affect the comparisons of the Bayesian evidences between different models in our paper.} Like in that analysis, we observe two main peaks at $\omega\sim 10$ and $\omega\sim18$, but the frequency of the oscillations  is unconstrained and data allow also larger (and smaller) values of $\omega$. Similar contours are observed for EG20. In this case, though, the mode at $\omega\sim18$ is more pronounced than the others. This is also observed in the posterior distribution for the restricted CPSC model (compare purple and blue contours in Fig.~\ref{fig:CPSC_vs_TEMPLATE_LOG}).

The contours also show that the amplitude of the SiRe model is much more constrained than the CPSC one. This is expected, as the oscillations in the SiRe extend to all scales and therefore, while they can provide a better fit to the CMB residuals at some scales, at the same time they spoil the fit to the spectra at other multipoles. On the other hand, the amplitude parameter in the CPSC model represents the {\em maximum} amplitude the resonant signal takes before rapidly dying off, so it is natural to get constraints that are significantly relaxed with resepct to the SiRe model.

The Bayes factors listed in Table~\ref{tab:evidence_nodip_log} show that the restricted CPSC model provides a slightly better fit to the anamolies than the SiRe model.

\begin{table}
	\centering
	\begin{tabular}{|l|l|l|}
		\hline
		Dataset          & P18& EG20  \\ \hline
		$\ln B\, ({\rm CPSC})$     &      $-0.28\pm0.38$           & $-0.66\pm0.36$         \\         \hline			$\ln B\, ({\rm SiRe})$     &      $-2.29\pm0.38$           & $-2.24\pm0.38$         \\         \hline
	\end{tabular}
	\caption{Bayes factors ($\ln B$) obtained for the restricted CPSC model and the SiRe model {\it w.r.t.}~the baseline model. }
	\label{tab:evidence_nodip_log}
\end{table}

\subsection{Best-fit candidates}

\begin{table}[t]
	\begin{center}
		\begin{tabular}{|c||c|c|}
			\hline CPSC
			& P18  & EG20  \\ \hline
			$\log_{10} P_{\zeta *}$    &-8.640& -8.640 \\ \hline 
			$N_T$  &1.19 &  1.19\\ \hline  	
			$C_\Theta$ & 0.0181 & 0.0184 \\ \hline 
			${m_\sigma}/{H}$ & 18.19 & 18.18 \\ \hline
			$\Delta P_{\rm clock}/P$  &0.038    &0.038 \\  
			\hline
			\hline 
			$k_0\,\times10^{3} {\rm Mpc}$ & 2.030 & 2.030\\ \hline
			$\xi\sigma_f$       &0.058 & 0.058\\  
			\hline 
		\end{tabular}

		\begin{tabular}{|c||c|c|}
			\hline SiRe
			& P18  & EG20  \\ \hline
			${\rm{ln}}(10^{10} A_s)$       & 3.046&3.042 \\  \hline  
			$n_s$       &0.9648 & 0.9656\\  \hline  
			$\omega$  &18.02 & 18.13\\ \hline  
			$\varphi$  & 5.80& 5.92\\ \hline  
			$\Delta P/P$  &0.016 & 0.015\\	
			\hline
		\end{tabular}
		\begin{tabular}{|c||c|c|c|c|c|c||}
			\hline
			& \multicolumn{5}{c|}{P18 (Plik bin1)} \\ \hline
			& $\Delta \chi^2_{\rm TOT}$   & $\Delta\chi^2_\mathrm{prior}$    & $\Delta\chi^2_\mathrm{high-\ell}$  & $\Delta\chi^2_\mathrm{low-T}$ & $\Delta\chi^2_\mathrm{low-E}$  \\ \hline  
			CPSC&   12.9    & -0.10    &11.07 & 0.62 & 1.34 \\ \hline
			SiRe  &  12.0  &   0.05  &10.56 &0.11  &1.20  \\ \hline\hline
			& \multicolumn{5}{c|}{EG20 (CamSpec v12.5HMcln)} \\ \hline
			CPSC      & 13.4 & -0.11   &13.28    & 0.30   &-0.03   \\\hline						SiRe  & 10.6   &-0.27     & 10.85& 0.09 & -0.04 \\ \hline
			
		\end{tabular}
		\caption{[Top] Best-fit candidates for the restricted CPSC model found in P18 and EG20 likelihood using TTTEEE+lowT+lowE. $N_0$ is fixed to $N_0=14.3$ (see Sec.~\ref{sec:restricted model}) and $k_0$ is now defined as $k_0\equiv a(N_0+N_T) H(N_0+N_T)/2$. In the first table, the two parameters below the double horizontal line are not independent of the first five parameters.
			[Bottom] Individual $\chi^2$'s for each candidate.}
		\label{tab:Nodipcandidate}
	\end{center}
\end{table}

\begin{figure}
	\includegraphics[width=\columnwidth]{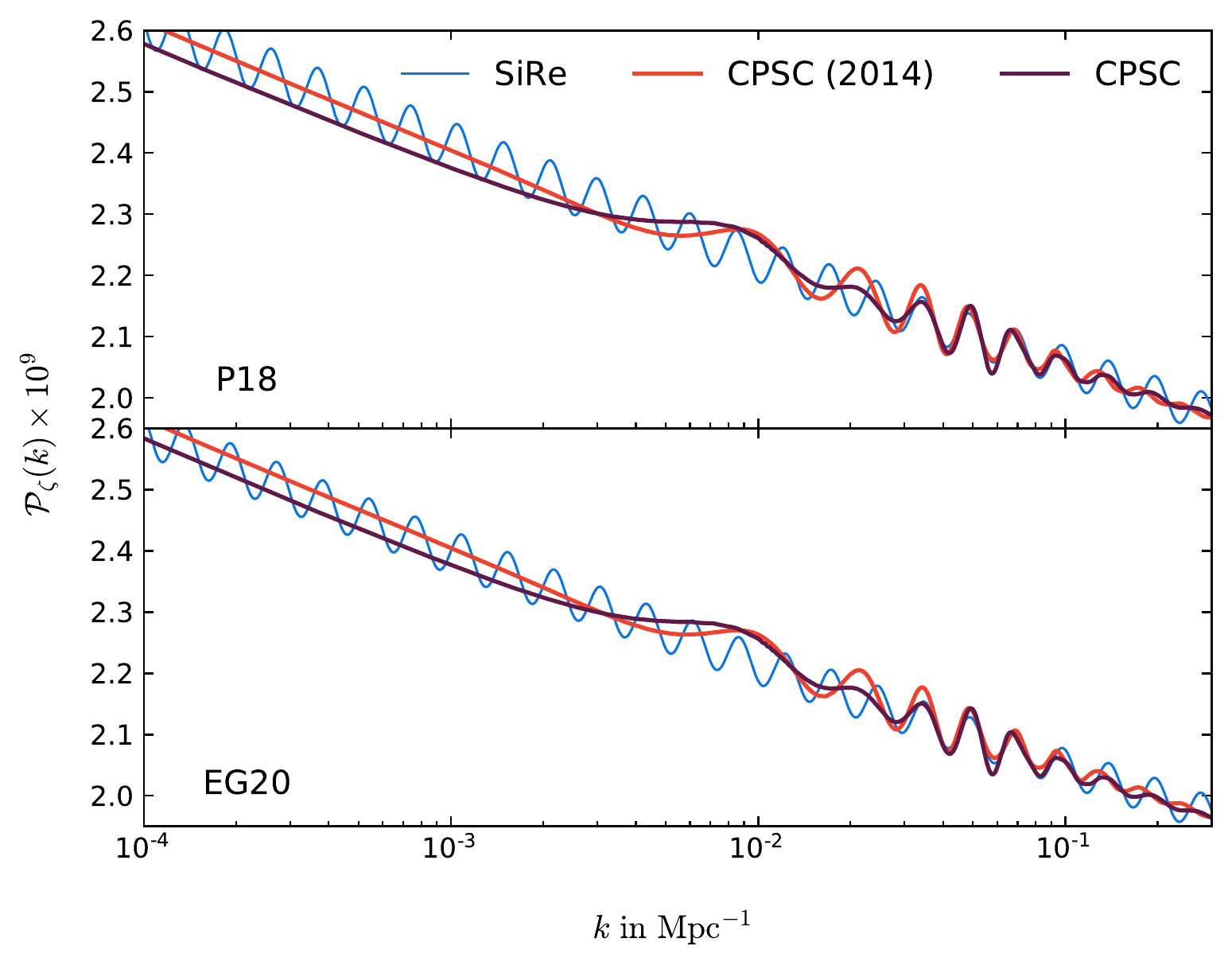}
	\caption{\label{fig:LFC-nodip} Best-fit candidates for the restricted CPSC model and the SiRe model. The candidates for P18 and EG20 are shown in the upper and lower panel respectively. For a comparison, we also plot a best-fit candidate found in \cite{Braglia:2021ckn} for another CPSC model \cite{Chen:2014joa,Chen:2014cwa}.}
\end{figure}

\begin{figure*}
	\centering
	\includegraphics[width=\columnwidth]{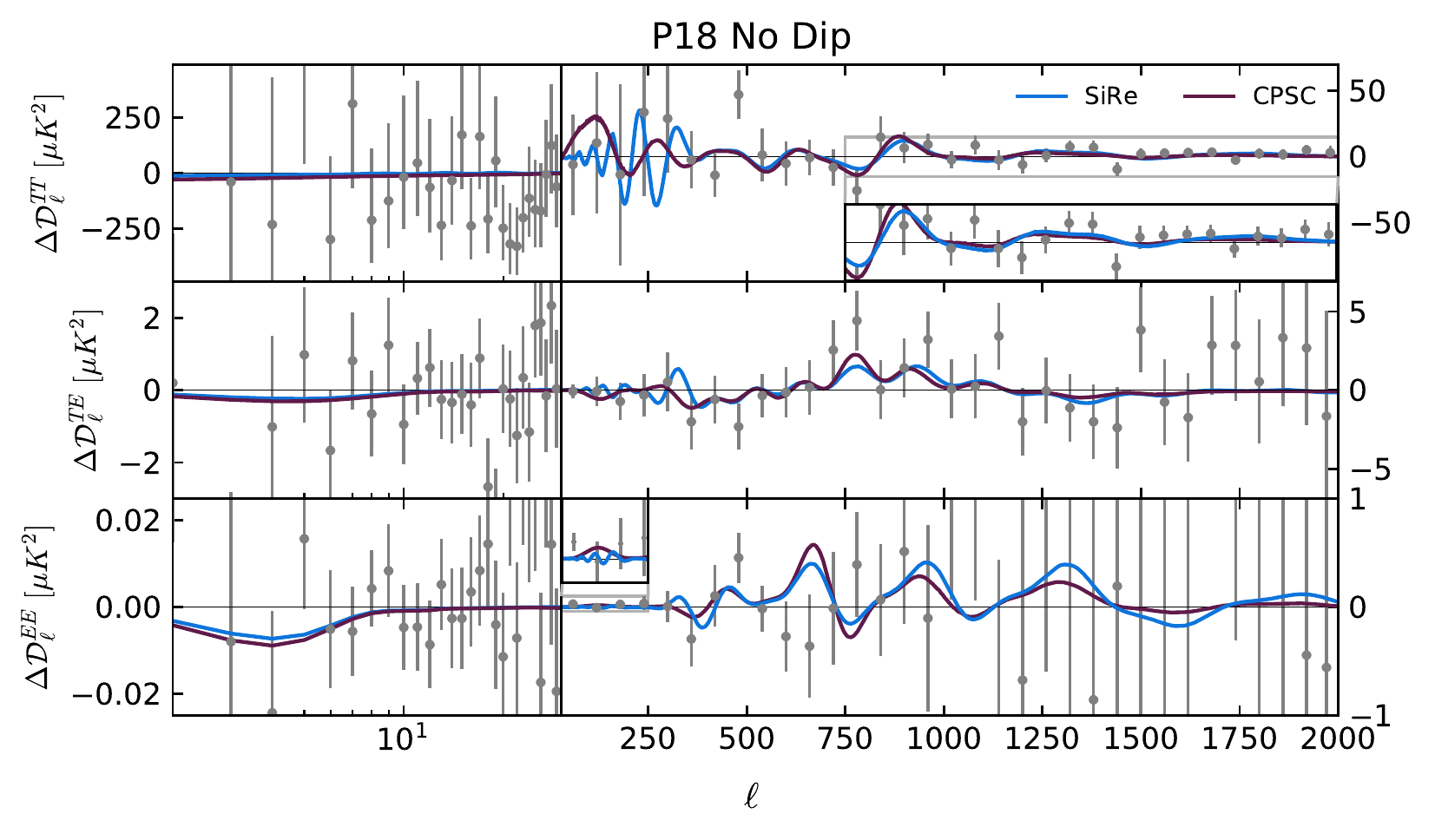}
	\includegraphics[width=\columnwidth]{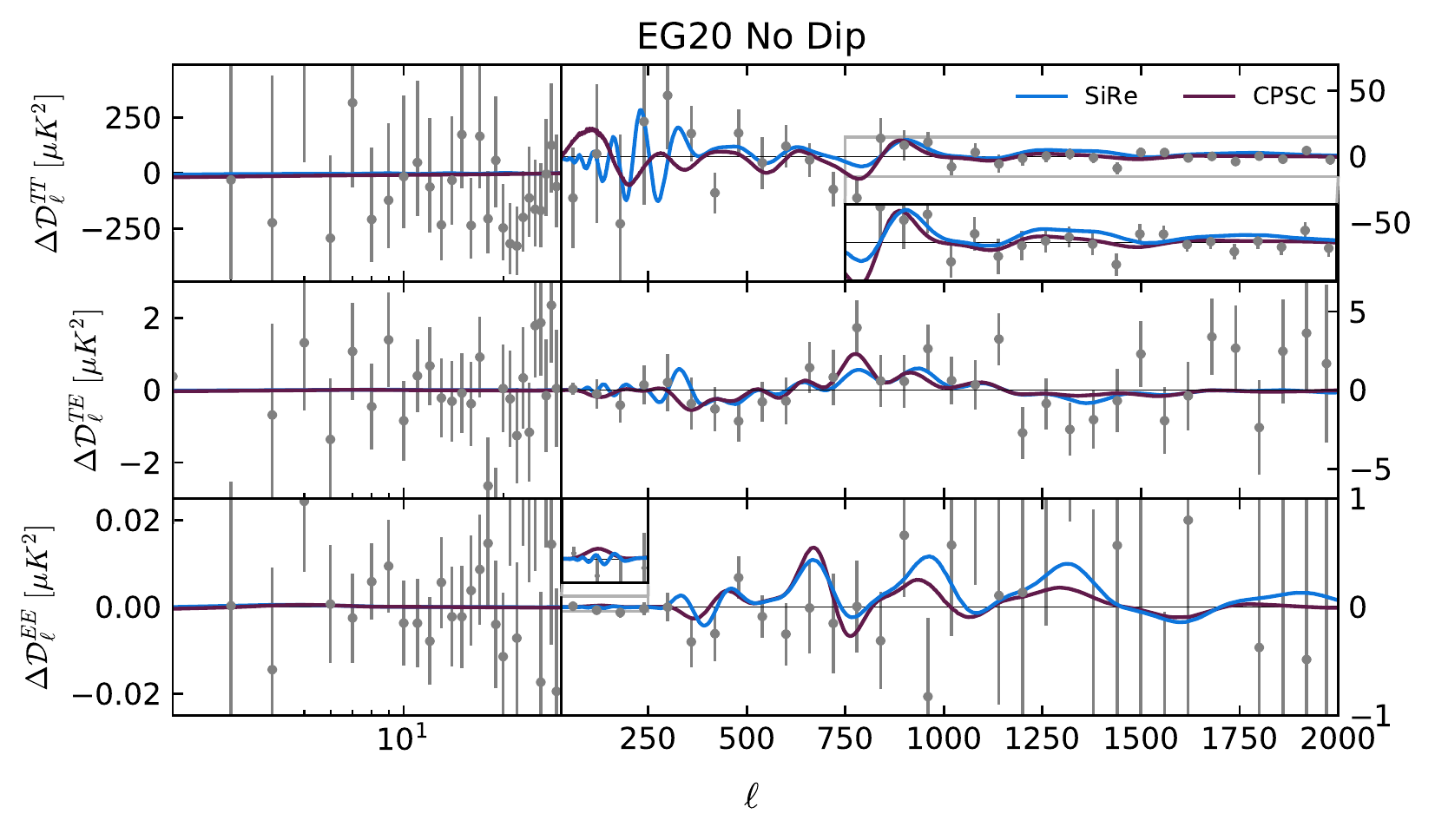}	\caption{\footnotesize\label{fig:Res_nodip} Residual plots for the best-fit candidates to P18 [top] and EG20 [bottom] for the restricted CPSC model (purple) and for the SiRe model (blue).}
\end{figure*}

\begin{figure}
	\includegraphics[width=\columnwidth]{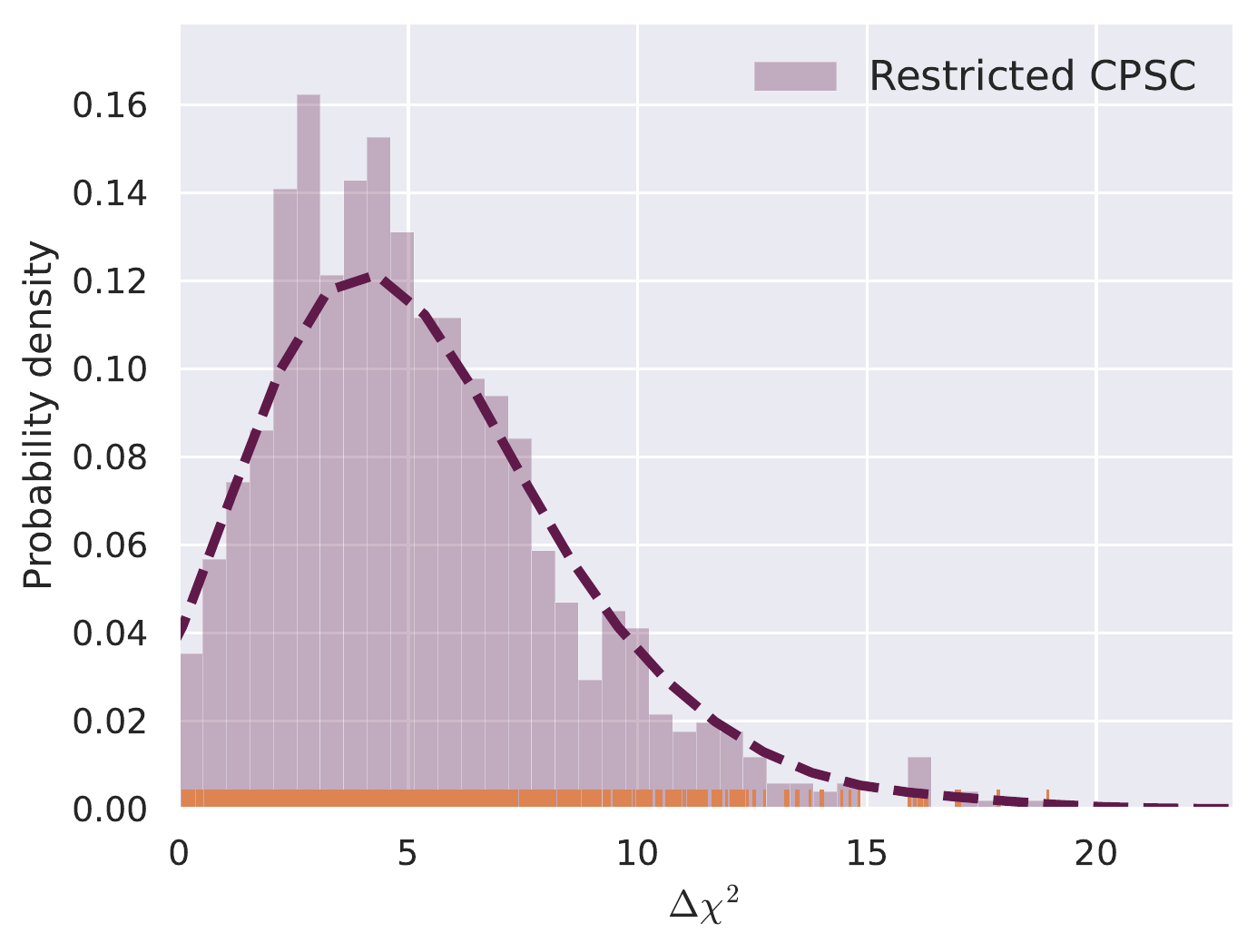}
	\caption{\footnotesize\label{fig:null_3P_model} $\Delta\chi^2$ histogram for the restricted model from fitting 1000 simulations of Planck noise. The dashed line is the probability distribution function estimated from the data points, that are shown with yellow ticks on the x-axis.  }
\end{figure}

We now search for best-fit candidates for the restricted model and compare them with the candidates for the SiRe model. As in the full model case, spectra with different values of the frequency can improve the fit with respect to the featureless model. However, for simplicity, for this restricted model we only show the global best-fit candidates for both P18 and EG20. The corresponding best-fit parameters, PPS and residual plots are shown in Table~\ref{tab:Nodipcandidate}, Fig.~\ref{fig:LFC-nodip} and \ref{fig:Res_nodip}, respectively.

Several conclusions can be drawn by looking at the table and the plots. First of all, since none of the models addresses the low-$\ell$ dip, the  fit to lowT and lowE is very similar to that of a featureless spectrum, resulting in an overall $\Delta\chi^2$ which is lower than that of the full CPSC model. Nevertheless, the fit of the restricted CPSC candidate to high-$\ell$ is very similar to that of the full CPSC model. The reason is that the part of the feature signal ascribed to the oscillations of the massive field is essentially the same as that of LFC II, except for some small residual oscillations induced by the step in the potential in LFC II. Indeed, a comparison of Fig.~\ref{fig:LFC_res} with Fig.~\ref{fig:Res_nodip} shows that the two signals induce exactly the same residuals for $\ell\gtrsim500$, but different residuals for $\ell\lesssim500$, leading to a fit to high-$\ell$ data which is slightly worse (better) for P18 (EG20) than LFC II. 
The slight difference in the best-fit $\chi^2$ between P18 and EG20 can be understood as the residuals around $\ell\lesssim350$ being slightly different between P18 and EG20. As mentioned in Section~\ref{sec:BF_full}, this is also the reason why P18 is better fitted by LFC IIa and EG20 by LFC IIb, because both show the same clock signal but slightly different dip feature signals. In this case, the restricted CPSC candidate fails to fit the data point around $\ell\sim60$ in the EE P18 residuals which was instead fitted by LFC II. On the other hand it does provide a very good fit to the binned TT EG20 data points around  $\ell\sim60$ and $360$ which qualitatively explains why the fit to EG20 is slightly better than LFC II. 

Regarding the SiRe model candidates, it is interesting to see that they share the same frequency as the CPSC candidates, 
as was also reported in the analysis of Ref.~\cite{Akrami:2018odb} (see also Footnote~\ref{ft:SiRe}). Moreover, not only are the logarithmic oscillations exactly in phase with those in the resonant part of the CPSC signal, but also their amplitude is nearly the same around $k\sim0.03-0.06\,{\rm Mpc}^{-1}$, before the clock oscillations slowly decay away towards smaller scales.
Although for P18 the $\Delta\chi^2$'s of both models are very similar, for EG20 the CPSC model fits better than SiRe. This is because larger oscillation amplitudes are allowed in CPSC than SiRe near $\ell \sim 750$ to better fit the anomalies in both TT and TE spectra, for reasons explained previously.

We also note that both the Bayesian evidence and the best-fit $\chi^2$ improvement of the restricted model are slightly better than those of another CPSC model proposed in \cite{Chen:2014joa,Chen:2014cwa} and compared with the Planck data in
\cite{Braglia:2021ckn}. As reviewed in Sec.~\ref{Sec:Sensitivity}, the main difference of these two models is the arrangement of the step location, and the model in \cite{Chen:2014joa,Chen:2014cwa} has one more parameter than the restricted model here. It is an interesting subject to study how well the future data can tell such fine details between models (see Fig.~\ref{fig:LFC-nodip} for the current best-fits).

Two main lessons can be learned from this Section. On one hand, it seems that the full CPSC signal, composed of a non-trivial sequence of sharp and resonant features, is better than a simple resonant model in fitting the Planck data (although the difference is not  significant). On the other hand, and most importantly, we observe that three models, i.e.~the full and restricted CPSC and the SiRe one, hint at a series of logarithmic oscillations with a frequency $\omega\sim18$ in the range of wavenumbers $k\sim0.03-0.06\,{\rm Mpc}^{-1}$. In the context of primordial standard clock, this would amount to hinting that a massive particle, with mass 18 times the Hubble scale of inflation and oscillating during the inflationary period, could be observed through CMB anisotropies observations.

	Finally, like what we did for the full model in Sec.~\ref{sec:BF_full}, we would like to understand what is the typical $\Delta\chi^2$ magnitude expected from noise fitting. The results are shown in Fig.~\ref{fig:null_3P_model}. The $\Delta\chi^2$  distribution peaks around $4$ and only $21$  out of  our $1000$ realizations display $\Delta\chi^2\ge13$, which is the improvment to the fit from our bestfit candidate. As in our analysis for the full model, although such an improvement can in principle be attributed to noise fitting, it seems quite unlikely. Also in this case, for the same reasons explained in Sec.~\ref{sec:BF_full}, our results are more optimistic than Ref.~\cite{Meerburg:2013dla}.

\section{Forecasts for future CMB experiments}
\label{sec:forecasts}

Given the best-fit candidates found in the previous Sections, it is now our goal to explore the prospects of their detection with future CMB surveys.
In this decade, one space based and several ground based CMB missions will start observing the CMB sky. LiteBIRD~\cite{Hazumi:2021yqq} is a recently approved JAXA space CMB mission dedicated to large scale observations. There are several approved ground based CMB observations (Simons Observatory~\cite{Ade:2018sbj}, POLARBEAR~\cite{POLARBEAR:2016wwl} 
and BICEP3-Keck array~\cite{Moncelsi:2020ppj}) that also will be operational in a few years.  More recently, CMB S4, an ambitious next generation ground based experiment consisting in a network of dedicated telescopes in the South Pole and Atacama~\cite{CMB-S4:2016ple,CMB-S4:2017uhf} has  received strong support in the  ASTRO2020 report\footnote{\href{https://www.nap.edu/resource/26141/interactive/?utm_source=NASEM+Space+and+Physics&utm_campaign=faf962de05-EMAIL_CAMPAIGN_2018_11_28_05_46_COPY_01&utm_medium=email&utm_term=0_6f3f7595c9-faf962de05-522397405}{https://www.nap.edu/resource/26141/}}. Finally, 
near-ultimate CMB space missions such as proposed CORE~\cite{Delabrouille:2017rct} (not funded), ECHO~\cite{CMBBHARAT} (CMBBHARAT proposal), Probe of Inflation and Cosmic Origins (PICO)~\cite{Hanany:2019lle} (NASA funded study) are designed to explore the full sky with great sensitivity and resolution. 

A primary goal of these observations is to constrain or detect the primordial B-mode polarization. To reach this goal, the E-mode polarization, which is a significant background for the B-mode, will have to be measured very precisely along the way. In this work, we will be making use of their E-mode sensitivity.

While single ground based survey cannot cover the entire sky as done in space missions and therefore cannot constrain the physics at very large scales, they accurately map intermediate to small scale sky as they have many more detectors and relatively small beams.
Therefore, the presence of features at both large and small scales in our candidates provides an unique opportunity to explore our abilities to detect features with upcoming and proposed observations. 
Our goal is thus to build a roadmap of how they will improve constraints on our model and its statistical significance over the next decades. To do so, we will forecast the constraints on our models for four different observations, namely, SO, LiteBIRD, CMB S4 and PICO, and their combinations.

We now describe in details the setup of our forecast analysis. The angular power spectrum of noise associated to a survey can be expressed as~\cite{Knox:1995dq,Tegmark:1997vs},
\begin{equation}
{\cal C}_l^{\rm Noise}=\frac{4\pi f_{\rm sky}\sigma^2}{N b_\ell(\theta)^2},  
\end{equation}
with $f_{\rm sky}$ denoting the fraction of the sky covered by the survey, and $\sigma$ being the r.m.s.~noise per pixel with $N$ pixels covering the sky. $b_\ell$ represents the beam function for observation that scales as,
\begin{equation}
    b_\ell(\theta)=\exp(-\ell(\ell+1)\theta^2/2),
\end{equation}
with $\theta$ being the standard deviation of the assumed Gaussian beam. Denoting the Full Width at Half Maxima as $\theta_{\rm FWHM}$, $\theta$ can be expressed  as $\theta_{\rm FWHM}=\sqrt{8\ln 2}\theta$. Using these relations, the noise power spectra can be obtained as

\begin{eqnarray}
{\cal C}^{\rm Noise}&=&\frac{4\pi f_{\rm sky}\sigma^2}{N}\exp\left[\frac{\ell(\ell+1)\theta_{\rm FWHM}^2}{8\ln 2}\right],\nonumber\\
&=&s^2\exp\left[\frac{\ell(\ell+1)\theta_{\rm FWHM}^2}{8\ln 2}\right],
\end{eqnarray}
with $s^2=\frac{4\pi f_{\rm sky}\sigma^2}{N}$ reflecting the measure of the sensitivity (inverse of the weight per solid angle). 

On top of the instrumental noise that increases at small scales (${\cal C}^{\rm Noise}\sim\exp(\ell^2)$ for large $\ell$), the cosmic variance noise plays the major part at large and intermediate scales. Since we have $2\ell+1$ modes averaging to give the band power ${\cal C}_\ell^{\rm CMB}$, the $\Delta{\cal C}_\ell$ due to the cosmic variance is given by,

\begin{equation}
    \Delta{\cal C}_\ell=\sqrt{\frac{2}{(2\ell+1)f_{\rm sky}}}\left[{\cal C}_\ell^{\rm CMB}+{\cal C}_l^{\rm Noise}\right] ~.
\end{equation}

We now describe the experimental setups adopted in our forecast analysis. Of the experiments mentioned above, SO is the only one  that is already operating and will start taking data in early 2023. A first data release can be reasonably  expected well within the time of a decade. SO has 6 frequency channels between 27-280 GHz covering 40\% of the full sky in the large aperture telescope (LAT). We use the baseline noise configuration in the LAT for our analysis.\footnote{The temperature and polarization noise power spectra for SO that we use in our analysis are available at \href{https://github.com/simonsobs}{https://github.com/simonsobs}.} LAT has a temperature noise level of 6$\mu K-arcmin$ (combined noise levels in the 93 GHz and 145 GHz channels). SO is expected to have a resolution better then $3'$. 
Note the temperature noise of SO is very much affected by atmospheric noise, which makes it quite large at multipoles $\ell\lesssim1000$. To obviate this, we combine the noise of SO with that of Planck using inverse weighted sum at $40 \leq\ell\leq 1500$ for  the same sky fraction. Furthermore, we add Planck noise for the remaining $f_{\rm sky}=0.3$  and include information from large scales using Planck noise at $\ell<40$ with an effective sky fraction of $80\%$.

Second in line is LiteBIRD, 
a JAXA full-sky CMB mission scheduled to launch around 2029
 which will operate in 15 frequency bands between 34-448 GHz. It will reach a polarization sensitivity of 2.2 $\mu K-arcmin$ with an angular resolution of $0.5^\circ$ at 100 GHz. Note that LiteBIRD has worse resolution compared to SO and therefore the signal-to-noise ratio drops drastically at large multipoles ($> 800$). However, being a full sky survey (effective sky fraction $\sim 70\%$) with high sensitivity, LiteBIRD is expected to perform significantly better than SO at low and intermediate multipoles. At the multipoles where the noise is cosmic variance dominated, the $\Delta{\cal C}_\ell$ will be 1.32 times smaller in LiteBIRD compared to SO.  Therefore, the two surveys will be highly complementary. In our analysis, we combine LiteBIRD to Planck and SO by substituting Planck's noise with the one for LiteBIRD at $\ell<800$.

\begin{table}[]
\begin{tabular}{|l|l|l|}
\hline
Datasets           & Multipole range                                                                                                               & $f_{\rm sky}$                                                 \\ \hline
Planck+SO          & \begin{tabular}[c]{@{}l@{}}Planck: $2-39$\\              $40-2500$\\ SO: $40-2500$\end{tabular}                               & \begin{tabular}[c]{@{}l@{}}0.8\\ 0.2\\ 0.4\end{tabular}       \\ \hline
Planck+SO+LiteBIRD & \begin{tabular}[c]{@{}l@{}}Planck: $800-2500$\\ SO: $40-2500$\\ LiteBIRD: $2-39$  \\                   $40-1350$\end{tabular} & \begin{tabular}[c]{@{}l@{}}0.2\\ 0.4\\ 0.8\\ 0.2\end{tabular} \\ \hline
Planck+S4+LiteBIRD & \begin{tabular}[c]{@{}l@{}}Planck: $800-2500$\\ S4: $40-2500$\\ LiteBIRD: $2-39$  \\                   $40-1350$\end{tabular} & \begin{tabular}[c]{@{}l@{}}0.2\\ 0.4\\ 0.8\\ 0.2\end{tabular} \\ \hline
PICO               & PICO: $2-2500$                                                                                                                & 0.7                                                           \\ \hline
\end{tabular}
	\caption{Dataset combination for the forecast. We provide the multipole ranges and sky fraction used for each of the datasets for the forecast. Note that the multipole ranges are chosen ensuring insignificant overlap between datasets where the signal-to-noise ratios are compareable. }
	\label{tab:forecase_combination}
\end{table}

CMB-S4~\cite{CMB-S4:2016ple} is a proposal for ground based stage-IV CMB observation with nearly 500,000 detectors. It is designed to reach $1\mu K$-arcmin sensitivity over the span of four years covering 40\% of the sky. With this sensitivity it can provide bounds on the tensor-to-scalar ratio with a standard deviation of 0.0005. However, for our purpose of feature forecast we are mainly interested in the Large Aperture Telescope of S4 that will constrain the intermediate to small scale CMB polarization to an unprecedented accuracy (1-3 arcmin resolution). Detection of small scale CMB peaks with better accuracy than Planck and LiteBIRD will provide a complementary angle in constraining primordial features.
Since LiteBIRD will already be operative by the time CMB S4 starts operations, in our analysis we combine Planck, LiteBIRD and CMB S4 using the procedure outlined above for Planck+LiteBIRD+SO.

\begin{figure}
\includegraphics[width=\columnwidth]{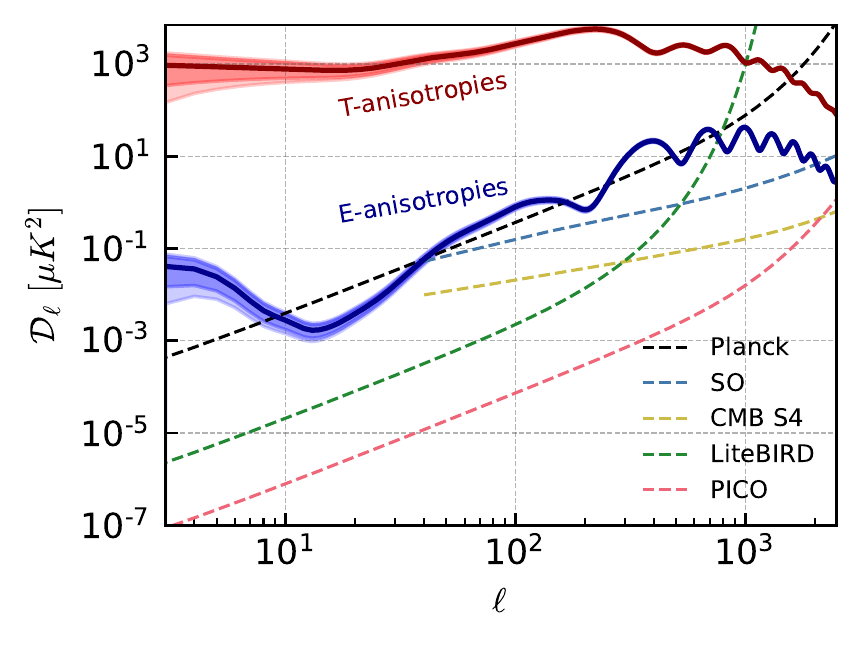}
\caption{\label{fig:Noisecls} The noise power spectra for E-modes from four observations -- Planck, LiteBIRD, SO, CMB S4  and PICO are plotted in dashed lines. The $\Delta {\cal C}_{\ell}$ due to cosmic variance are also plotted as contours around the ${\cal C}_{\ell}$. The cosmic variance differs only by the $f_{\rm sky}$. Here we only plot it for $f_{\rm sky}=0.7$.  }
\end{figure}

PICO is a NASA funded mission concept study. It is designed to be a near-ultimate full sky CMB mission. It will have 21 frequency bands between 21-799 GHz with a polarization sensitivity of $0.65\mu K-arcmin$ with an arcminute resolution. With these specifications we will be having the strongest constraints on the angular power spectra from largest to arcminute angular scales. 
The noise spectra for Planck, LiteBIRD, SO, CMB S4 and PICO are plotted in Fig.~\ref{fig:Noisecls} and a summary of our data combinations is provided in Table~\ref{tab:forecase_combination}.

\begin{figure*}
\includegraphics[width=\columnwidth]{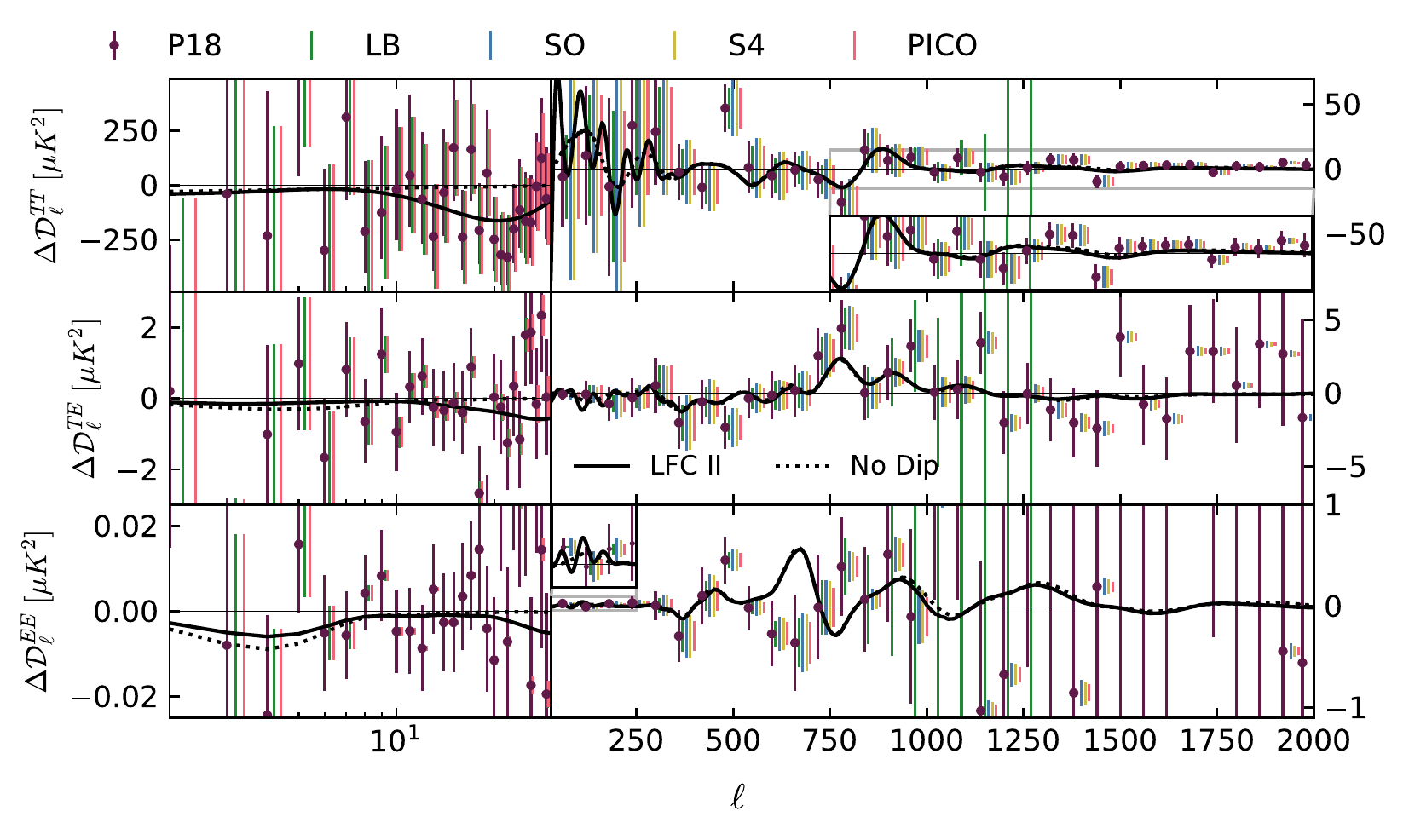}
\caption{\label{fig:LFC_forecast_errors} Residual plot of the two best-fit candidates showing P18 error bars together with the error bars expected by future experiments such as LiteBIRD (green), the Simons Observatory (blue), CMB S4 (gold) and PICO (pink). In the EE panel, where data points are outside the y-axis range, we have moved the errors for SO and PICO near the upper or lower frame edge in order to demonstrate their sizes.  }
\end{figure*}

The features our model generates in PPS exhibit three different characteristic properties, namely a dip at the largest angular scales followed by sinusoidal oscillations at the intermediate scales, both contributed by the sharp feature; and a clock signal at the small scales, contributed by an oscillating massive field. Therefore this model utilizes the full potentials of these future observations in constraining features at different characteristic length scales. 

 We use a likelihood corresponding to an inverted Wishart distribution~\cite{Hamimeche:2008ai} that is represented as
\begin{align}
&-2\ln{\cal L} (\boldsymbol{{\cal C}_{\ell}|{\hat{\cal C}}_{\ell}})=
\notag \\
&\sum_{\ell}(2\ell+1)\left\{{\rm Tr}[\boldsymbol{{\hat{\cal C}}_{\ell}{\cal C}_{\ell}}^{-1}]-\ln|\boldsymbol{{\hat{\cal C}}_{\ell}{\cal C}_{\ell}}^{-1}|-n\right\}~,
\end{align}
with $\boldsymbol{{\cal C}}_{\ell}=\langle\boldsymbol{a}_{\ell m}\boldsymbol{a}_{\ell m}^{\dagger}\rangle$ being the covariance matrix defined with n-dimensional vector $\boldsymbol{a}_{\ell m}$ and we use the estimator $\boldsymbol{\hat{{\cal C}}}_{\ell}=\frac{1}{2\ell+1}\sum_m\boldsymbol{a}_{\ell m}\boldsymbol{a}_{\ell m}^{\dagger}$. The components of the spherical harmonics define this vector $\boldsymbol{a}_{\ell m}={\rm Transpose}\left[a_{\ell m}^{X1},a_{\ell m}^{X2},\dots,a_{\ell m}^{Xn}\right]$ where $Xi$'s are fields and in our case we are using temperature (T) and polarization (E) fields. 

The E-mode polarization contains correlated feature signals to those in the temperature measurement. Although, in principle, there will be correlated feature signals in the B-mode, their amplitude will be very small even if the overall tensor-to-scalar ratio of our model is detected; and so features in the B-mode will not play any role here and we do not use B-mode polarization in the forecast.

 In our analysis we always assume $\ell_{\rm max}=2500$. Note that, while SO, S4 and PICO can explore much smaller scales, we use the cutoff $\ell_{\rm max}=2500$, as the fiducial angular power spectra have the features located mainly at scales larger than the cutoff. Also for this forecast analysis, we use {\tt CosmoChord} for parameter estimation and marginal likelihood estimation. We note that, given the multimodality which we observe in our model, the nested sampling forecast is the correct technique to be used and a simple Fisher forecast, only suitable for Gaussian constraints, would not produce reliable results. Given the noise power spectrum for one observation, we compare the model with feature against the fiducial power spectra. The baseline model is then compared against the same fiducial. The Bayes' factor is computed from the marginal likelihoods obtained from these two analyses.

Since we expect PICO to perform best in constraining all main characteristics of this model, in addition to the analysis using the full information from PICO, we also perform an analysis using the PICO survey specifications but limiting to large multipoles only, $\ell_{\rm min}\ge 600$ ($k\gtrsim 0.044~{\rm MPc^{-1}}$). These are the scales where the clock signal appears in this model. If detected, the clock signal has a particularly important implication for the primordial universe as it measures the scale factor $a(t)$ directly and can be used to distinguish inflation and alternative scenarios model-independently. Such a separate forecast allows us to assess how well this part of the full signal can be constrained. We also note that, even if the clock signal of this inflation model can be detected in high confidence, additional analyses are needed to conclude whether it can be distinguished from standard clock signals of alternative-to-inflation scenarios, and more generally from other oscillatory feature signals. We leave this step to future studies.

In the following, we will adopt two different fiducial candidates, i.e.~the LFC IIa in Table~\ref{tab:candidates} and the \emph{No Dip} candidate in Table~\ref{tab:Nodipcandidate}. As discussed above, despite sharing the same resonance signal, they significantly differ in their sharp feature signals and in the number of free parameters. Given the high sensitivity of the future surveys we explore in our analysis, these details play a prominent role in the strength of the constraints on the parameters and the evidence for detection. In Fig.~\ref{fig:LFC_forecast_errors}, we plot the CMB residuals against P18 errors and forecast errors for LiteBIRD, SO, S4  and PICO for a comparison. As already discussed, we clearly see that LiteBIRD with wide beams cannot resolve small scales. On the other hand, at large and intermediate scales, due to lower sky coverage SO and S4  have larger uncertainty. Furthermore, their temperature noise at $\ell\lesssim 1000$ is larger than that of Planck due to atmospheric noise.    

\subsection{Full model}
\label{Sec:Forecast_full_model}

\begin{figure*}
	\includegraphics[width=\columnwidth]{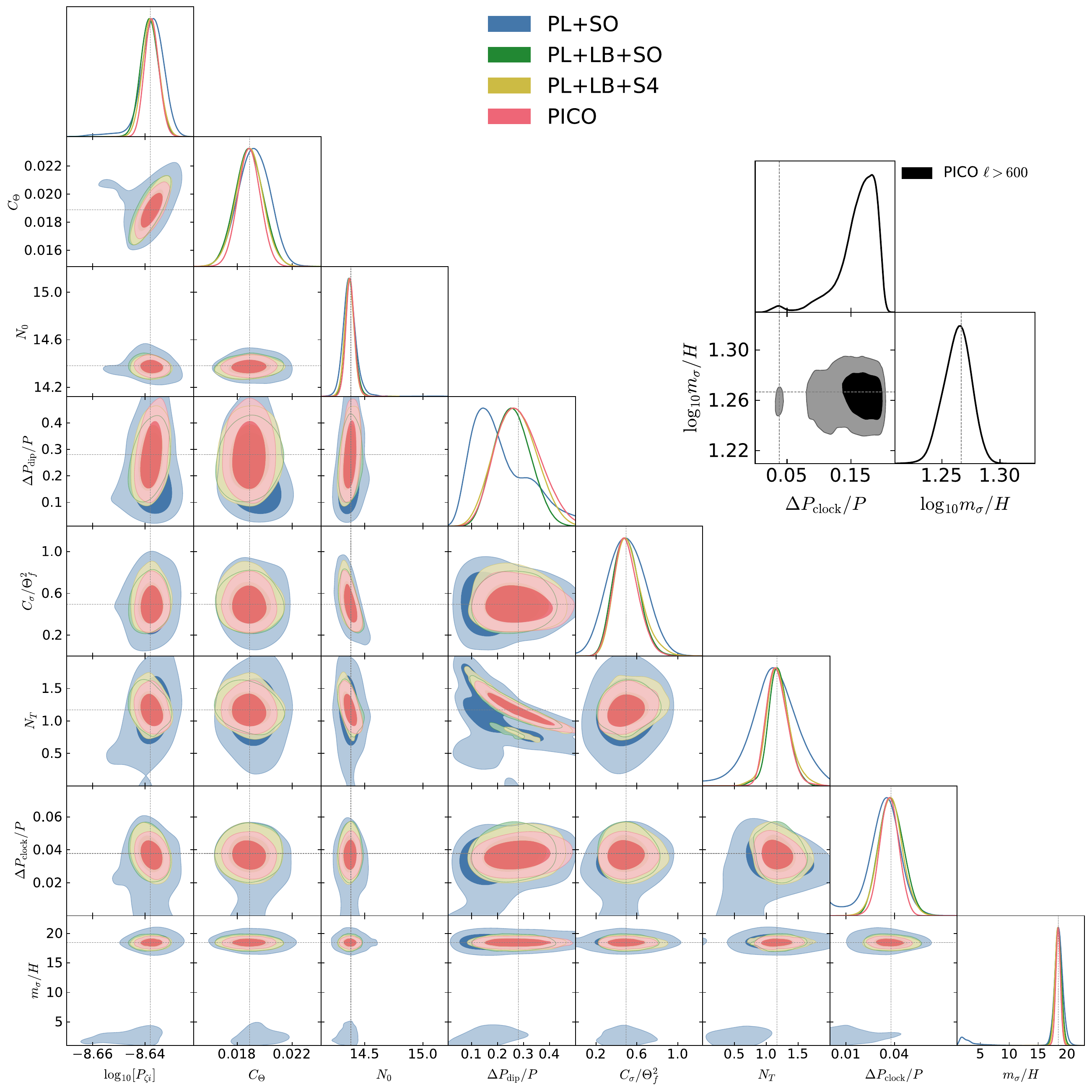}
	\caption{\label{fig:LFC_forecast} Projected constraints on model parameters with future experiments  using LFC-IIa in Fig.~\ref{fig:bestfit} as fiducial model. In the top-right corner, we plot constraints on the frequency and amplitude of the clock signal obtained using only $\ell>600$ information from PICO. In order to show the accuracy to which the frequency of the signal can be determined, we plot derived constraints on $m_\sigma/H$ rather than on $\log_{10}m_\sigma/H$, which is the actual parameter varied in the analysis. }
\end{figure*}

We first present the results of our analysis adopting LFC IIa as fiducial cosmology.
Constraints on the model parameters are summarized in the triangle plots in Fig.~\ref{fig:LFC_forecast} and the Bayes factors with respect to the baseline model are reported in Table~\ref{tab:evidence_forecasts}.

\begin{table}
	\centering
	\begin{tabular}{|l|l|l|}
		\hline
		Experiment          & Full & Restricted \\ \hline
		Planck+SO    & $4.90\pm0.32$               &        $1.77\pm 0.31$ \\         \hline
		Planck+SO+LiteBIRD   & $22.68\pm0.32$               &   $2.38\pm 0.31$ \\         \hline
		Planck+S4+LiteBIRD    &  $24.34\pm 0.33$              &   $4.81\pm 0.32$  \\         \hline
		PICO    &      $28.83\pm0.34$           & $9.12\pm0.33$        \\         \hline
	\end{tabular}
	\caption{Projected Bayes factors ($\ln B$) obtained for different experiments {\it w.r.t.}~the baseline (featureless) model. The analyses are done using single experiment at a time without joining the Planck data.  }
	\label{tab:evidence_forecasts}
\end{table}

The amplitude of the scalar perturbations (represented by parameter $P_{\zeta *}$) is expected to be constrained similarly by 
 all the combinations of experiments considered (with PICO of course being the most constraining). Note, however, that SO and S4 alone could not constrain the optical depth at reionization $\tau_{\rm reio}$ and will provide looser constraints on $P_{\zeta *}$ \cite{Mortonson:2009qv}. This is why we had to complement them with large scale information. 
Also the tilt of the spectrum, controlled by the parameter $C_\theta$, is strongly constrained in all the cases we consider
 due to accurate measurements at small scales. 

Let us now discuss the parameters related to features, starting from the dip feature. The amplitude of the dip feature is characterized by $\Delta P_{\rm dip}/P$. LiteBIRD and PICO are able to find two-tailed distribution on this parameter indicating the strong possibility of detection of this feature, if present. The location of the feature, parameterized by $N_0$, and the extension of the sinusoidal running, parameterized by $C_\sigma/\Theta_f^2$, are also stringently constrained by these experiments. We observe that, from the 1-dimensional posterior distribution of $\Delta P_{\rm dip}/P$, a spectrum without a dip can be excluded at more than $4\sigma$ level by PL+LB+SO, PL+LB+S4 or PICO, as also suggested by previous studies~\cite{Miranda:2014fwa}. Physically, our findings imply that future experiments will be able to test  a step in the inflaton trajectory. However, the errors on $N_0$ and $C_\sigma/\Theta_f^2$ are not small enough to distinguish LFCIIa from LFCIIb. 
On the other hand, none of these parameters can be constrained as well by SO because of its poor resolution at large scales.  For this reason, the dip feature remains undetected by the combination PL+SO. Still, we observe a slight improvement in the constraints because of a reduction in the degeneracy with the clock parameters, which can be more accurately constrained (see below). Let us note that S4 suffers the same problems as SO. This, however, cannot be observed in Fig.~\ref{fig:LFC_forecast}, where we consider it in combination with LiteBIRD, which is expected to become operative before S4.

We now go on and discuss constraints on the clock signal, starting from its location, parameterized by $N_T$\footnote{We note that, being the best-fit $N_T$ very close to the upper limit of the prior adopted in the P18 data analysis, we have adopted a larger prior $N_T\in[0,\,2]$ in our forecast analysis.}. The parameter is well constrained within its prior range, except for PL+SO. In the PL+SO case, even if the constraining power at high-$\ell$, where the fiducial clock signal shows up, is better than the one of Planck alone, $N_T$ cannot be well constrained because of its degeneracy with $\Delta P_{\rm dip}/P$, as evident from Fig.~\ref{fig:LFC_forecast} and Eqs.~\eqref{DP_dip} and~\eqref{Eq:NT}. The latter is not detected and has a very broad posterior distribution, thus inducing a quite large error on $N_T$. Let us stress, though, that even in this case the posterior clearly peaks at the fiducial value.

\begin{figure*}
	\includegraphics[width=.75\columnwidth]{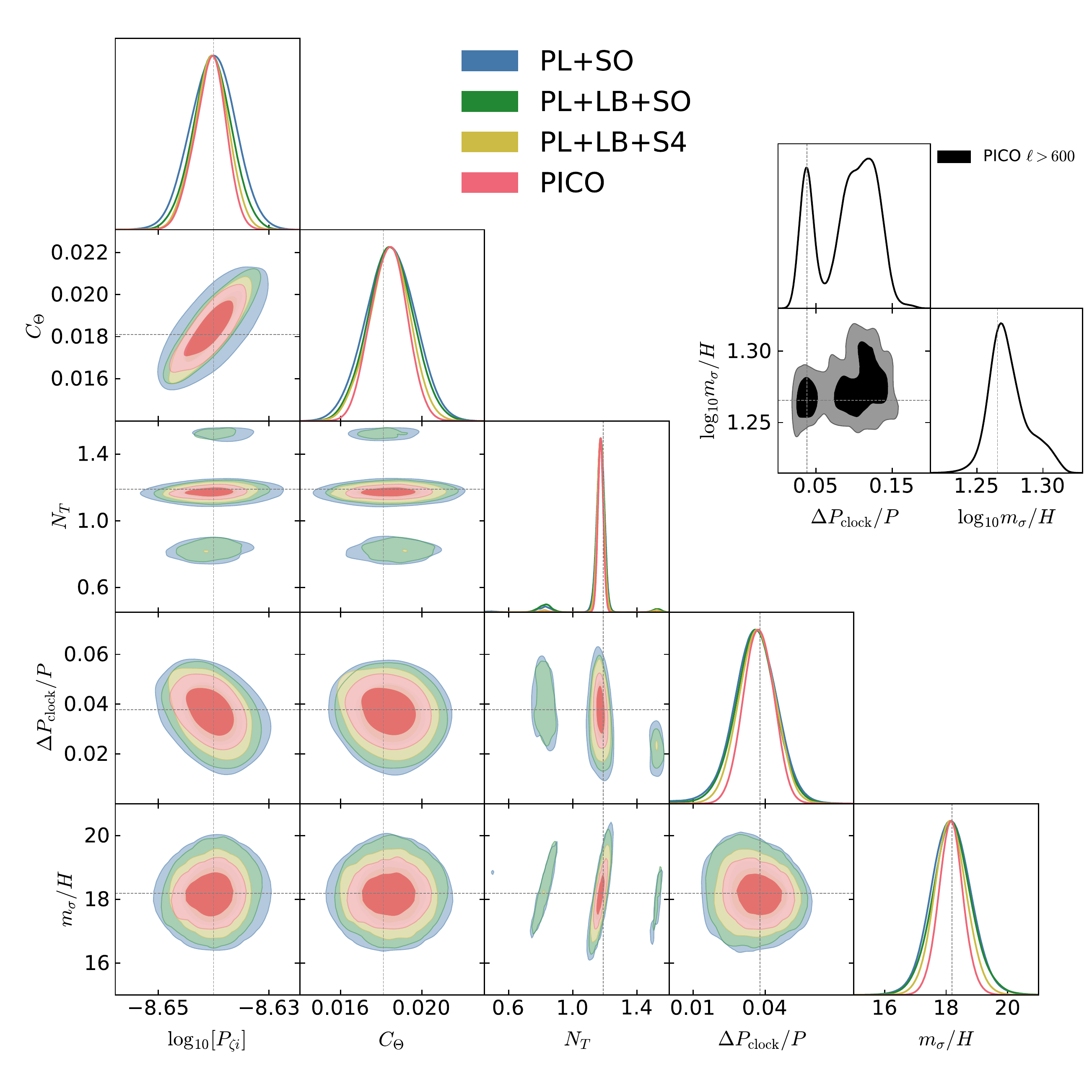}
	\caption{\label{fig:LFC_forecast_3params} Projected constraints on model parameters with future observations using the restricted model candidate in Fig.~\ref{fig:LFC-nodip} as fiducial model.  In the top-right corner, we plot constraints on the frequency and amplitude of the clock signal obtained using only $\ell>600$ information from PICO. In order to show the accuracy to which the frequency of the signal can be determined, we plot derived constraints on $m_\sigma/H$ rather than on $\log_{10}m_\sigma/H$, which is the actual parameter varied in the analysis.}
\end{figure*}

 As discussed in the previous Sections, several values of the clock frequency are compatible with current CMB data. Fig.~\ref{fig:LFC_forecast} shows that, assuming LFCII as the true model of the Universe, future experiments  will be able to pinpoint that frequency with exquisite accuracy. Indeed, all the  combinations of experiments show a projected posterior distribution for $m_\sigma/H$ that is very narrow around the fiducial value, except for Pl+SO which also admits a secondary peak (although much less significant) sourced by degeneracies with the dip feature parameters. However, even in the latter case, the constraints around the main mode are quite stringent, i.e. $m_{\sigma}/H = 17.2^{+2.2}_{-0.59}$ at 68\% CL. With large scale information from LiteBIRD, the identification of the dip feature will help reduce degeneracies and tigthen the constraints to $m_{\sigma}/H = 18.46\pm 0.65$ at 68\% CL  (PL+LB+SO). PL+LB+S4 and PICO will further reduce the  68\% CL error down to $\sigma\left(m_\sigma/H\right)=0.59$ and $0.41$ respectively.
	All of these constraints on the clock frequency are so tight that they will be able to rule out other feature candidates such as LFCI and HFC. 

Concerning the amplitude of the clock signal $\Delta P_{\rm clock}/P$, we observe only an upper limit from PL+SO.
As before, this is due to degeneracies with the dip feature and does not happen if we take the restricted model candidate as fiducial (see next subsection). When such a degeneracy is removed thanks to accurate large scale measurements LiteBIRD or PICO, we find that the amplitude of the clock signal can be constrained very well. The projected 68\% CL constraints are $\Delta P_{\rm clock}/P = 0.0375\pm 0.0074$  for PL+LB+SO and PL+LB+S4 and $\Delta P_{\rm clock}/P = 0.0368\pm 0.0059$ corresponding to detections at $5\sigma$ and $6\sigma$ levels respectively. Together with the constraints on the clock frequency, such findings will have profound impact on our understanding of the primordial universe physics, as they will imply the existence of a classically oscillating massive particle $18$ times heavier than the Hubble scale.  
 
Let us now comment on the projected Bayes factors against the baseline models, which we summarize in Table~\ref{tab:evidence_forecasts}. The tighter constraints will all lead to very strong (from Planck + SO) to decisive evidence (from the other combinations of experiments). Our results therefore imply that, assuming that LFCII is indeed the true model of the Universe, future experiments will be able to rule out the baseline near-scale-invariant power spectrum at very high statistical significance. Note that the question of whether the components of the signal (dip and/or resonant feature) will be detected has been discussed above.

To summarise our discussion on the abilities of each observation toward constraining different features: Individually, the dip feature signal and the clock frequency are constrained well by LiteBIRD and SO (or S4), respectively; but other feature properties are not well constrained due to the different focuses on large versus small scale measurement and due to the limiting uncertainties of the experiments.
Interestingly, when such experiments are combined with other data to add complementary information at the scales to which they are not sensitive, the constraints on our CPSC model will significantly improved compared to the ones presented in the previous Sections. SO will increase the evidence in favor of the CPSC model, but we have to wait for LiteBIRD to get a detetction of the feature signals it produces. Indeed, combined with Planck and SO, LiteBIRD will be able to detect both the dip and the clock feature at the $3.4\sigma$ and $5\sigma$ levels respectively. As SO is going to start data taking in 2023 and LiteBIRD is planned to be launched in 2029, we may be able to detect such a bestfit candidate in less than a decade. Next generations of CMB experiments such as S4 and PICO will be able to further tigthen the constraints on our model.

Finally, we would like to check how well PICO can probe the scale factor evolution during inflation
by measuring the clock signal at small scales, because of the special importance of this signal as explained in Section~\ref{Sec:CPSC models}. 
Although this signal has been analyzed above as part of the full CPSC signal, to minimize the interference between the clock signal and the rest of the full signal, here we forecast constraints on our model using only PICO's small scale information at $\ell>600$, since these are the scales where the clock signal shows up in LFC II. We plot constraints from such analysis in the top-right corner of Fig.~\ref{fig:LFC_forecast}. Given that we use no information about large scales, the only feature parameters that can be meaningful constrained are $\Delta P_{\rm clock}/P$ and $m_\sigma/H$, the other parameters of our model are hardly constrained. Although constraints on $\Delta P_{\rm clock}/P$ are not very tight, we observe that $\Delta P_{\rm clock}/P=0$ is ruled out, as the posterior distribution vanishes there. Values of $\Delta P_{\rm clock}/P$ larger than the fiducial input are also allowed as they correspond to a clock signal beginning at larger scales than the fiducial therefore matching very well the fiducial spectrum for $\ell>600$. Importantly, we see that the small scale measurement of the CMB angular spectra alone places a very tight constraint on the frequency, suggesting that the constraint on the frequency obtained with the full scale PICO mainly comes from that on the resonant part of the clock signal, hinting at a real chance to confirm or rule out inflation as the physical scenario operating during the primordial Universe. 

\subsection{Restricted model}

We now change the fiducial to the best-fit candidate for the restricted model (see Table~\ref{tab:Nodipcandidate}). Constraints on model parameters are shown in Fig.~\ref{fig:LFC_forecast_3params}.

This model has only 3 extra parameters that describe a standard clock signal with a much milder sharp feature part. As such, constraints from the combination PL+SO are not affected by any of the degeneracies with the large scale dip feature, to which SO is not sensitive. Without the dip signal, the posterior distributions for the feature parameters are all peaked around the fiducial restricted model. The signal location, i.e. $N_T$, is better constrained than in the full case, due to the absence of the degeneracy with the dip amplitude. However, for PL+SO and PL+LB+SO we observe two other modes in the posterior distribution although much less significant. The posterior for the clock frequency posterior, on the other hand, is not multimodal even in the PL+SO combination. At 68 \% CL, we obtain $m_{\sigma}/H = 18.1\pm 2.4$ for PL+SO, $m_{\sigma}/H = 18.2\pm 1.7$ for PL+LB+SO, $m_{\sigma}/H = 18.15\pm 0.83$ for PL+LB+S4 and $m_{\sigma}/H = 18.16\pm 0.41$ for PICO. These constraints are slightly less stringent than those obtained for the full model, perhaps because of the different fiducial clock mass. Nevertheless, they are very tight constraints which can rule out other bestfit candidates. 

Finally, as opposed to the full model fiducial, if restricted LFC is really the true model of the Universe, we will not need to wait for LiteBIRD, because already SO will be able to detect it. This can be seen by 68 \% CL constraints on the clock amplitude that we find to be $\Delta P_{\rm clock}/P = 0.0358\pm 0.0092$  for PL+SO, $\Delta P_{\rm clock}/P = 0.0359\pm 0.0086$ for PL+LB+SO,  $\Delta P_{\rm clock}/P = 0.0366\pm 0.0077$ for PL+LB+S4 and $\Delta P_{\rm clock}/P = 0.0372\pm 0.0063$ for PICO, corresponding to $3.9\sigma$, $4.2\sigma$, $5.2\sigma$ and   $5.9\sigma$ level detections respectively. Therefore, within the next five years, we may be able to detect the signature of an oscillating massive field with the inflationary fingerprint.

As before, we repeat the excercise of  using only PICO small scale information at $\ell>600$. We obtain essentially the same results as in the case of the full model. This can be easily seen by comparing the top-right corner of Fig.~\ref{fig:LFC_forecast_3params} and \ref{fig:LFC_forecast}. For the same reason, here the only remaining feature parameter, i.e.~$N_T$, cannot be well constrained.

Compared to the full model case, the Bayes factors are now lower (see Table~\ref{tab:evidence_forecasts}). The differences are due to the dip feature at the large scale, which significantly contributes to the Bayes factors as soon as LiteBIRD joins.  The conclusion is that in order to find very strong to decisive evidence for the clock signal we will have to wait till S4.

\section{Conclusion and discussions}
\label{Sec:Conclusion}

	Since WMAP's first precise measurement of the CMB temperature anisotropies, several anomalies at different scales have been observed or confirmed in the more accurate Planck data. Although none of them is statistically significant, these anomalies are extremely intriguing for model builders as, if taken at face value, they may hint at new physics.  In this paper, we have proposed a classical primordial standard clock model in which an adventurous inflationary trajectory imprints features in the primordial power spectrum that address anomalies at both large and small scales.
	A typical trajectory of our model consists in the inflaton that falls down a step in the potential, leading to a sharp feature signal with a large dip at large scales in the primordial power spectrum, and subsequently enters a curved path exciting the oscillations of a massive (clock) field in the perpendicular direction, giving rise to a special resonance feature at smaller scales.

	With a recently developed pipeline, we have compared our model to CMB anisotropies data from Planck in temperature (TT), E-mode polarization (EE) and their cross-correlation (TE). Although we find our model to be statistically indistinguishable from the baseline $\Lambda$CDM, several interesting candidates arise that provide better fits to data. Indeed, the global best-fits of our theory provide impressive improvements in the fit to data with a $\Delta\chi^2= 19.8$ or $17.7$, depending on the high-$\ell$ likelihood used in the analysis (Plik bin1 or Camspec respectively), for a full model with six extra parameters, and $12.9$ or $13.4$ for a restricted model with three extra parameters. Not only can the full model provide an excellent fit to the dip around $\ell\sim20-30$ in the TT data, but also the oscillations of a clock field with $m_\sigma/H\sim18$ produce a clock signal that simultaneously fits some major residuals in the TT and TE data around $500\lesssim\ell\lesssim1500$. The importance of the clock signal to fit the high-$\ell$ anomalies is confirmed by the study of the restricted model where we remove the dip feature in the primordial power spectrum by removing the step in the potential. In this case the data points to a best-fit with the same clock signal as the one in the full model candidate.
	
	Motivated by these candidates, we have forecast the capability of future CMB experiments to detect them. As examples of CMB missions aimed at a precise mesaurement of CMB polarization, we have considered the Simons Observatory (SO), LiteBIRD, CMB-S4 and PICO. While the first two missions have already been approved and will map small and large scales respectively, CMB S4, which will map short small scales even more accurately than SO, has not been approved yet, though it received strong support from ASTRO2020. PICO differs in that it is only a funded study and, if approved, will provide cosmic-variance limited measurements of temperature and E-mode polarization on all scales.  While SO and LiteBIRD alone will only be able to characterize the small scale clock signal and the large scale dip feature respectively, their combination, together with existing Planck data to add information at intermediate scales, will detect or rule out our model with definitive statistical evidence.

	\begin{figure}
		\includegraphics[width=\columnwidth]{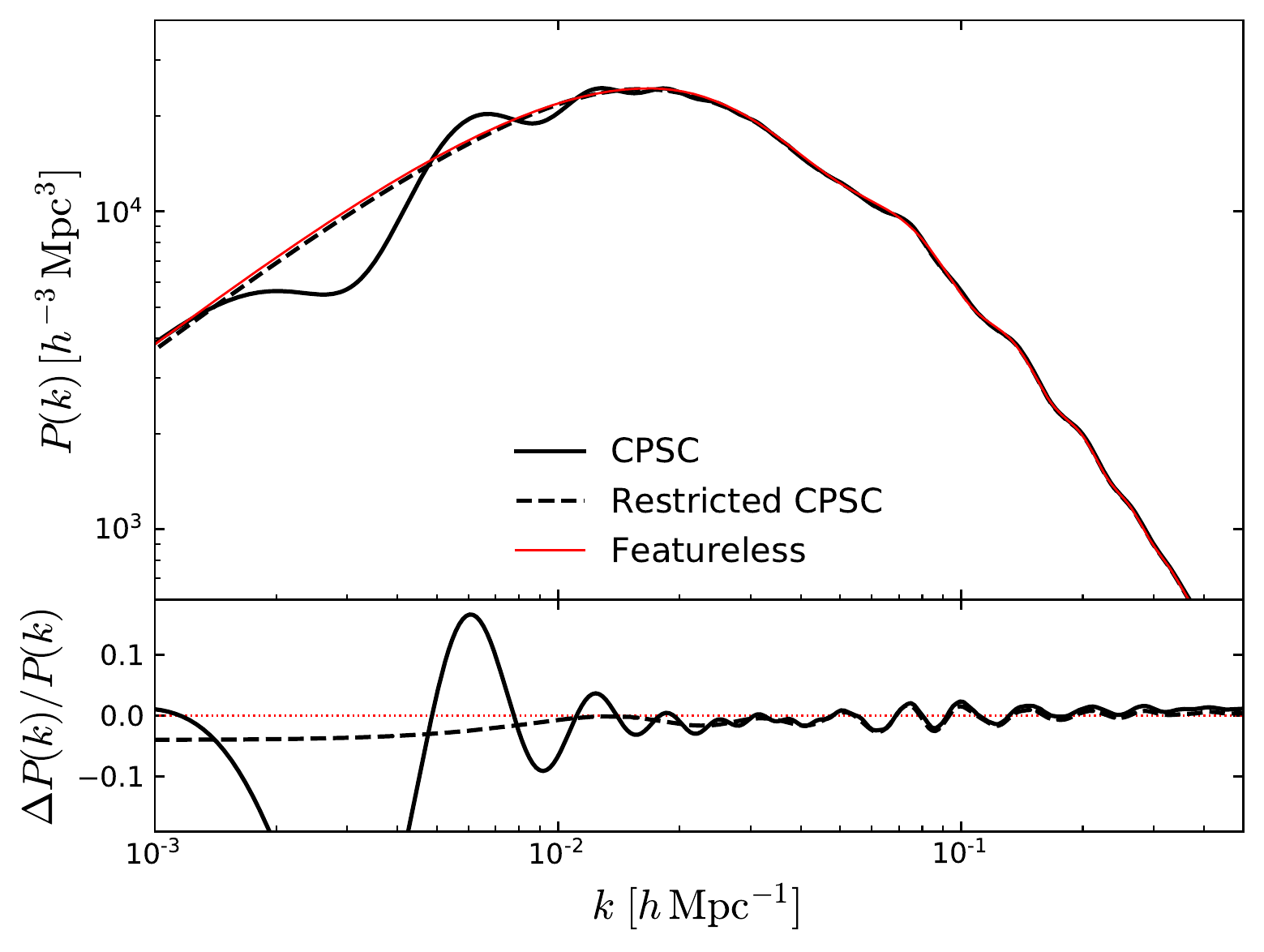}
		\caption{\label{fig:LSS} Linear matter power spectrum at $z=0$ for our bestfit candidates. }
	\end{figure}
	
	Our results demonstrate the capability of 
	such experiments to probe the detailed history of the Universe in its infancy, before the onset of the Hot Big Bang scenario. Indeed, if any of our best-fit candidates is detected, this will point to a step-like feature in the inflationary potential (only in the case of the full model candidate) and to a detection of a massive particle with mass $\sim20$ times larger than the Hubble scale of inflation.  Even more interestingly, the detection of the small scale clock signal will make it possible to rule out alternative-to-inflation scenarios model-independently as they predict different running for such a signal. We believe that such profound physical implications provide 
	 strong additional support for the science case of a stage  CMB experiment as well as a full-scale cosmic-variance limited  one.

	While at the present stage CMB anisotropies data from Planck are the best data to constrain our model, the feature signal also leaves correlated imprints in the large-scale distributions of galaxies~\cite{Huang:2012mr,Hazra:2012vs,Chen:2016vvw,Ballardini:2016hpi,Palma:2017wxu,LHuillier:2017lgm,Ballardini:2017qwq,Vasudevan:2019ewf,Beutler:2019ojk,Ballardini:2019tuc,Debono:2020emh,Chen:2020ckc,Li:2021jvz} (see Fig.~\ref{fig:LSS}) and atomic hydrogen~\cite{Chen:2016zuu,Xu:2016kwz}, which will help stringently constrain the intermediate and small scale signals of our model. Feature models are also known to generate correlated signals in primordial non-Gaussianities~\cite{Chen:2006xjb,Chen:2008wn,Adshead:2011jq,Hazra:2012yn, Bartolo:2013exa,Fergusson:2014hya,Fergusson:2014tza} that may be used as independent constraints.
	It is also important to study how well different feature signals, including features in both inflation and alternative-to-inflation scenarios, can be observationally separated.
	
	As more data come in, the models may be ruled out, modified or improved, but we hope that the possible hints in data and rich connections between feature signals and fundamental physics, which we demonstrated in this paper, illustrate some exciting prospects of probing the history of the primordial Universe using data from near future observations.

While at the present stage CMB anisotropies data from Planck are the best data to constrain our model, the feature signal also leaves correlated imprints in the large-scale distributions of galaxies~\cite{Huang:2012mr,Hazra:2012vs,Chen:2016vvw,Ballardini:2016hpi,Palma:2017wxu,LHuillier:2017lgm,Ballardini:2017qwq,Vasudevan:2019ewf,Beutler:2019ojk,Ballardini:2019tuc,Debono:2020emh,Chen:2020ckc,Li:2021jvz} (see Fig.~\ref{fig:LSS}) and atomic hydrogen~\cite{Chen:2016zuu,Xu:2016kwz}, which will help stringently constrain the intermediate and small scale signals of our model. Feature models are also known to generate correlated signals in primordial non-Gaussianities~\cite{Chen:2006xjb,Chen:2008wn,Adshead:2011jq,Hazra:2012yn, Bartolo:2013exa,Fergusson:2014hya,Fergusson:2014tza} that may be used as independent constraints.
It is also important to study how well different feature signals, including features in both inflation and alternative-to-inflation scenarios, can be observationally separated.

As more data come in, the models may be ruled out, modified or improved, but we hope that the possible hints in data and rich connections between feature signals and fundamental physics, which we demonstrated in this paper, illustrate some exciting prospects of probing the history of the primordial Universe using data from near future observations.

\medskip
\section*{Acknowledgments}

We thank Mario Ballardini, Karim Benabed, Josquin Errard, Steven Gratton, John Kovac, Umberto Natale, Luca Pagano, Daniela Paoletti and Daniela Saadeh for very helpful discussions and the anonymous referees for suggestions that helped improve the paper. We acknowledge the use of the GetDist package \cite{Lewis:2019xzd} to produce some of the figures in this paper. The computations in this paper were run on the FASRC Cannon cluster supported by the FAS Division of Science Research Computing Group at Harvard University. MB is supported by the Atracción de Talento contract no. 2019-T1/TIC-13177 granted by the Comunidad de Madrid in Spain.

\appendix

\section{Several details of the model}
\label{Sec:several details}
\begin{figure}
	\includegraphics[width=.9\columnwidth]{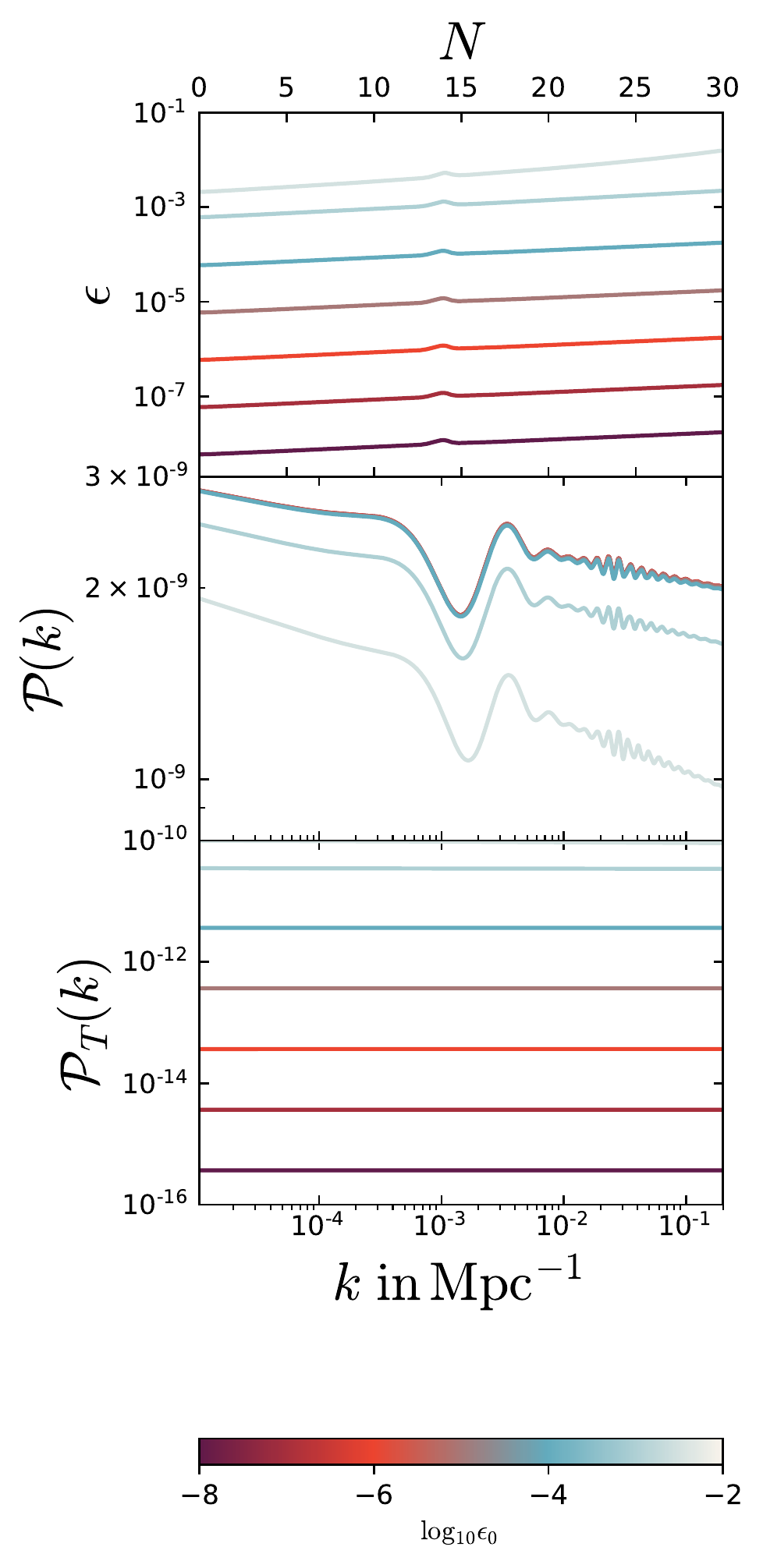}
	\caption{\label{fig:eps0} Effects of varying the $\epsilon$ parameter. We plot the background slow-roll parameter $\epsilon$ and the scalar and tensor power spectra in the top, center and bottom panel respectively. The parameters for the case with $\epsilon=10^{-7}$ are the same used in Fig.~\ref{fig:PK_intro}. For other values of $\epsilon$, we perform the rescaling mentioned in the text. }
\end{figure}

\subsection{Feature signal dependence on the value of $\epsilon$}

Here we provide some details on the dependence of the model predictions on the value of the slow-roll parameter $\epsilon$, mentioned in Sec.~\ref{sec:effectiveparameters}.

As mentioned in Sec.~\ref{sec:effectiveparameters}, the predictions on the scalar power spectrum are independent of the precise value of $\epsilon$ if the rescaling 
 $\epsilon \to C^2 \epsilon$ is accompanied by $V_{\rm inf} \to C^2 V_{\rm inf}$, $C_\sigma \to C^2 C_\sigma$, $\sigma_f \to C \sigma_f$, $\Theta_{0} \to C \Theta_{0}$, $\Theta_{T} \to C \Theta_{T}$,  $\xi \to C^{-1} \xi$ and finally $\Theta_f \to C \Theta_f$. However, since the energy scale of inflation does change, the tensor power spectrum also changes by a factor of $C^2$. This is explicitly shown in Fig.~\ref{fig:eps0}.

As can be seen, as long as $\epsilon<10^{-3}$ the only difference is in the amplitude of the power spectrum of tensor fluctuations, whereas curvature ones are unaffected. For larger values of $\epsilon$, the formulae at first order in slow-roll that we described in Sec.~\ref{sec:effectiveparameters} become less accurate and the simple scaling mentioned above is not enough to keep the $P(k)$ unchanged and the amplitude and tilt parameters ($V_{\rm inf}$ and $C_\Theta$) need to be further adjusted after the scaling.

\subsection{Cases with $m_\sigma\lesssim 3H/2$}
\begin{figure}
	\includegraphics[width=\columnwidth]{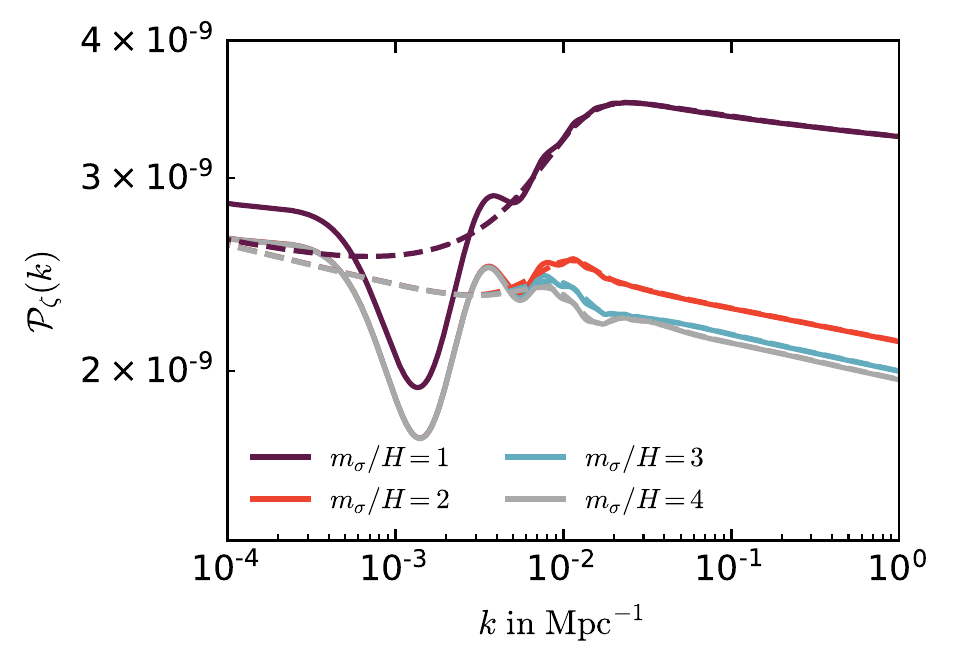}
	\caption{\label{fig:VLFC} Scalar power spectra for very small values of $m_\sigma/H$. The parameters are the same as used in Fig.~\ref{fig:PK_intro}, except for $N_T=1$ and for $m_\sigma/H$ which is varied according to the legend. We also plot the power spectra for $\Theta_f\to\infty$ in dashed lines. }
\end{figure}

In our discussion of the CPSC in Section~\ref{sec:effectiveparameters}, we focused on cases in which the spectator scalar has large mass $m_\sigma>3H/2$ so that it oscillates in the inflationary background. However, the parameter space explored in our data analysis also includes values of $m_\sigma/H$ as small as $m_\sigma/H=1$. Therefore,  we find it useful to provide the reader with some insights into the feature signal for  $m_\sigma<3H/2$. 

For a scalar field with $m_\sigma\le 3H/2$, once excited classically, it behaves as a critically or over damped oscillator following
\bea
\ddot\sigma + 3H \dot\sigma +m_\sigma^2 \sigma =0 ~.
\eea
The amplitude decays as
\bea
\sigma\sim e^{\left[ -\frac{3}{2} \pm \sqrt{\frac{9}{4}-\frac{m^2}{H^2}} \right] Ht}
\eea
without oscillation. If the coupling between this massive field and the inflaton is constant, this induces a $k$-dependent bump or dip feature signal in the primordial power spectrum that behaves as
\bea
\frac{\Delta P_\zeta}{P_\zeta} \sim \left( \frac{k}{k_0} \right)^{ -\frac{3}{2}+\sqrt{\frac{9}{4}-\frac{m^2}{H^2}} } ~,
\eea
where $k>k_0$. This feature is superimposed with the sharp feature signal.

Here we have assumed inflation, so $H$ is approximately a constant. Results for an arbitrary non-inflationary background can be obtained in a similar way. Comparing to the $m \gg 3H/2$ case that is chosen in \cite{Chen:2011zf,Chen:2011tu} to extract the information of the background $a(t)$, the $k$-dependence of the feature in this case may be more easily mimicked by other effects.
Nonetheless, for the purpose of the primordial feature models, the case with lighter massive fields $m_\sigma\lesssim 3H/2$ provides an example of a different type of feature signals.

We plot a few examples for small values of $m_\sigma/H$ (not necessarily smaller than $3/2$) in Fig.~\ref{fig:VLFC}, for both the full model (solid lines) and the specialized case with $\Theta_f\to\infty$ (dashed lines). As described above, the few, very damped oscillations, are now superimposed with the sharp feature signal, as can be clearly seen by comparing the predictions of the full and restricted models. Another interesting property of the power spectra is that lowering  $m_\sigma/H$ we observe a progressive suppression of the large scale power, as also observed in Ref.~\cite{Braglia:2021ckn}, which is independent of the step in the $\Theta$ potential. This, as mentioned in Section~\ref{sec:results_restricted},  explains why we see a smooth peak in the posterior of $m_\sigma/H$.

Another property to note is that in the main text we have identified $P_\zeta=H^2/(4\pi^2 \dot\Theta^2)$ and  $\epsilon=\dot{\Theta}^2/2H^2$. 
When $m_\sigma/H$ is very small, although being a good approximation for small $\xi\sigma_f$, these two formulae receive non-negligible correction when $\xi\sigma_f\gtrsim\mathcal{O}(0.1)$, as in Fig.~\ref{fig:VLFC}. 
This is because $V(\sigma)$ is now less steep; and so, after entering the curved path, the inflaton shifts more in the $\sigma$-direction due to centrifugal force, resulting a larger correction in the kinetic term for $\Theta$.
The correct formula is then $\epsilon\simeq\dot{\Theta}^2(1+\xi\sigma_f)^2/2H^2$ leading to  $P_\zeta^{\rm physical}=(1+\xi\sigma_f)^2 P_\zeta^{\rm main\, text}$, which shows that the power spectrum is enhanced by a factor of $(1+\xi\sigma_f)^2$ in agreement with the step-behavior in the PPS shown in Fig.~\ref{fig:VLFC}. This step-behavior is superimposed with the sharp feature signal and the bump signal mentioned above.

\begin{figure}
	\includegraphics[width=.9\columnwidth]{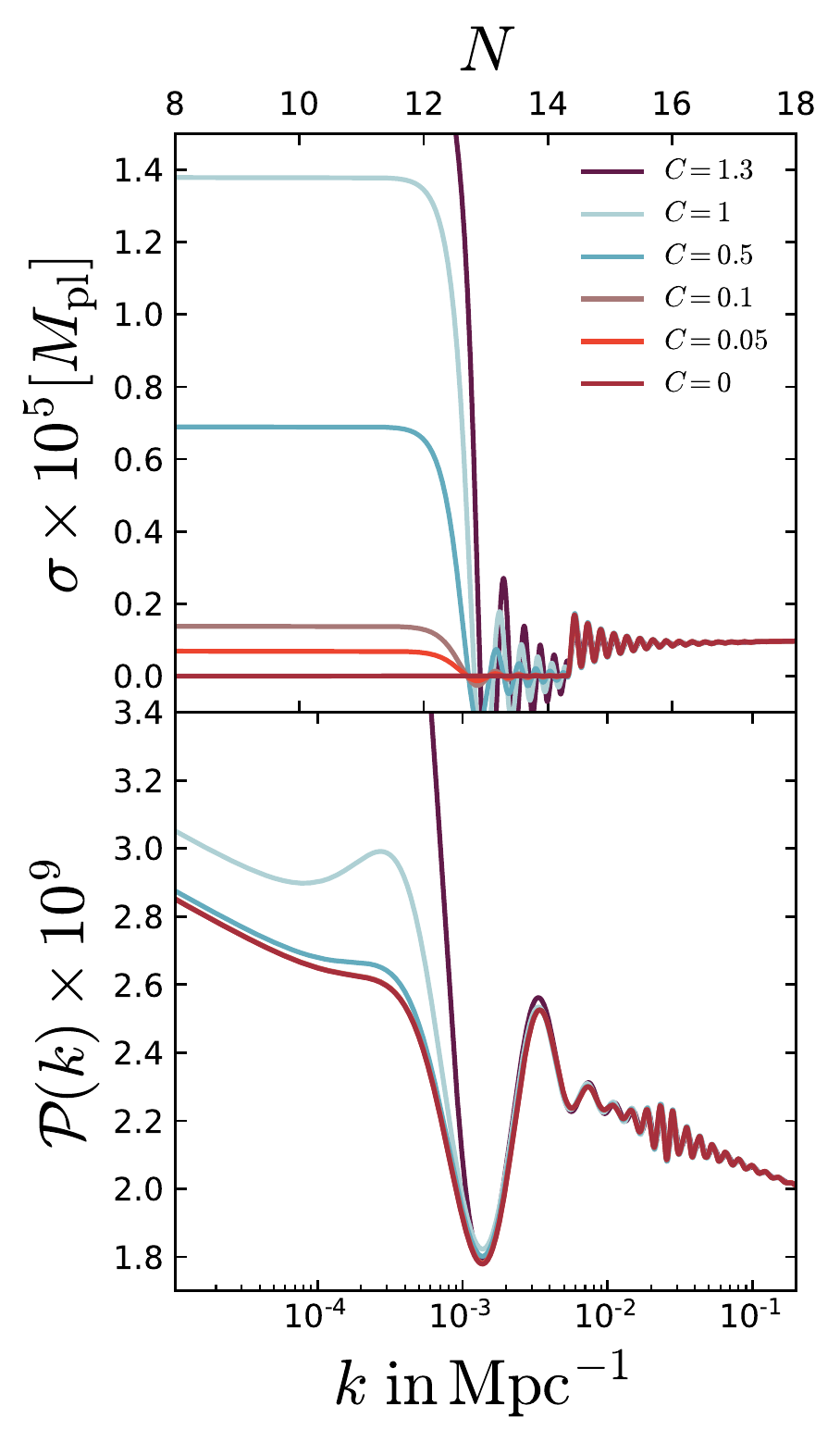}
	\caption{\label{fig:sigmai} [Top] Evolution of the clock field and [bottom] predictions on the PPS for a set of different values of $C\equiv\sigma_i/\sigma_f$.  }
\end{figure}
\subsection{Initial condition on the clock field}

Finally, we provide some details on the dependence of the model predictions on the initial value of the scalar field $\sigma_i$. We show the background evolution of $\sigma$ and the corresponding predictions for the PPS for a set of different values of $C\equiv\sigma_i/\sigma_f$ in Fig.~\ref{fig:sigmai}. For initial conditions $\sigma_i\neq0$, the clock field remains initially frozen. The reason is that, because of the step, the derivative of the potential with respect to $\sigma$ is suppressed by a factor $\exp\left(-(\Theta-\Theta_0)^2/\Theta_f^2\right)$. As $\Theta$ approaches the step, this suppression factor becomes smaller and the clock field eventually rolls down towards the bottom of the potential at $\sigma=0$ and starts to oscillate around it. When the turning trajectory is introduced, the clock field gets displaced to its new effective minimum as explained in the main text and enters a second oscillatory phase producing the clock signal. The first oscillatory stage does not produce any detectable signal, since the the inflaton and the clock field are not directly coupled. On the other hand, since $\sigma$ is not at rest, the clock field contributes to the total potential energy and slightly enhances the Hubble parameter while it is frozen and we observe an enhancement of power spectrum at large scales, before the dip. Although such large scales are the ones mostly affected by cosmic variance, for $C\gtrsim0.7$ the enhancement becomes significant and the predictions at those scales are different from those obtained by setting $C=0$ as we did in the main text.

\section{Addition of small scale ACT data  }
\label{Appendix:ACT}

In this Appendix, we complement our data set with measurements of small scale anisotropies. As a representative example, we consider data from the ACT collaboration~\cite{ACT:2020gnv}. Since small scale data cannot set any meaningful constraints on the dip feature signal, we limit the analysis to the restricted model of Section~\ref{sec:restricted model}. 

We consider TT, TE and EE anisotropies power spectra from the ACT DR4 dataset. The likelihood delivered by the ACT collaboration\footnote{\href{https://lambda.gsfc.nasa.gov/product/act/act_dr4_likelihood_cmb_info.cfm}{https://lambda.gsfc.nasa.gov/product/act/act\_dr4\_likelihood\_cmb\_info.cfm}} \cite{ACT:2020frw} only depends on one nuisance parameter, i.e.~the overall polarization efficiency $y_p$, while all the other contributions from foregrounds are already marginalized over. Unlike the data used in the main text, which were unbinned, ACT data are thus binned in bandpowers whose covariance includes the noise and foregrounds effects. The multiple range covered by DR4 is $600\leq\ell\leq4125$ in TT and $350\leq\ell\leq4125$ in TE and EE. Note that the lack of information at large angular leads to bad constraints on the optical depth $\tau_{\rm reio}$. A simple way to fix it, used in \cite{ACT:2020frw}, is to include a Gaussian prior on it. However, we decide not to use it and to first simply explore ACT-only constraints. 

As ACT is highly complementary to Planck, which has a better sensitivity to anisotropies at larger scales, we also explore constraints on our feature models from their combination. In order not to double count information from the two datasets, we restrict the range of ACT data following the procedure outlined in Ref.~\cite{ACT:2020frw}, i.e.~we cut the TT likelihood at $\ell_{\rm min}=1800$ and use only higher multipoles. 
We only present results obtained for the combination of ACT and EG20, since running the P18 likelihood is computationally much more expensive. At low-$\ell$, we use both lowT and lowE\footnote{Note that lowE data are usually trade for a prior on $\tau_{\rm reio}$ when combining ACT and Planck. However, we use lowE to be consistent with the results in the main text.}.

To further speed up the computation we set the flag $\tt{lsample\_boost}=50$ in CosmoMC which allows for a faster computation of the CMB power spectra. We have checked that this produces differences in the $\chi^2$ of the order $\Delta\chi^2\sim0.01$, which does not affect our analysis.

	\begin{figure*}[t]
	\includegraphics[width=\columnwidth]{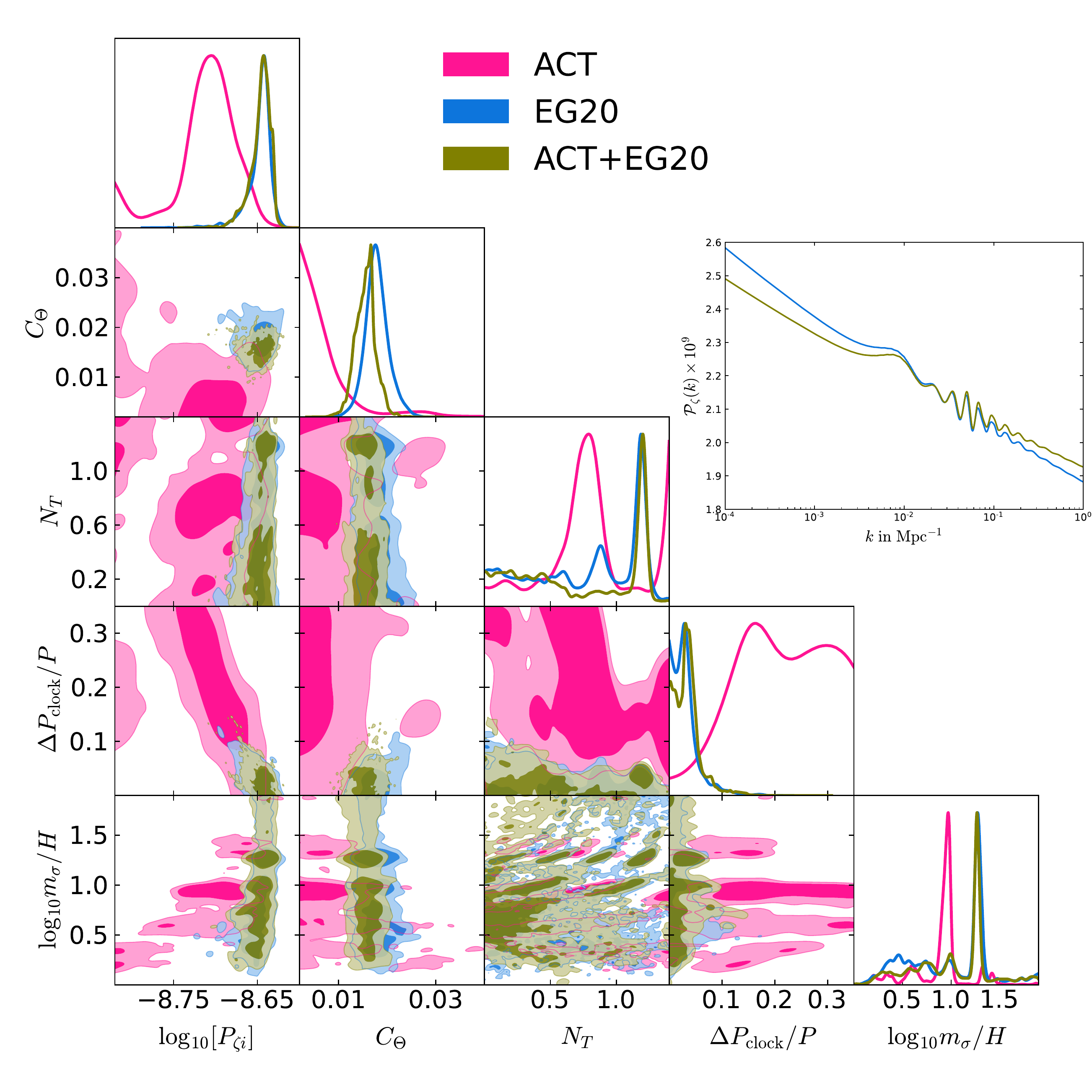}
	\caption{\footnotesize\label{fig:ACT}  Triangle plot for the restricted model for Atacama Cosmology Telescope (ACT, pink), Clean CamSpec v12.5HMcln (EG20, light blue) and their combination (olive). In the inset, we also show the best-fit for the combined dataset and compare it to the one for EG20. }
\end{figure*}

Our results are presented in Fig.~\ref{fig:ACT}. ACT data cannot place any sensible constrain on our model if they are not supplemented by other data at large scales. In particular, it is interesting to notice that the parameter governing the spectral index, i.e.~$C_\Theta$, is only bounded from above. It is well known that ACT data constrains the PPS to be scale invariant \cite{ACT:2020gnv}. However, adopting the prior in Table~\ref{tab:priors},  our model does not admit this possibility by construction since $n_s\simeq1-2 C_\Theta$ with $C_\Theta>0$. This leads to an evidence of $\ln B=+4.5$ in favor of the CPSC model over the featureless one (the one with $\Delta P_{\rm clock}/P=0$) because oscillatory features can improve the fit compared to the a red-tilted featureless spectrum, which is not preferred by ACT data. 

Constraints change when EG20 is added to ACT. As can be seen, now the overall PPS amplitude can be constrained owing to the inclusion of lowE data. Furthermore, constraints on $C_\Theta$ are very similar to those from EG20, though slightly shifted towards a less red tilt. However, as also mentioned in the main text, the addition of ACT does not significantly change constraints of the feature parameters, as can be seen from their posterior distributions. The reasons are the following. First, the bestfit feature signals that provide a better fit to Planck data significantly deviate from scale invariance  at $\ell\lesssim1500$. 
 Although from $900\lesssim\ell\lesssim1500$ the error bars on the EE power spectra are smaller than Planck ones, the ones on TT (and thus TE) are worse. Planck's TT and TE currently have the most constraining power on our model.  Secondly, as mentioned above, ACT data are binned, making it hard to use them to sizeably improve constraints on highly oscillatory feature signals.  As a result, the Bayes factor is a bit worse than the one for EG20 only, but still inconclusive, i.e. $\ln B=-1.80\pm0.38$.
 
 To see this point more clearly, let us discuss the bestfit candidate of the combined dataset. As shown by the inset in Fig.~\ref{fig:ACT}, the feature signal is the same as the restricted LFC presented in Table~\ref{tab:Nodipcandidate}. The total $\Delta\chi^2$ to data does not change much and is broken down into individual $\chi^2$'s as follows
 
 \begin{table}[h]
 	\centering
 	\begin{tabular}{|c|c|c|c|c|c|c|}
 		\hline
 		$\Delta \chi^2_{\rm TOT}$   & $\Delta\chi^2_\mathrm{prior}$    & $\Delta\chi^2_\mathrm{high-\ell}$  & $\Delta\chi^2_\mathrm{low-T}$ & $\Delta\chi^2_\mathrm{low-E}$ &$\Delta\chi^2_{\rm ACT} $\\  \hline  
 		14.0     & 0.63& 12.84      & 0.70 & -0.04 & -0.23  \\\hline 
 	\end{tabular}
 \end{table}
 The fit to ACT data is essentially as good as the baseline bestfit model, as the amplitude of the oscillations is very small at the multipoles tested by ACT. The slightly less red tilt degrades a bit the fit to high-$\ell$ and lowE Planck data, but the effect is compensated by the fit to lowT which is slightly improved.

\section{Comparison of different parameterizations}
\label{sec:LinPar}

	\begin{figure*}[t]
		\includegraphics[width=\columnwidth]{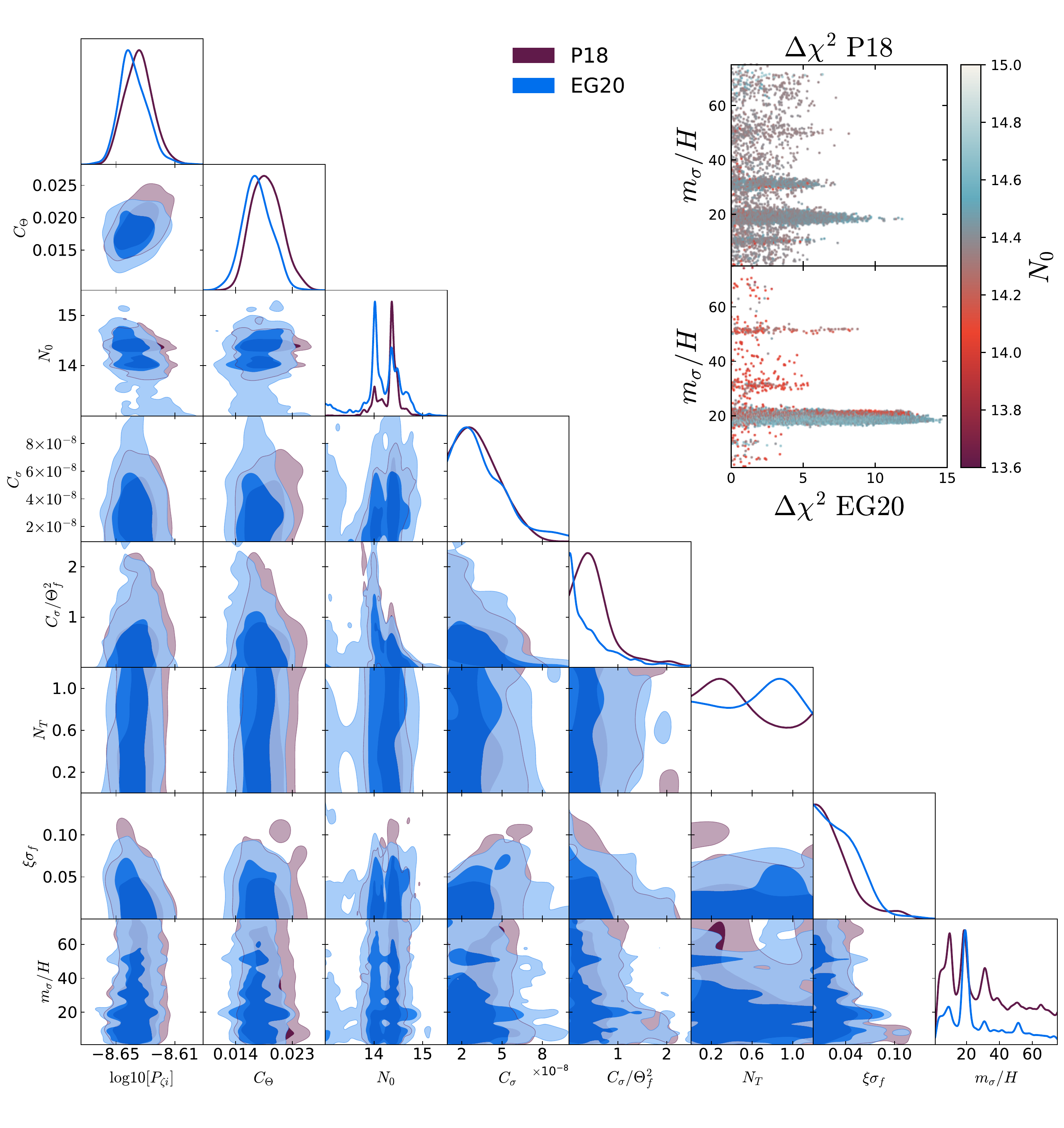}
		\caption{\footnotesize\label{fig:triangle} Same as Fig.~\ref{fig:triangle_log} but with a different choice of priors and parameterization described in App.~\ref{sec:LinPar}.}
	\end{figure*}

    \medskip
	\begin{table}
		\centering
		\begin{tabular}{|l|l|l|}
			\hline
			Dataset          & P18 (Plik bin1) & EG20 (CamSpec v12.5HMcln) \\ \hline
			$\ln B$     &      $-0.69\pm0.38$           & $-1.47     \pm0.36$        \\         \hline
		\end{tabular}
		\caption{Same as Table~\ref{tab:evidence_log} but with a different choice of priors and parameterization described in App.~\ref{sec:LinPar}.}
		\label{tab:evidence}
	\end{table}

	\begin{table}
		\centering
		\begin{tabular}{|l|l|l|}
			\hline
			Dataset          & P18 (Plik bin1) & EG20 (CamSpec v12.5HMcln) \\ \hline
			$\ln B$     &      $-0.38\pm0.38$           & $-0.42\pm0.35$         \\         \hline
		\end{tabular}
		\caption{Same as Table~\ref{tab:evidence_nodip_log} but with a different choice of priors and parameterization described in App.~\ref{sec:LinPar}.}
		\label{tab:evidence_nodip}
	\end{table}

	\begin{figure*}[t]
		\includegraphics[width=\columnwidth]{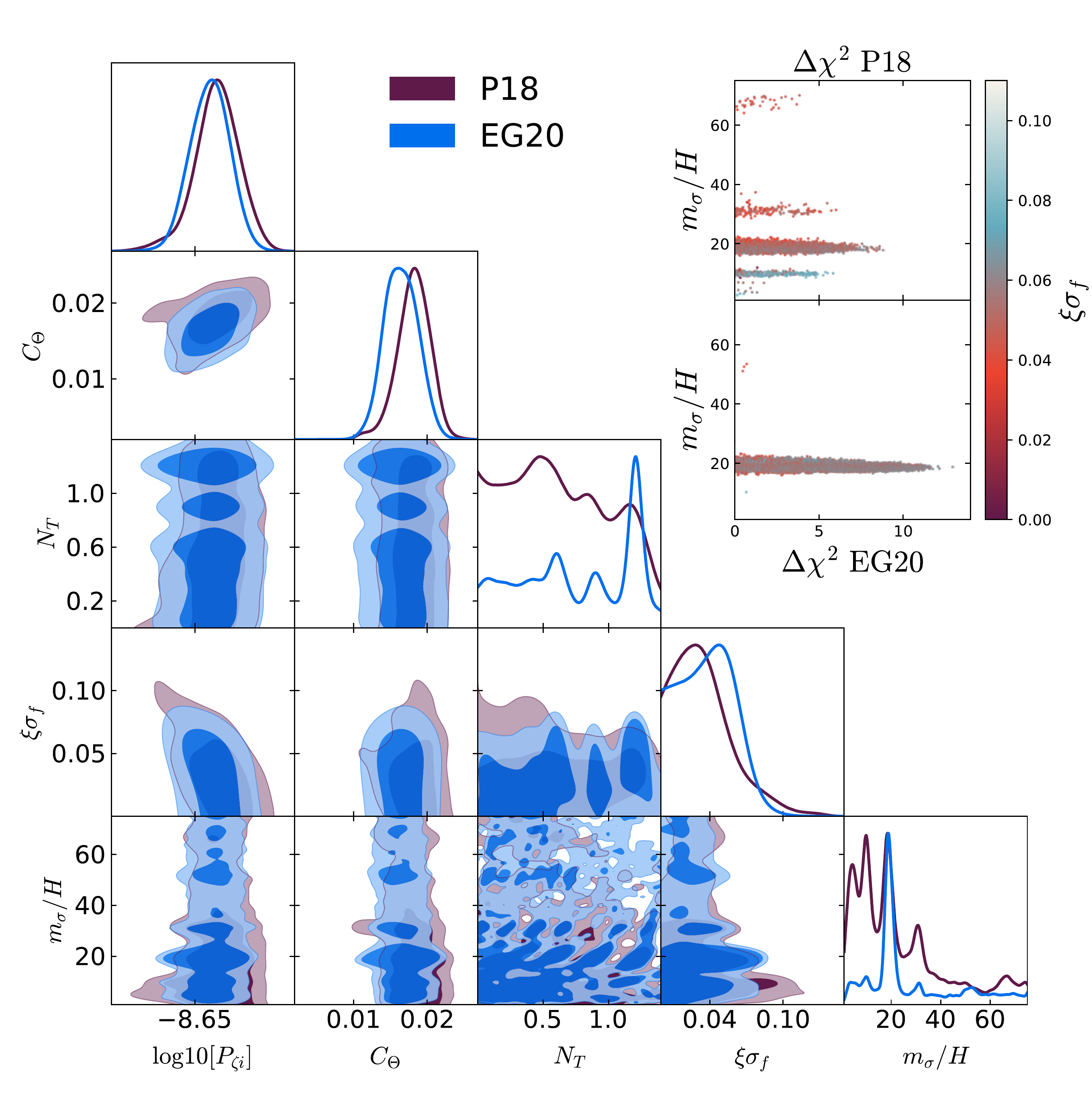}
		\caption{\footnotesize\label{fig:triangle_nodip} Same as Fig.~\ref{fig:triangle_log_3params} but with a different choice of priors and parameterization described in App.~\ref{sec:LinPar}.}
	\end{figure*}
	
	\begin{figure}
		\includegraphics[width=.9\columnwidth]{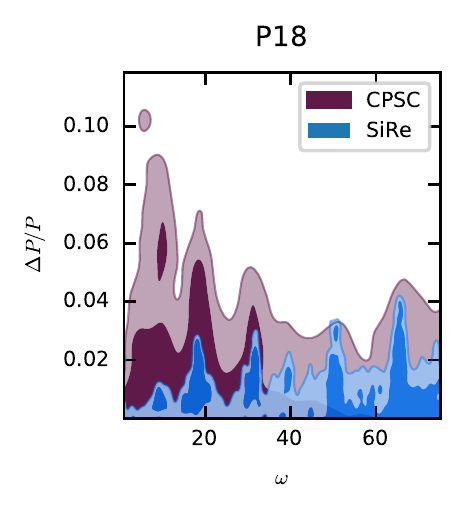}
		\includegraphics[width=.9\columnwidth]{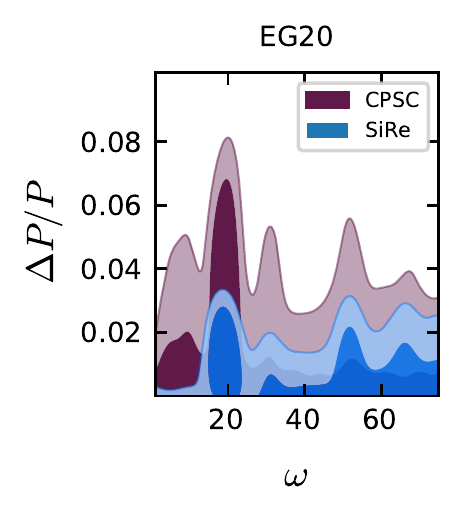}
		\caption{\label{fig:CPSC_vs_TEMPLATE_LIN} Same as Fig.~\ref{fig:CPSC_vs_TEMPLATE_LOG} but with a different choice of priors and parameterization described in App.~\ref{sec:LinPar}.}
	\end{figure}

In order to test the robustness of our results, in this Appendix we test our model using a different parameterization. We trade the parameters  $(\Delta P_{\rm dip}/P,\,\Delta P_{\rm clock}/P)$ for $(C_\sigma,\,\xi\sigma_f)$ respectively and assume a linear prior on $m_\sigma/H$ instead of a logarithmic one. Indeed, since the prior range for $m_\sigma/H$ is quite large and we have already observed that several frequencies are allowed by data, it is possible, at least in principle, that prominent peak around $m_\sigma/H\sim18$, which is one of the main results of this paper, is an artifact of the particular choice of the prior on the frequency parameter $m_\sigma/H$. It is therefore important to test the robustness of our results against a different choice of priors and parameterization.
We adopt the following priors on these parameters: $C_\sigma\times10^7\in[0.09,\,1]$,  $\xi\sigma_f\in[0.0,\,0.15]$ and $m_\sigma/H\in[1,\, 75]$. 
	
We show the results obtained using this parameterization in Fig.~\ref{fig:triangle}. The Bayes factors are given in Table~\ref{tab:evidence}. Note that, since the parameterization as well as the prior volume have changed, the Bayes factor is also expected to change. Nevertheless, within the errors, the Bayes factors for each data set are consistent with those quoted in Table~\ref{tab:evidence} and the model selection is inconclusive.

What is crucial to observe is that both posteriors again show  a peak around $m_\sigma/H\sim18$, which confirms LFCII as our best-fit candidate, no matter what prior is adopted in the analysis. This is also corroborated by the sample plots in Fig.~\ref{fig:triangle}, that visually show that the better likelihood samples  concentrate around $m_\sigma/H\sim18$ for both datasets. Since the linear prior gives more weight to large values of $m_\sigma/H$ we see that high frequencies are now better explored by the nested sampling, causing the sample plots to be more populated in the high frequency region. Overall, the constraints are consistent with our findings in the main text, confirming the robustness of our analysis. 
	
For completeness, we also quote the constraints on the restricted CPSC model and on the SiRe model in Fig.~\ref{fig:triangle_nodip} and \ref{fig:CPSC_vs_TEMPLATE_LIN} as well as quote the evidence for the restricted model using this parameterization in Table~\ref{tab:evidence_nodip}. Also in this case, the results are consistent with the main text.

\end{document}